\documentclass[useAMS,usenatbib,breaklinks=true]{mnras}
\usepackage{graphicx}
\usepackage{times}
\usepackage{mn2e-breakabs}
\usepackage{amssymb}
\usepackage{amsmath}
\usepackage[dvipsnames]{xcolor}
\usepackage{comment}

\newcommand\LCDM{$\Lambda$CDM}
\newcommand\Mpc{\, \mathrm{Mpc}}
\newcommand\Mpch{\, \mathrm{Mpc}/h}
\newcommand\Mpchc{\, (\mathrm{Mpc}/h)^{3}}
\newcommand\hMpc{\, h/\mathrm{Mpc}}
\newcommand\hMpcc{\, (h/\mathrm{Mpc})^{3}}
\newcommand\Gpch{\, \mathrm{Gpc}/h}
\newcommand\Gpchc{\, (\mathrm{Gpc}/h)^{3}}
\newcommand\eV{\, \mathrm{eV}}
\newcommand\kmps{\, \mathrm{km}/\mathrm{s}}
\newcommand\apar{\alpha_{\parallel}}
\newcommand\aper{\alpha_{\perp}}
\newcommand\sig{\sigma_{8}}
\newcommand\fsig{f\sig}
\newcommand\DHu{D_{\mathrm{H}}}
\newcommand\DM{D_{\mathrm{M}}}
\newcommand\DV{D_{\mathrm{V}}}
\newcommand\fid{\mathrm{fid}}
\newcommand\rdrag{r_{\mathrm{drag}}}
\newcommand\Leg[1]{\mathcal{L}_{#1}}
\newcommand\eff{\mathrm{eff}}
\newcommand\zeff{z_{\eff}}
\newcommand\vc[1]{\mathbf{#1}}
\newcommand\vk{\vc{k}}
\newcommand\vr{\vc{r}}
\newcommand\veta{\vc{\eta}}

\newcommand\aver[1]{\left\langle#1\right\rangle}
\newcommand\wsys{w_{\mathrm{sys,i}}}
\newcommand\wcp{w_{\mathrm{cp,i}}}
\newcommand\wnoz{w_{\mathrm{noz,i}}}
\newcommand\wcomp{w_{\mathrm{comp,i}}}
\newcommand\wfkp{w_{\mathrm{FKP,i}}}
\newcommand\wtot{w_{\mathrm{tot},i}}
\newcommand\chunkz{\texttt{chunk\_z}}
\DeclareMathOperator{\median}{median}

\newcommand\Eq[1]{Eq.~\eqref{eq:#1}}
\newcommand\Fig[1]{Figure~\ref{fig:#1}}
\newcommand\Tab[1]{Table~\ref{tab:#1}}
\newcommand\Sec[1]{Section~\ref{sec:#1}}
\newcommand\App[1]{Appendix~\ref{app:#1}}

\usepackage{hyperref}
\hypersetup{
    colorlinks=true,
    linkcolor=blue,
    citecolor=blue,
}

\begin{document}

\title[DR16 eBOSS ELG BAO and RSD measurements] 
{The Completed SDSS-IV extended Baryon Oscillation Spectroscopic Survey: measurement of the BAO and growth rate of structure of the emission line galaxy sample from the anisotropic power spectrum between redshift 0.6 and 1.1}

\author[A. de~Mattia et al.]{\parbox{\textwidth}{
Arnaud de~Mattia\thanks{Email: arnaud.de-mattia@cea.fr}$^{1}$,
Vanina Ruhlmann-Kleider$^{1}$,
Anand Raichoor$^{2}$,
Ashley J. Ross$^{3}$,
Am\'elie Tamone$^{2}$,
Cheng Zhao$^{2}$,
Shadab Alam$^{4}$,
Santiago Avila$^{5,6}$,
Etienne Burtin$^{1}$,
Julian Bautista$^{7}$,
Florian Beutler$^{4}$,
Jonathan Brinkmann$^{8}$,
Joel R. Brownstein$^{9}$,
Michael J. Chapman$^{10,11}$,
Chia-Hsun Chuang$^{12}$,
Johan Comparat$^{13}$,
H\'elion du~Mas~des~Bourboux$^{9}$,
Kyle S. Dawson$^{9}$,
Axel de la Macorra$^{14}$,
H\'ector Gil-Mar\'in$^{15,16}$,
Violeta Gonzalez-Perez$^{17}$,
Claudio Gorgoni$^{2}$,
Jiamin Hou$^{13}$,
Hui Kong$^{3}$,
Sicheng Lin$^{18}$,
Seshadri Nadathur$^{7}$,
Jeffrey A. Newman$^{19}$,
Eva-Maria Mueller$^{20}$,
Will J. Percival$^{10,11,21}$,
Mehdi Rezaie$^{22}$,
Graziano Rossi$^{23}$,
Donald P. Schneider$^{24}$,
Prabhakar Tiwari$^{25}$,
M. Vivek$^{26}$,
Yuting Wang$^{25}$,
Gong-Bo Zhao$^{25,27}$} \vspace*{4pt} \\ 
\scriptsize $^{1}$ IRFU, CEA, Universit\'e Paris-Saclay, F-91191 Gif-sur-Yvette, France\vspace*{-2pt} \\ 
\scriptsize $^{2}$ Institute of Physics, Laboratory of Astrophysics, \'Ecole Polytechnique F\'ed\'erale de Lausanne (EPFL), Observatoire de Sauverny, 1290 Versoix, Switzerland\vspace*{-2pt} \\ 
\scriptsize $^{3}$ Center for Cosmology and Astro-Particle Physics, Ohio State University, Columbus, OH 43210, USA\vspace*{-2pt} \\ 
\scriptsize $^{4}$ Institute for Astronomy, University of Edinburgh, Royal Observatory, Blackford Hill, Edinburgh, EH9 3HJ, UK\vspace*{-2pt} \\ 
\scriptsize $^{5}$ Departamento de F\'isica Te\'orica M8, Universidad Aut\'onoma de Madrid, E-28049 Cantoblanco, Madrid, Spain\vspace*{-2pt} \\ 
\scriptsize $^{6}$ Instituto de Fisica Teorica UAM/CSIC, Universidad Autonoma de Madrid, 28049 Madrid, Spain\vspace*{-2pt} \\ 
\scriptsize $^{7}$ Institute of Cosmology \& Gravitation, Dennis Sciama Building, University of Portsmouth, Portsmouth, PO1 3FX, UK\vspace*{-2pt} \\ 
\scriptsize $^{8}$ Apache Point Observatory, P.O. Box 59, Sunspot, NM 88349, USA\vspace*{-2pt} \\ 
\scriptsize $^{9}$ Department of Physics and Astronomy, University of Utah, 115 S 1400 E, Salt Lake City, UT 84112, USA\vspace*{-2pt} \\ 
\scriptsize $^{10}$ Waterloo Centre for Astrophysics, University of Waterloo, Waterloo, ON N2L 3G1, Canada\vspace*{-2pt} \\ 
\scriptsize $^{11}$ Department of Physics and Astronomy, University of Waterloo, Waterloo, ON N2L 3G1, Canada\vspace*{-2pt} \\ 
\scriptsize $^{12}$ Kavli Institute for Particle Astrophysics and Cosmology, Stanford University, 452 Lomita Mall, Stanford, CA 94305, USA\vspace*{-2pt} \\ 
\scriptsize $^{13}$ Max-Planck-Institut f\"ur Extraterrestrische Physik, Postfach 1312, Giessenbachstr., 85748 Garching bei M\"unchen, Germany\vspace*{-2pt} \\ 
\scriptsize $^{14}$ Instituto de F\'isica, Universidad Nacional Aut\'onoma de M\'exico, Apdo. Postal 20-364 Ciudad de M\'exico, M\'exico\vspace*{-2pt} \\ 
\scriptsize $^{15}$ Institut de Ci\`encies del Cosmos, Universitat de Barcelona, ICCUB, Mart\'i i Franqu\`es 1, E08028 Barcelona, Spain\vspace*{-2pt} \\ 
\scriptsize $^{16}$ Institut  d’Estudis  Espacials  de  Catalunya  (IEEC),  E08034  Barcelona,  Spain\vspace*{-2pt} \\ 
\scriptsize $^{17}$ Astrophysics Research Institute, Liverpool John Moores University, 146 Brownlow Hill, Liverpool L3 5RF, UK\vspace*{-2pt} \\ 
\scriptsize $^{18}$ Center for Cosmology and Particle Physics, Department of Physics, New York University, 726 Broadway, New York, NY 10003, USA\vspace*{-2pt} \\ 
\scriptsize $^{19}$ PITT PACC, Department of Physics and Astronomy, University of Pittsburgh, Pittsburgh, PA 15260, USA\vspace*{-2pt} \\ 
\scriptsize $^{20}$ Sub-department of Astrophysics, Department of Physics, University of Oxford, Denys Wilkinson Building, Keble Road, Oxford OX1 3RH, UK\vspace*{-2pt} \\ 
\scriptsize $^{21}$ Perimeter Institute for Theoretical Physics, 31 Caroline St. North, Waterloo, ON N2L 2Y5, Canada\vspace*{-2pt} \\ 
\scriptsize $^{22}$ Department of Physics and Astronomy, Ohio University, 251B Clippinger Labs, Athens, OH 45701, USA\vspace*{-2pt} \\ 
\scriptsize $^{23}$ Department of Physics and Astronomy, Sejong University, Seoul 143-747, Korea\vspace*{-2pt} \\ 
\scriptsize $^{24}$ Institute for Gravitation and the Cosmos, The Pennsylvania State University, University Park, PA 16802, USA\vspace*{-2pt} \\ 
\scriptsize $^{25}$ National Astronomy Observatories, Chinese Academy of Science, Beijing, 100101, P.R. China\vspace*{-2pt} \\ 
\scriptsize $^{26}$ Indian Institute of Astrophysics, Koramangala, Bangalore 560034, India\vspace*{-2pt} \\ 
\scriptsize $^{27}$ School of Astronomy and Space Science, University of Chinese Academy of Sciences, Beijing 100049, P.R. China\vspace*{-2pt} \\ 
}

\date{To be submitted to MNRAS} 

\pagerange{\pageref{firstpage}--\pageref{lastpage}} \pubyear{2020}
\maketitle
\label{firstpage}

\begin{abstract}
We analyse the large-scale clustering in Fourier space of emission line galaxies (ELG) from the Data Release 16 of the Sloan Digital Sky Survey IV extended Baryon Oscillation Spectroscopic Survey. The ELG sample contains 173,736 galaxies covering 1,170 square degrees in the redshift range $0.6 < z < 1.1$. We perform a BAO measurement from the post-reconstruction power spectrum monopole, and study redshift space distortions (RSD) in the first three even multipoles. Photometric variations yield fluctuations of both the angular and radial survey selection functions. Those are directly inferred from data, imposing integral constraints which we model consistently. The full data set has only a weak preference for a BAO feature ($1.4\sigma$). At the effective redshift $z_{\rm eff} = 0.845$ we measure $D_{\rm V}(z_{\rm eff})/r_{\rm drag} = 18.33_{-0.62}^{+0.57}$, with $D_{\rm V}$ the volume-averaged distance and $r_{\rm drag}$ the comoving sound horizon at the drag epoch. In combination with the RSD measurement, at $z_{\rm eff} = 0.85$ we find $f\sigma_8(z_{\rm eff}) = 0.289_{-0.096}^{+0.085}$, with $f$ the growth rate of structure and $\sigma_8$ the normalisation of the linear power spectrum, $D_{\rm H}(z_{\rm eff})/r_{\rm drag} = 20.0_{-2.2}^{+2.4}$ and $D_{\rm M}(z_{\rm eff})/r_{\rm drag} = 19.17 \pm 0.99$ with $D_{\rm H}$ and $D_{\rm M}$ the Hubble and comoving angular distances, respectively. These results are in agreement with those obtained in configuration space, thus allowing a consensus measurement of $f\sigma_8(z_{\rm eff}) = 0.315 \pm 0.095$, $D_{\rm H}(z_{\rm eff})/r_{\rm drag} = 19.6_{-2.1}^{+2.2}$ and $D_{\rm M}(z_{\rm eff})/r_{\rm drag} = 19.5 \pm 1.0$. This measurement is consistent with a flat $\Lambda$CDM model with Planck parameters.
\end{abstract}


\newpage
\section{Introduction}
\label{sec:introduction}

Why the Universe expansion is accelerating has been one of the most pressing questions of cosmology in the last two decades. The Universe expansion history is most naturally probed through the properties of the large scale structure. In particular, the distribution of galaxies as measured by spectroscopic redshift surveys can be studied through two types of clustering analyses, which we carry out in this paper. The first type of analysis relies on the baryon acoustic oscillation (BAO) feature~\citep{Eisenstein1998:astro-ph/9709112v1} to measure the distance-redshift relation. The second type of analysis is based on galaxy peculiar velocities. Indeed, redshifts of galaxies are not only due to Hubble expansion but also depend on their peculiar velocities. Thus, converting redshifts into comoving distances assuming only the former contribution leads to galaxy coordinates being biased along the line-of-sight~\citep{Kaiser1987}. As peculiar velocities trace the gravitational potential field due to matter, these so-called redshift space distortions (RSD) make clustering measurements a way to test gravity and to measure the matter content of the Universe.

In this work we study the clustering properties of the emission line galaxy (ELG) sample, which is part of the extended Baryon Oscillation Spectroscopic Survey (eBOSS,~\citealt{Dawson2016:1508.04473v2}) Data Release~16 (DR16,~\citealt{Ahumada2019:1912.02905v1}) of the Sloan Digital Sky Survey~IV~\citep{Blanton2017:1703.00052v2}. ELG spectra were collected by the BOSS (Baryon Oscillation Spectroscopic Survey) spectrograph~\citep{Smee2013:1208.2233v2} located at Apache Point Observatory~\citep{Gunn2006:astro-ph/0602326v1}, New Mexico.

This study is part of a coordinated release of the final eBOSS measurements, also including BAO and RSD in the clustering of luminous red galaxies~\citep{GilMarin2020,Bautista2020} and quasars \citep{Hou2020,Neveux2020}. An essential component of these studies is the construction of data catalogues \citep{Ross2020,Lyke2020}, mock catalogues \citep{Lin2020,Zhao2020}, and N-body simulations for assessing systematic errors \citep{Alam2020,Avila2020,Rossi2020,Smith2020}. At the highest redshifts ($z>2.1$), the coordinated release of final eBOSS measurements includes measurements of BAO in the Lyman-$\alpha$ forest \citep{duMasDesBourboux2020}. The cosmological interpretation of these results in combination with the final BOSS results and other probes is found in \citet{eBOSS2020}\footnote{A summary of all SDSS BAO and RSD measurements with accompanying legacy figures can be found at \url{https://sdss.org/science/final-bao-and-rsd-measurements/}. The full cosmological interpretation of these measurements can be found at \url{https://sdss.org/science/cosmology-results-from-eboss/}.}. Multi-tracer analyses to measure BAO and RSD using LRG and ELG samples are presented in \citet{Wang2020,ZhaoGB2020}.

Star-forming ELGs are an interesting tracer for clustering analyses. Indeed, the star formation rate increases up to $z \sim 2$, where red galaxies are rarer. Also, the strong emission lines, such as H$\alpha$ or the [OII] doublet, ease the redshift measurement --- thus allowing reduced spectroscopic observing time. The eBOSS ELG sample, with $173,736$ galaxies distributed in the redshift range $0.6 < z < 1.1$ (effective redshift of $\zeff = 0.845$), is the largest and highest redshift ELG spectroscopic sample ever assembled and the third one to be used for cosmology~\citep{Blake2011:1104.2948v1,Contreras2013:1302.5178v1,Okumura2016:1511.08083v2}. The eBOSS ELG sample has already been used to study the evolution of star forming galaxies \citep{Guo2019:1810.05318v3} and the circumgalactic medium of ELGs \citep{Lan2018:1806.05786v2}. Details about the eBOSS ELG target selection can be found in~\citet{Raichoor2016:1505.01797v2,Raichoor2017:1704.00338v1}. The DR16 ELG clustering catalogue is described in~\citet{Raichoor2020}, which also includes a measurement of the isotropic BAO feature in configuration space. \citet{Tamone2020} presents the RSD analysis in configuration space.

In what follows, we present and analyse the clustering of galaxies of the DR16 ELG catalogue in Fourier space. We perform a RSD measurement using the observed galaxy density field, and an isotropic BAO measurement after reconstructing the density field to remove non-linear damping of the BAO signal.

The RSD analysis of the observed galaxy power spectrum allows joint constraints to be derived %
on the product $f(\zeff)\sig(\zeff)$, and ratios $\DHu(\zeff)/\rdrag$ and $\DM(\zeff)/\rdrag$,
at the effective redshift of the sample $\zeff$. In these combinations, $f(z)$ is the logarithmic derivative of the linear growth factor with respect to the scale factor $a = 1/\left(1+z\right)$ (hereafter referred to as the growth rate of structure) and $\sig(z)$ is the amplitude of the linear matter power spectrum measured in spheres of radius $8 \Mpch$ at redshift $z$. We also use $\DHu(z) = c/H(z)$, the Hubble distance related to the Hubble expansion rate $H(z)$, $\DM = (1+z)D_{\mathrm{A}}(z)$, the comoving angular diameter distance related to the proper angular diameter distance $D_{\mathrm{A}}$ %
and $\rdrag$ the comoving sound horizon when the baryon-drag optical depth equals unity. The BAO only analysis is sensitive to $\DM/\rdrag$, $\DHu/\rdrag$ or a combination of both terms. In this work, we measure the ratio $\DV/\rdrag$ with the volume-averaged distance $\DV(z) = \left(\DM^{2}(z)\DHu(z)z\right)^{1/3}$.

 The RSD model we use is based on state-of-the-art two-loop order perturbation theory to ensure reliable modelling of the power spectrum up to mildly non-linear scales. Besides the standard survey window function, we also model the radial integral constraint generated by the survey radial selection function being estimated from observed data. %
 We carefully review potential sources of systematic errors, and apply correction schemes after validation based on mock catalogues. After mitigation of systematic effects, we measure the first three even Legendre multipoles of the pre-reconstruction power spectrum and the post-reconstruction monopole and compare these measurements with model predictions to derive cosmological constraints. We combine isotropic BAO and RSD analyses at the likelihood level in order to release the Gaussian assumption on the posteriors and strengthen our cosmological measurement. 

The paper is organised as follows. \Sec{power_spectrum_estimation} presents the power spectrum estimators used to compute multipoles in a periodic box and within a real survey geometry. All components of the power spectrum RSD and BAO models are described in \Sec{model} and the fitting methodology is introduced in \Sec{fitting_methodology}. Model validation against N-body simulations is detailed in \Sec{mock_challenge}. \Sec{data_mocks} briefly describes the eBOSS DR16 ELG sample and the adopted scheme to correct for known systematic effects, as well as approximate mocks used to test this procedure. We show the impact of residual systematics on clustering measurements, and introduce techniques to mitigate them in \Sec{mocks_systematics}. 
Cosmological fits and their implications are presented in \Sec{results}. These measurements are combined with configuration space results of~\citet{Tamone2020} in \Sec{consensus}.
We conclude in \Sec{conclusions}.

\section{Power spectrum estimation}
\label{sec:power_spectrum_estimation}

In this section we detail our measurements of the power spectrum multipoles of the galaxy density field in a periodic box (used in \Sec{mock_challenge_outerrim}) and within a real, sky-cut geometry (used in \Sec{data_mocks} and beyond).

\subsection{Periodic box}
\label{sec:power_spectrum_estimation_box}

We define the density contrast:
\begin{equation}
\delta_{g}(\vr) = \frac{n_{g}(\vr)}{\bar{n}_{g}} - 1
\label{eq:density_contrast}
\end{equation}
where $n_{g}(\vr)$ is the galaxy density at 
comoving position $\vr$, and $\bar{n}_{g}$ its average over the whole box of volume $V$. Taking the Fourier transform $\delta_{g}(\vk)$ of this field, power spectrum multipoles are calculated as:
\begin{equation}
P_{\ell}(k) = \frac{2\ell+1}{V} \int \frac{d\Omega_{k}}{4\pi} \delta_{g}(\vk)\delta_{g}(-\vk)\Leg{\ell}(\hat{\vk} \cdot \hat{\veta}) - P_{\ell}^{\mathrm{noise}}(k)
\label{eq:power_spectrum_estimator_box}
\end{equation}
$\Leg{\ell}$ being the Legendre polynomial of order $\ell$ and $\hat{\veta}$ the unit global line-of-sight $\veta$ vector, which we choose to be one axis of the box. The shot noise term is non-zero for the monopole only:
\begin{equation}
P_{0}^{\mathrm{noise}} = \frac{1}{\bar{n}_{g}}.
\end{equation}

We use the implementation of the periodic box power spectrum estimator in the Python toolkit \texttt{nbodykit}~\citep{Hand2017:1712.05834v1}. The density contrast field $\delta_{g}(\vr)$ in \Eq{density_contrast} is interpolated on a $512^{3}$ mesh following the triangular shaped cloud (TSC) scheme. In the following (see \Sec{mock_challenge}), the box size is $3000 \Mpch$ and thus the Nyquist frequency is $k_{N} \simeq 0.5 \hMpc$, more than twice larger than the maximum wavenumber used in the RSD analysis ($k = 0.2 \hMpc$). We checked that using a $700^{3}$ mesh ($k_{N} \simeq 0.7 \hMpc$) does not change our measurement in a detectable way. Then, the term $\delta_{g}(\vk)$ in \Eq{power_spectrum_estimator_box} is calculated with a fast Fourier transform (FFT) of the interpolated density contrast and the interlacing technique of~\citet{Sefusatti2015:1512.07295v2} is used to mitigate aliasing effects. 

The integral over the solid angle $d\Omega_{k}$ in \Eq{power_spectrum_estimator_box} is performed in spherical shells of $\Delta k = 0.01 \hMpc$, from $k = 0 \hMpc$. The discrete $k$-space grid makes the angular mode distribution irregular at large scales, an effect which we account for in the model (see \Sec{model_irregular_mu}).

\subsection{Real survey geometry}
\label{sec:power_spectrum_estimation_survey}

Following~\cite{Feldman1993:astro-ph/9304022v1} (FKP) the power spectrum estimator of~\citet{Yamamoto2005:astro-ph/0505115v2} makes use of the FKP field:
\begin{equation}
F(\vr) = n_{g}(\vr) - \alpha_{s} n_{s}(\vr)
\label{eq:fkp_field}
\end{equation}
where $n_{g}(\vr)$ and $n_{s}(\vr)$ denote the density of observed and random galaxies, respectively, at %
comoving position $\vr$. Random galaxies come from a Poisson-sampled synthetic catalogue accounting for the survey selection function.
Observed and random galaxy densities include weights, i.e. corrections for systematics effects and FKP weights~\citep{Feldman1993:astro-ph/9304022v1}. The scaling $\alpha_{s}$ is defined by:
\begin{equation}
\alpha_{s} = \frac{\sum_{i=1}^{N_{g}} w_{g,i}}{\sum_{i=1}^{N_{s}} w_{s,i}}.
\label{eq:alpha_global}
\end{equation}
with $N_{g}$, $N_{s}$ and $w_{g}$, $w_{s}$ the number and weights of observed and random galaxies, respectively. Then, the power spectrum multipoles are given by~\citet{Bianchi2015:1505.05341v2}:
\begin{equation}
P_{\ell}(k) = \frac{2\ell+1}{I} \int \frac{d\Omega_{k}}{4\pi} F_{0}(\vk)F_{\ell}(-\vk) - P_{\ell}^{\mathrm{noise}}(k)
\label{eq:power_spectrum_estimator_survey}
\end{equation}
with:
\begin{equation}
F_{\ell}(\vk) = \int d^{3} r F(\vr) \Leg{\ell}(\hat{\vk} \cdot \hat{\veta}) e^{i \vk \cdot \vr},
\label{eq:fourier_fkp_field}
\end{equation}
following the same notations as in \Sec{power_spectrum_estimation_box}. In this estimator, we take for line-of-sight $\hat{\veta} = \hat{\vr}$ where $\hat{\vr}$ is the direction to the second galaxy of the pair. This approximation introduces wide-angle effects in the power spectrum multipoles as well as the associated window function. However, these effects have been shown to not impact current RSD and BAO studies significantly~\citep{Castorina2018:1709.09730v2,Beutler2019:1810.05051v3}.

The normalisation term $I$ is given by:
\begin{equation}
I = \alpha_{s} \sum_{i=1}^{N_{s}} w_{s,i} n_{g,i}.
\label{eq:power_spectrum_normalisation}
\end{equation}
and the shot noise, which only contributes to the monopole in a scale-independent way, is:
\begin{equation}
P_{0}^{\mathrm{noise}} = \frac{1}{I} \left[\sum_{i=1}^{N_{g}} w_{g,i}^{2} + \alpha_{s}^{2} \sum_{i=1}^{N_{s}} w_{s,i}^{2} \right]
\label{eq:power_spectrum_shot_noise}
\end{equation}
For the density $n_{g,i}$, we take the redshift density $n(z)$, computed by binning (weighted) data in redshift slices of $\Delta z = 0.005$ (see \Sec{data} and~\citealt{Raichoor2020}).

We use the implementation of the Yamamoto estimator in the Python toolkit \texttt{nbodykit}~\citep{Hand2017:1712.05834v1} to measure power spectra. Again, the FKP field~\eqref{eq:fkp_field} is interpolated on a $512^{2}$ mesh with the TSC scheme. Terms $F_{\ell}(\vk)$ of \Eq{fourier_fkp_field} are calculated with FFTs and the interlacing technique of~\citet{Sefusatti2015:1512.07295v2} is used to mitigate aliasing effects. Here we use a box size of $4000 \Mpch$, so the Nyquist frequency is $k_{N} \simeq 0.4 \hMpc$. We again checked that using a $700^{3}$ mesh ($k_{N} \simeq 0.55 \hMpc$) does not change our measurement significantly.

The integral over the solid angle $d\Omega_{k}$ in \Eq{power_spectrum_estimator_survey} is also performed in spherical shells of $\Delta k = 0.01 \hMpc$, starting from $k = 0 \hMpc$, unless otherwise stated. 

\subsection{Fiducial cosmology}
\label{sec:fiducial_cosmology}

To obtain the FKP field as a function of comoving position~$\vr$ we turn galaxy redshifts into distances assuming a fiducial cosmology.
This fiducial cosmology will be also used (unless otherwise stated) to compute the linear matter power spectrum for the RSD and BAO analyses in \Sec{model}. For both purposes, we utilised the same fiducial cosmology as in BOSS DR12 analyses~\citep{Alam2017:1607.03155v1}\footnote{Note that only $\Omega_{m}$ (and $\Omega_{\Lambda}$) matter for the redshift to comoving distance (in $\Mpch$) conversion.}:
\begin{equation}
\begin{split}
h = 0.676, \,\, \Omega_{m} = 0.31, \,\, \Omega_{\Lambda} = 0.69, \,\, \Omega_{b}h^{2} = 0.022,\\
\sigma_{8} = 0.80, \,\, n_{s} = 0.97, \,\, \sum m_{\nu} = 0.06 \eV. \qquad
\end{split}
\label{eq:fiducial_cosmology}
\end{equation}
Within this fiducial cosmology, that will be used throughout this paper (unless otherwise stated), $\rdrag^{\fid} = 147.77\Mpc$.

\section{Model}
\label{sec:model}

In this section we review the different ingredients of the RSD model (Sections~\ref{sec:model_tns} to~\ref{sec:model_integral_constraints}) and the isotropic BAO template (\Sec{model_isotropic_bao}), which will be used throughout this paper.

\subsection{Redshift space distortions}
\label{sec:model_tns}

The RSD model we use follows closely that of~\citet{Taruya2010:1006.0699v1,Taruya2013:1301.3624v1}, hereafter referred to as the TNS model, as used in~\citet{Beutler2016:1607.03150v1}. The redshift-space %
galaxy power spectrum is expressed as a function of $k$ the norm of the wavenumber $\vc{k}$ and its cosine angle to the line-of-sight $\mu$:
\begin{align}
P_{\mathrm{g}}(k,\mu) &= D_{\mathrm{FoG}}(k,\mu,\sigma_{v}) \left[P_{\mathrm{g},\delta\delta}(k) + 2f\mu^{2}P_{\mathrm{g},\delta\theta}(k) \right. \nonumber \\
& \left. + f^{2}\mu^{4}P_{\theta\theta}(k) + b_{1}^{3}A(k,\mu,\beta) + b_{1}^{4}B(k,\mu,\beta)\right],
\label{eq:power_galaxy_rsd}
\end{align}
with $\beta = f/b_{1}$ and $b_{1}$ is the linear bias. We adopt a Lorentzian form for the Finger-of-God effect~\citep{Jackson1972,Cole1995:astro-ph/9412062v1}: 
\begin{equation}
D_{\mathrm{FoG}}(k,\mu,\sigma_{v}) = \left[1+\frac{\left(k\mu\sigma_{v}\right)^{2}}{2}\right]^{-2},
\end{equation}
with $\sigma_{v}$ the velocity dispersion.

Galaxy-galaxy and galaxy-velocity power spectra $P_{\mathrm{g},\delta\delta}(k)$ and $P_{\mathrm{g},\delta\theta}(k)$ are given by:
\begin{align}
P_{\mathrm{g},\delta\delta}(k) & = b_{1}^{2}P_{\delta\delta}(k) + 2b_{2}b_{1} P_{b2,\delta}(k) + 2b_{s2}b_{1}P_{bs2,\delta}(k) \nonumber \\
& + 2b_{3\mathrm{nl}}b_{1}\sigma_{3}^2(k)P_{\mathrm{m}}^{\mathrm{lin}}(k) + b_{2}^{2}P_{b22}(k) \nonumber \\
& + 2b_{2}b_{s2}P_{b2s2}(k) + b_{s2}^{2}P_{bs22}(k) + N_{g},
\label{eq:power_galaxy_galaxy}
\end{align}
and:
\begin{align}
P_{\mathrm{g},\delta\theta}(k) &= b_{1}P_{\delta\theta}(k) + b_{2}P_{b2,\theta}(k) \nonumber \\
& + b_{s2}P_{bs2,\theta}(k) + b_{3\mathrm{nl}}\sigma_{3}^{2}(k)P_{\mathrm{m}}^{\mathrm{lin}}(k),
\label{eq:power_galaxy_velocity}
\end{align}
where $1$-loop bias terms $P_{b2,\delta}(k)$, $P_{bs2,\delta}(k)$, $\sigma_{3}^2(k)$, $P_{b2,\theta}(k)$ and $P_{bs2,\theta}(k)$ are provided in~\citet{McDonald2009:0902.0991v1,Beutler2016:1607.03150v1}. In this paper, power spectra $P_{\delta\delta}(k)$, $P_{\delta\theta}(k)$ and $P_{\theta\theta}(k)$, as well as RSD correction terms $A(k,\mu,\beta)$ and $B(k,\mu,\beta)$, 
are calculated at $2$-loop order, following the \texttt{RegPT} scheme~\citep{Taruya2012:1208.1191}.
We compute the linear matter power spectrum $P_{\mathrm{m}}^{\mathrm{lin}}(k)$ in the fiducial cosmology~\eqref{eq:fiducial_cosmology} (except otherwise stated) with the Boltzmann code \texttt{CLASS}~\citep{Class2011:1104.2933v3} and keep it fixed in the cosmological fits.

Second and third order non-local biases $b_{s2}$ and $b_{3\mathrm{nl}}$ are fixed assuming local Lagrangian bias~\citep{Chan2012:1201.3614v2,Baldauf2012:1201.4827v1,Saito2014:1405.1447v4}: %
\begin{align}
b_{s2} &= -\frac{4}{7}\left(b_{1} - 1\right), \\
b_{3\mathrm{nl}} &= \frac{32}{315}\left(b_{1} - 1\right).
\end{align}

\subsection{The distance-redshift relationship}
\label{sec:model_ap_test}

The fiducial cosmology used to turn angular positions and redshifts into distances may differ from the underlying cosmology of the data (or mock) galaxy sample. This leads to distortions in the $(k,\mu)$ space which can be detected through the so-called Alcock-Paczynski (AP) test~\citep{Alcock1979}. We define the scaling parameters $(\apar,\aper)$ to relate the observed radial and transverse wavenumbers $(k_{\parallel},k_{\perp})$ to the true ones $(k_{\parallel}^{\prime},k_{\perp}^{\prime}) = (k_{\parallel}/\apar,k_{\perp}/\aper)$. In the $(k,\mu) = (\sqrt{k_{\parallel}^{2}+k_{\perp}^{2}},k_{\parallel}/k)$ space, the corresponding mapping $(k,\mu) \rightarrow (k^{\prime},\mu^{\prime})$ is given by~\citet{Ballinger1996:astro-ph/9605017v1}:
\begin{align}
k^{\prime} &= \frac{k}{\aper} \left[1 + \mu^{2} \left(\frac{\aper^{2}}{\apar^{2}}-1\right) \right] ^ {1/2}, \\
\mu^{\prime} &= \frac{\mu \aper}{\apar} \left[1 + \mu^{2} \left(\frac{\aper^{2}}{\apar^{2}}-1\right) \right] ^ {-1/2}.
\label{eq:transform_ap}
\end{align}
Then, the power spectrum multipoles including the AP effect are expressed as:
\begin{equation}
P_{\ell}(k) = \frac{2 \ell + 1}{2 \apar \aper^{2}} \int_{-1}^{1} d\mu P_{\mathrm{g}}\left(k^{\prime}(k,\mu),\mu^{\prime}(\mu)\right) \Leg{\ell}(\mu).
\label{eq:power_ap}
\end{equation}
In practice, the AP effect is  mostly sensitive to the change in the position of the BAO feature imprinted in the power spectrum at $\rdrag$, the comoving sound horizon at the redshift at which the baryon-drag optical depth equals unity \citep{Hu1995:astro-ph/9510117v2}. Thus, scaling parameters are related to the true and fiducial cosmologies ($\fid$):
\begin{align}
\apar &= \frac{\DHu(\zeff)\rdrag^{\fid}}{\DHu^{\fid}(\zeff)\rdrag} \\
\aper &= \frac{\DM(\zeff)\rdrag^{\fid}}{\DM^{\fid}(\zeff)\rdrag},
\label{eq:scaling_parameters}
\end{align}
with $\DHu^{\fid}(\zeff)$ the Hubble distance and $\DM^{\fid}(\zeff)$ the comoving angular diameter distance given at the effective redshift of the galaxy sample $\zeff$, the superscript $\fid$ denoting quantities in the fiducial cosmology.
Note that the $\apar$ and $\aper$ dependence on $\rdrag$ is an approximation, and assumes that the scale-constraint in the power spectrum on the distance-redshift relationship only comes from BAO. Including this dependence in the $(\apar,\aper)$ parameters makes the power spectrum amplitude rescaling in the AP transform~\eqref{eq:power_ap} formally incorrect.
We expect this effect to be small for cosmological models that also fit the Planck constraints of~\citet{Planck2018:1807.06209v2}, who robustly measured $\rdrag = 147.09 \pm 0.26 \Mpc$ (from TT, TE, EE, lowE, CMB lensing), close to the value of our fiducial cosmology in \Eq{fiducial_cosmology}, $\rdrag^{\fid} = 147.77\Mpc$. The impact of the fiducial cosmology on data cosmological measurements will be tested in \Sec{results}.

\subsection{Irregular $\mu$ sampling}
\label{sec:model_irregular_mu}

Power spectrum multipoles are calculated on a discrete $\vc{k}$-space grid (\Sec{power_spectrum_estimation}), making the angular modes distribution irregular at large scales. We account for this effect in the model using the technique employed in~\cite{Beutler2016:1607.03150v1} which weights each $(k,\mu)$ mode according to its sampling rate $N(k,\mu)$ in the $\vc{k}$-space grid:
\begin{equation}
P_{\mathrm{g}}^{\mathrm{grid}}(k,\mu) = \frac{N(k,\mu)}{\int_{0}^{1} d\mu N(k,\mu)} P_{\mathrm{g}}\left(k,\mu\right).
\end{equation}
Though this correction should be applied after the convolution by the window function (discussed hereafter), as in~\cite{Beutler2016:1607.03150v1}, for the sake of computing time we include it when integrating the galaxy power spectrum over the Legendre polynomials in \Eq{power_ap}.
We checked that the impact of such a correction on the cosmological parameters measured on the eBOSS ELG data is of the order of $10^{-3}$, well below the statistical uncertainty.

\subsection{Survey window function}
\label{sec:model_window_function}

\begin{figure}
\centering
\includegraphics[width=0.8\columnwidth]{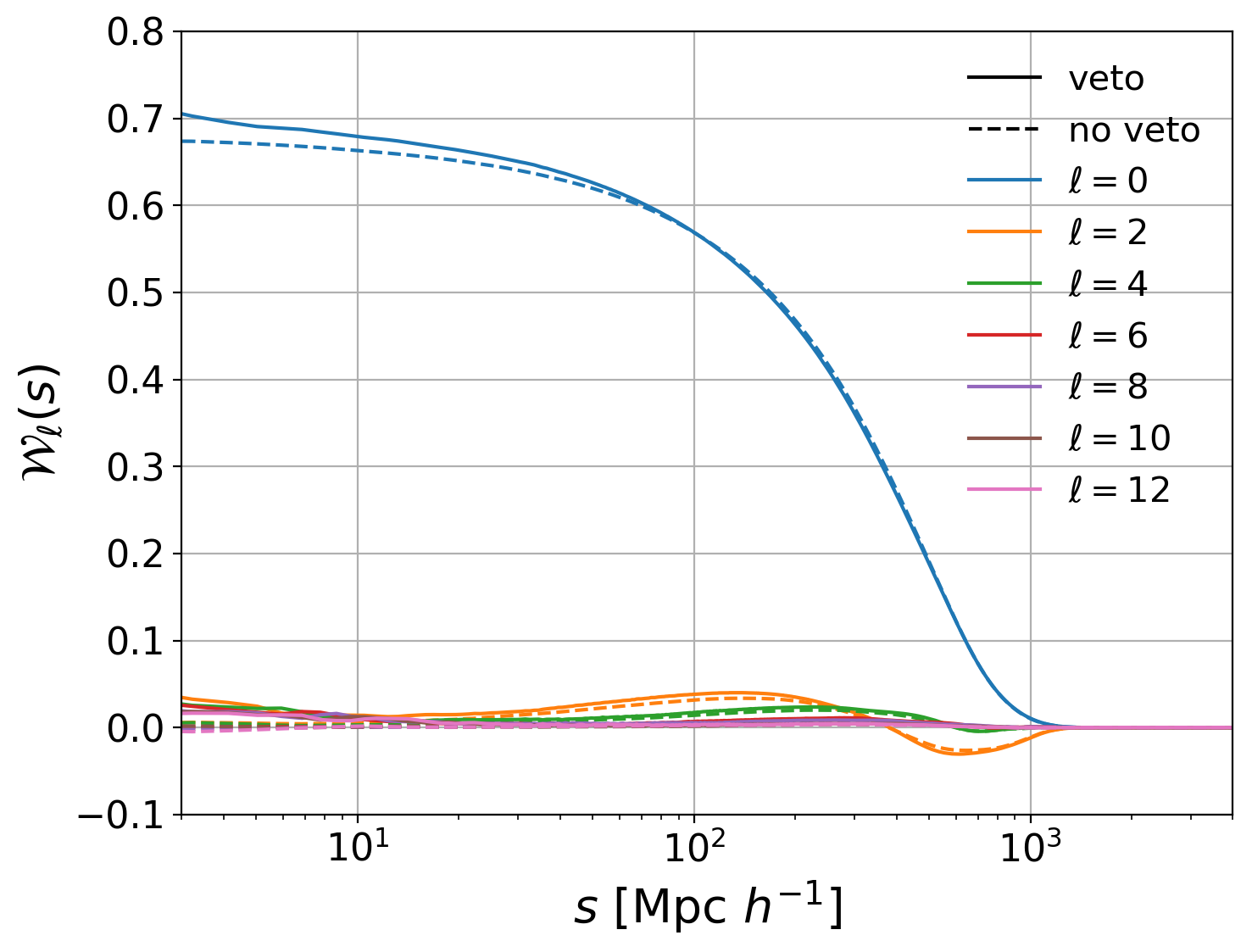}
\includegraphics[width=0.8\columnwidth]{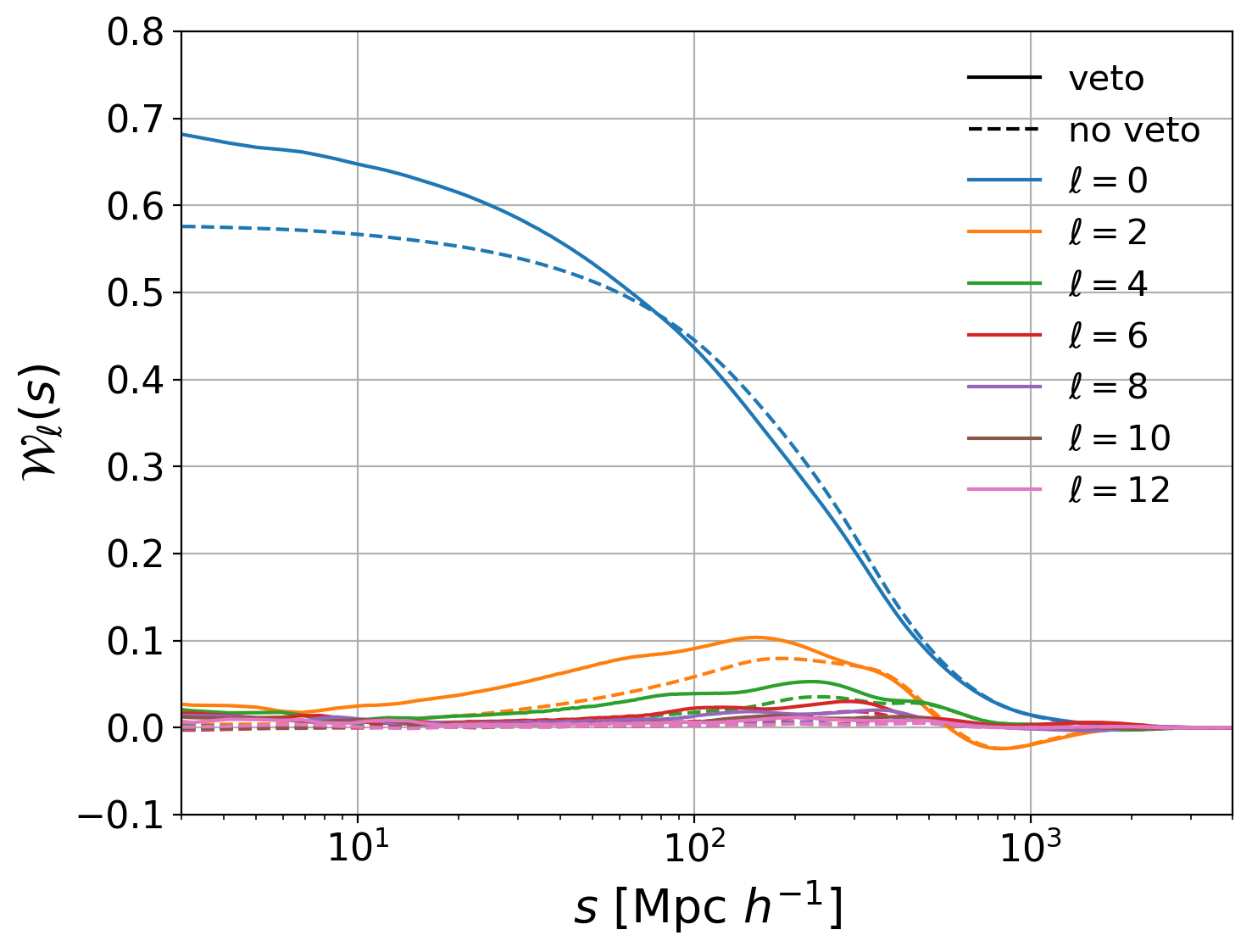}
\caption{Window function multipoles (top: NGC, bottom: SGC) of the EZ mocks (reproducing the eBOSS ELG sample), before (dashed lines) and after (continuous lines) application of the veto masks. Contrary to previous clustering analyses imposing window functions to converge to $1$ on small scales, we properly normalise these window functions by the same term as the power spectrum estimation. The height difference between the window function monopoles is explained by the area covered by veto masks (see text).}

\label{fig:window_function}
\end{figure}

The observed galaxy density is modulated by the survey selection function. The resulting window function effect is accounted for in the model using the formalism of~\citet{Wilson2015:1511.07799v2,Beutler2016:1607.03150v1}, and correctly normalised following~\citet{deMattia2019:1904.08851v3}. Indeed, because of the fine-grained veto masks of the eBOSS ELG survey and the conventional choice made to estimate the survey area entering the redshift density estimation, the value of $I$ in \Eq{power_spectrum_normalisation} used to normalise the power spectrum estimation in \Eq{power_spectrum_estimator_survey} is inaccurate. We thus use this value in the normalisation of window functions in the model, so that $I$ divides both the power spectrum measurements and model, and compensate. Therefore, the estimation of $I$ does not impact the recovered cosmological parameters.
\Fig{window_function} shows the window function multipoles of the EZ mocks (reproducing the eBOSS ELG sample, see \Sec{ezmocks}): the monopole has a non-zero slope even below $\lesssim 5 \Mpch$ due to the fine-grained veto masks. For comparison purposes, we also plot the window function without veto masks applied; in this case, the monopole stabilises faster on small scales. The area entering the estimation of $I$ used to normalise the unmasked window function has been kept fixed to the masked case. Since veto masks remove more area in the South Galactic Cap (SGC) than in the North Galactic Cap (NGC) \citep[see][]{Raichoor2017:1704.00338v1}, the unmasked SGC window function is relatively lower than the masked case compared to NGC.

In \Sec{mocks_systematics} we further check that veto masks do not bias cosmological measurements with our treatment of the window function.

The window function convolution requires to perform Hankel transforms between power spectrum and correlation function multipoles. We use for this purpose the \texttt{FFTLog} software~\citep{Hamilton2000:astro-ph/9905191v4}. As in~\cite{Beutler2016:1607.03150v1} we only consider correlation function multipoles $\xi_{\ell}(s)$ up to $\ell = 4$ in our calculations. We checked that adding $\xi_{6}(s)$ has a completely negligible impact on the model prediction.

\subsection{Integral constraints}
\label{sec:model_integral_constraints}

The mean of the observed density contrast $F(\vr)$ of \Eq{fkp_field} on the survey footprint is forced to $0$, as imposed by the definition of $\alpha_{s}$ in \Eq{alpha_global}, leading to a so-called global integral constraint (IC), which we model following~\citet{deMattia2019:1904.08851v3}.

\begin{figure}
\centering
\includegraphics[width=0.8\columnwidth]{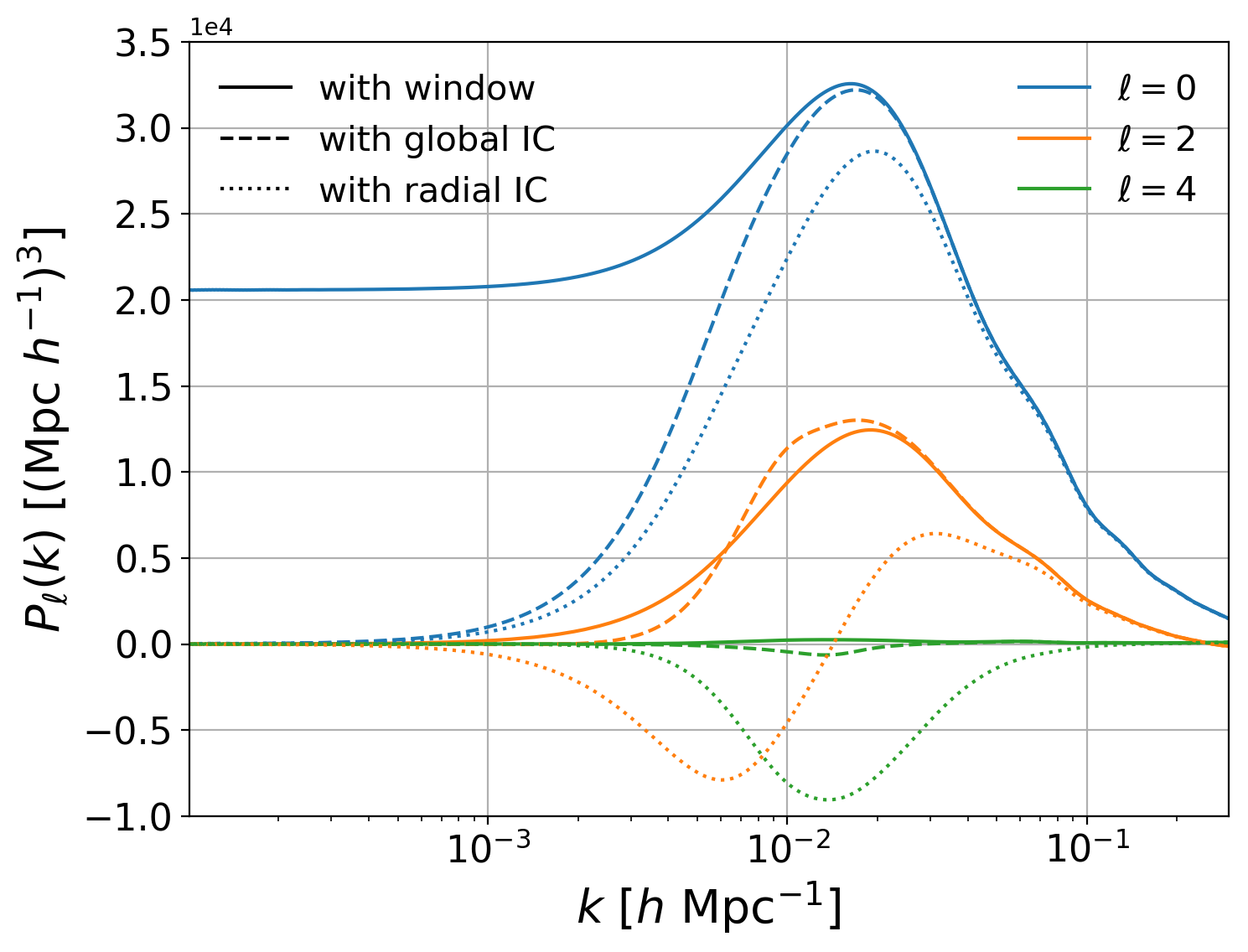}
\includegraphics[width=0.8\columnwidth]{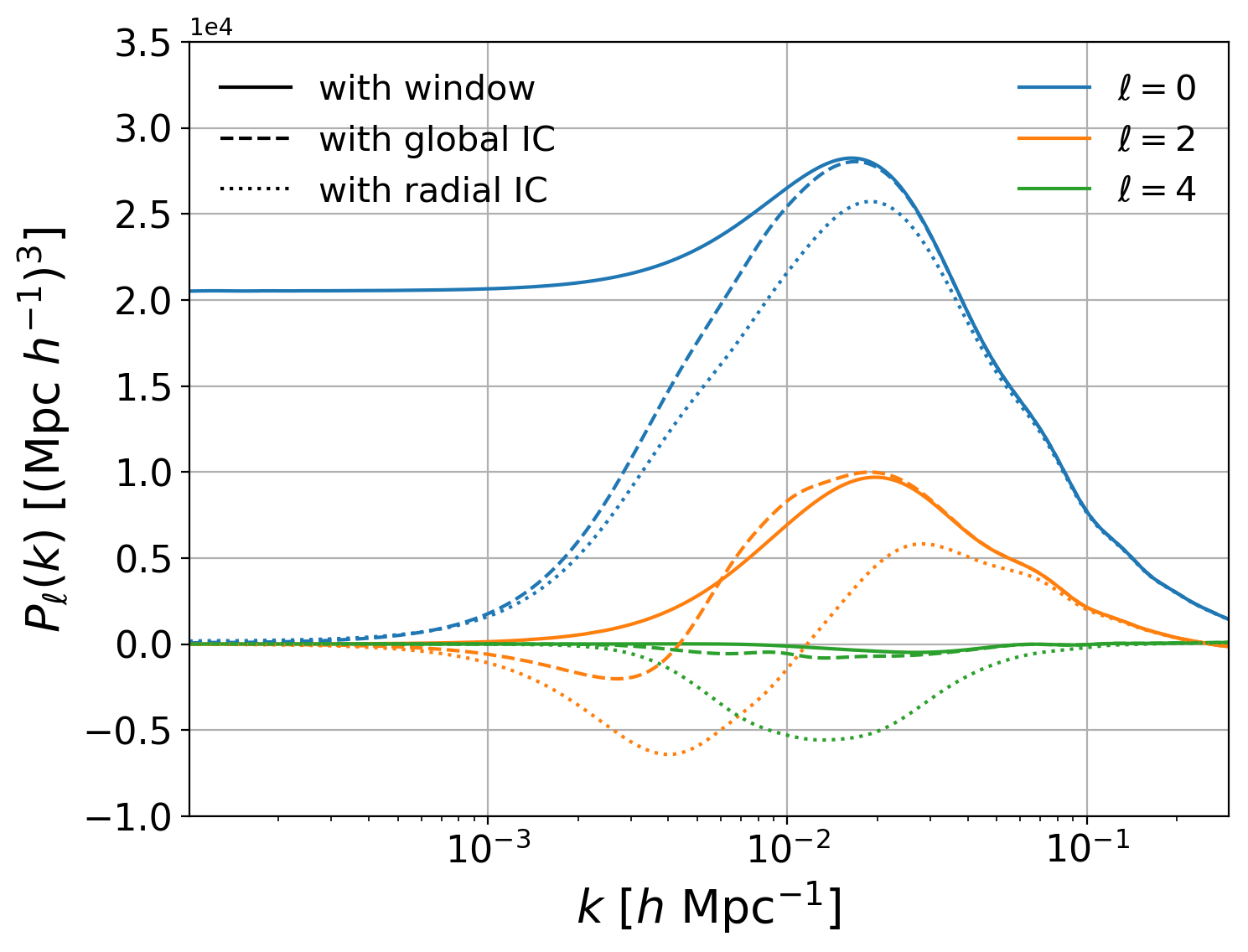}
\caption{Power spectrum multipoles (top: NGC, bottom: SGC; blue: monopole, red: quadrupole, green: hexadecapole) of the RSD model. The window function effect only is taken into account in continuous lines, and the additional impact of the global and radial integral constraints (IC) are shown in dashed and dotted lines, respectively. For this figure we choose $f=0.8$, $b_{1}=1.4$, $b_{2}=1$, $\sigma_{v}=4 \Mpch$.}
\label{fig:integral_constraints}
\end{figure}

Moreover, in this analysis, as well as in other BOSS and eBOSS clustering analyses~\cite[e.g.][]{Ross2012:1203.6499v3,Beutler2016:1607.03150v1,GilMarin2016:1509.06373v2}, redshifts of the random catalogue sampling the selection function are randomly drawn from the data (following the so-called \emph{shuffled} scheme). As discussed in~\citet{deMattia2019:1904.08851v3}, this leads to the suppression of radial modes and impacts the power spectrum multipoles on large scales. We will see in \Sec{mocks_systematics} that this effect, if not accounted for, would be one of the largest systematics in the eBOSS ELG sample. We thus model this effect following the prescription of~\citet{deMattia2019:1904.08851v3} by replacing the global integral constraint by the radial one for observed data and approximate mocks using the \emph{shuffled} scheme. The impact of the global and radial integral constraints on the power spectrum multipoles is shown in \Fig{integral_constraints}.
One can alternatively try to subtract the effect of the radial integral constraint from the data measurement \citep[see e.g.][]{Wang2020}.

Note that the integral constraint formalism will also be used to account for our mitigating angular observational systematics, as suggested in~\citet{deMattia2019:1904.08851v3} and detailed in \Sec{mocks_systematics}.

\subsection{Isotropic BAO}
\label{sec:model_isotropic_bao}

In this paper we perform an isotropic BAO measurement on the eBOSS ELG sample. We checked that the amplitude of the power spectrum measured at $k \simeq 0.1 \hMpc$ on post-reconstruction mock catalogues (see \Sec{ezmocks}) is roughly constant over $\mu$, suggesting that the BAO information is isotropically distributed. Thus, the monopole is optimal for single-parameter BAO-scale measurement, which can be used to constrain the following combination~\citep{Eisenstein2005:astro-ph/0501171v1,Ross2015:1501.05571v2}:
\begin{equation}
\alpha = \apar^{1/3} \aper^{2/3}.
\label{eq:bao_alpha_iso}
\end{equation}

To fit the isotropic BAO feature, we use the same power spectrum template (dubbed \emph{wiggle} template) as in previous analyses of BOSS and eBOSS~\cite[e.g.][]{Beutler2016:1607.03149v1,GilMarin2016:1509.06373v2,Ata2017:1705.06373v2}:
\begin{equation}
P(k,\alpha) = P_{\mathrm{sm}}(k) \mathcal{O}_{\mathrm{damp}}(k/\alpha) 
\label{eq:template_bao}
\end{equation}
where:
\begin{equation}
\mathcal{O}_{\mathrm{damp}}(k) = 1 + \left[\mathcal{O}(k) - 1 \right] e^{-\frac{1}{2}\Sigma_{\mathrm{nl}}^{2}k^2}.
\label{eq:template_bao_oscillations}
\end{equation}
$\mathcal{O}(k)$ is obtained by taking the ratio of the linear matter power spectrum $P_{\mathrm{m}}^{\mathrm{lin}}(k)$ to the no-wiggle power spectrum of~\citet{Eisenstein1998:astro-ph/9709112v1}, augmented by a five order polynomial term, fitted such that $\mathcal{O}(k)$ oscillates around $1$.
We take:
\begin{equation}
P_{\mathrm{sm}}(k) = B_{\mathrm{nw}}^{2} P_{\mathrm{nw}}(k) + \sum_{i=-2}^{i=2} A_{i}k^{i},
\label{eq:template_smooth}
\end{equation}
where $P_{\mathrm{nw}}(k) = P_{\mathrm{m}}^{\mathrm{lin}}(k)/\mathcal{O}(k)$. The number of broadband parameters $A_{i}$ is found such that the BAO template \Eq{template_bao} can reproduce the mean of the EZ mocks (see \Sec{ezmocks}) within $10\%$ of the uncertainty on the data power spectrum measurement.
To specify the BAO detection and for plotting purposes in \Sec{results_bao_data}, we will use the \emph{no-wiggle} template obtained by removing the oscillation pattern in \Eq{template_bao} (i.e. keeping only the $P_{\mathrm{sm}}(k)$ factor).

We cannot include a correction for the irregular $\mu$ sampling (\Sec{model_irregular_mu}) as the power spectrum template in \Eq{template_bao} is isotropic; this is not an issue since the correction seen in the case of the RSD model is very small.

We also neglect the integral constraints (\Sec{model_integral_constraints}), as their impact will be shown to be negligible in \Sec{mocks_systematics}.
The effect of the survey window function is accounted for according to \Sec{model_window_function} through the (dominant) monopole term only, since the power spectrum template is isotropic. This is legitimate since broadband terms typically absorb these smooth distortions of the power spectrum. We checked that totally ignoring the window function effect leads to a negligible change of $\simeq 10^{-3}$ in the $\alpha$ measurement with the eBOSS ELG data.

\section{Fitting methodology}
\label{sec:fitting_methodology}

This section details how the previously described RSD and BAO models are compared to the data to derive cosmological measurements.

\subsection{Reconstruction}
\label{sec:reconstruction}

For the isotropic BAO analysis, the galaxy field is reconstructed to enhance the BAO feature in its 2-point correlation function~\citep{Eisenstein2007:astro-ph/0604362v1}. This step (partially) removes RSD and non-linear evolution of the density field. We follow the procedure described in~\citet{Burden2015:1504.02591v2} and~\citet{Bautista2018:1712.08064v1}. 
We perform three reconstruction iterations, assuming the growth rate parameter $f = 0.82$ and the linear bias $b = 1.4$. The density contrast field is smoothed by a Gaussian kernel of width $15 \Mpch$. The choice of these reconstruction conditions and the assumed fiducial cosmology were shown to have very small impact on the BAO measurements in~\cite{VargasMagana:1610.03506v2} and~\cite{Carter2020:1906.03035v1}.

In this paper, isotropic BAO fits are performed on both pre- and post-reconstruction monopole power spectra, while the RSD analysis makes use of the monopole, quadrupole and hexadecapole of the pre-reconstruction power spectrum. As we will see in \Sec{results}, the posterior of the RSD only measurement is significantly non-Gaussian, making it hard to combine with the isotropic BAO posteriors. We thus also use jointly the above pre-reconstruction multipoles with the post-reconstruction monopole (taking into account their cross-covariance) to perform a combined RSD and isotropic BAO fit.
Further use of this new combination technique can be found in~\cite{GilMarin2020,ZhaoGB2021}. 

\subsection{Parameter estimation}
\label{sec:parameter_estimation}

In the RSD analysis, fitted cosmological parameters are the growth rate of structure $f$ and the scaling parameters $\apar$ and $\aper$.
Since $f$ is very degenerate with the power spectrum normalisation $\sig$, we quote the combination $\fsig$. As discussed in \cite{GilMarin2020,Bautista2020}, we take $\sig$ as the normalisation of the power spectrum at $8 \times \alpha \Mpch$ (instead of $8 \Mpch$), with $\alpha = \apar^{1/3} \aper^{2/3}$ as measured from the fit. We emphasise that the quoted $\fsig$ measurement can be straightforwardly compared to any $\fsig$ prediction, as usual. The sensitivity of our RSD measurements on the assumed fiducial cosmology is discussed in \Sec{mock_challenge_fiducial_cosmology}.
We consider $4$ nuisance parameters for the RSD fit: the linear and second order bias coefficients $b_{1}$ and $b_{2}$, the velocity dispersion $\sigma_{v}$ and $A_{g} = N_{g}/P_{0}^{\mathrm{noise}}$, with $N_{g}$ the constant galaxy stochastic term (see Eq.~\ref{eq:power_galaxy_galaxy}) and $P_{0}^{\mathrm{noise}}$ the measured Poisson shot noise (see Eq.~\ref{eq:power_spectrum_shot_noise}). Again, $b_{1}$ and $b_{2}$ are almost completely degenerate with $\sig$, so we quote $b_{1}\sig$ and $b_{2}\sig$.

For the isotropic BAO fit, the fitted cosmological parameter is $\alpha$. Nuisance parameters are $B_{\mathrm{nw}}$ and the broadband terms $\left(A_{i}\right)_{i \in [-2,2]}$ in \Eq{template_smooth}. These last terms are fixed by solving the least-squares problem for each value of $\alpha$, $B_{\mathrm{nw}}$. The non-linear damping scale $\Sigma_{\mathrm{nl}}$ is fixed using N-body simulations in \Sec{mock_challenge_isotropic_bao}.

For the combined RSD and post-reconstruction isotropic BAO fit, we use parameters from both analyses. We rely on \Eq{bao_alpha_iso} to relate $\alpha$ from the isotropic BAO fit to the $\apar$ and $\aper$ scaling parameters of the RSD fit.
We fix $B_{\mathrm{nw}}$ to $b_{1}$, as this choice introduced no detectable bias in the fits of the EZ mocks (see \Sec{mocks_systematics}). The varied parameters are reported in \Tab{fitting_parameters}.

The fitted $k$-range of the RSD measurement is $0.03 - 0.2 \hMpc$ for the monopole and quadrupole and $0.03 - 0.15 \hMpc$ for the hexadecapole. We choose such a minimum $k$ to avoid large scale systematics and non-Gaussianity. For the isotropic BAO fit we use the monopole between $0.03$ and $0.3 \hMpc$.

\subsection{Likelihood}
\label{sec:fitting_methodology_likelihood}

As is in some other eBOSS analyses~\cite[e.g.][]{Raichoor2020,Neveux2020,Bautista2020}, we use a frequentist approach to estimate the scaling parameter $\alpha$ for the isotropic BAO analysis. %
Bayesian inference is used to obtain posteriors for the eBOSS ELG RSD (and RSD~+~BAO) measurements. For the sake of computing time, we use a frequentist estimate of the cosmological parameters from the N-body based and approximate mocks and to perform data robustness tests.

In the frequentist approach, we perform a $\chi^{2}$ minimisation using the \texttt{Minuit}~\citep{Minuit1975} algorithm\footnote{ {\url{https://github.com/iminuit/iminuit}}}, taking large variation intervals for all parameters. We check that the fitted parameters do not reach the input boundaries. Errors are determined by likelihood profiling: the error on parameter $p_{i}$ is obtained by scanning the $p_{i} \rightarrow \min_{p_{j \neq i}} \chi^{2}(\vc{p})$ profile until the $\chi^{2}$ difference to the best fit reaches $\Delta \chi^{2} = 1$ (while minimising over other parameters $p_{j}$).

In the case of N-body mocks with periodic boundary conditions (see \Sec{mock_challenge_outerrim}), we compute an analytical covariance matrix following~\cite{Grieb2016:1509.04293v2}. In the case of data (see \Sec{data}) or sky-cut mocks (from N-body simulations in \Sec{mock_challenge} or approximate mocks in \Sec{data_mocks}), the power spectrum covariance matrix is estimated from approximate mocks (lognormal, EZ or GLAM-QPM mocks). We thus apply the Hartlap correction factor~\citep{Hartlap2007:astro-ph/0608064v2} to the inverse of the covariance matrix $C$ measured from the mocks:
\begin{equation}
\vc{\Psi} = \left(1-D\right) \vc{C}^{-1}, \quad D = \frac{n_{b}+1}{n_{m}-1}
\label{eq:covariance_hartlap}
\end{equation}
with $n_{b}$ the number of bins and $n_{m}$ the number of mocks. To propagate the uncertainty on the estimation of the covariance matrix, we rescale the parameter errors~\citep{Dodelson2013:1304.2593v2,Percival2014:1312.4841v1} by the square root of:
\begin{equation}
m_{1} = \frac{1 + B\left(n_{b} - n_{p}\right)}{1 + A + B\left(n_{p} + 1\right)},
\label{eq:covariance_m_1_naive}
\end{equation}
with $n_{p}$ the number of varied parameters and:
\begin{align}
A &= \frac{2}{\left(n_{m} - n_{b} - 1\right)\left(n_{m} - n_{b} - 4\right)}, \\
B &= \frac{n_{m} - n_{b} - 2}{\left(n_{m} - n_{b} - 1\right)\left(n_{m} - n_{b} - 4\right)}.
\end{align}
When cosmological fits are performed on the same mocks used to estimate the covariance matrix, the covariance of the obtained best fits should be rescaled by:
\begin{equation}
m_{2} = \left(1-D\right)^{-1} m_{1}.
\end{equation}
In \App{covariance_corrections} we propose a new version of this formula, accounting for a combined measurement of several independent likelihoods, which we use when fitting both the NGC and SGC. The magnitude of the rescaling~\eqref{eq:covariance_m_1} is of order $5.5\%$ at most (for the combined RSD~+~BAO measurements).

In the Bayesian approach, which we use to produce the posterior of the eBOSS ELG RSD and RSD~+~BAO measurements, the uncertainty on the covariance matrix estimation is marginalised over following~\cite{Sellentin2016:1511.05969v2}:
\begin{equation}
\mathcal{L}(\vc{x}^{d}\vert\vc{p}) \propto \left\lbrace 1 + \frac{1}{n_{m} - 1} \left[\vc{x}^{d} - \vc{x}^{t}(\vc{p})\right]^{T} \vc{C}^{-1} \left[\vc{x}^{d} - \vc{x}^{t}(\vc{p})\right] \right\rbrace^{-\frac{n_{m}}{2}}
\label{eq:likelihood_t}
\end{equation}
where we note the power spectrum measurements (data) $\vc{x}^{d}$ and the model (theory) $\vc{x}^{t}$ as a function of parameters $\vc{p}$. The combined NGC and SGC likelihood is trivially the product of NGC and SGC likelihoods.

Our posterior is the product of \Eq{likelihood_t} with flat priors on all parameters, infinite for all of them, except for $f$, $b_{1}$ and $\sigma_{v}$, which are lower-bounded by $0$ (see \Tab{fitting_parameters}). 

\begin{table*}
\caption{Varied parameters, and their priors in the case of Bayesian inference (MCMC). Priors are all flat, with infinite bounds, except for those mentioned below. No MCMC is run for the BAO only analysis. In all cases (including MCMC), parameters $\left(A_{i}\right)_{i \in [-2,2]}$ are solved analytically (see text)}.
\label{tab:fitting_parameters}
\centering
\begin{tabular}{lccc}
\hline
& RSD & BAO & RSD~+~BAO \\
\hline
varied parameters & $f$, $\apar$, $\aper$, $b_{1}$, $b_{2}$, $\sigma_{v}$, $A_{g}$ & $\alpha$, $B_{\mathrm{nw}}$, $\left(A_{i}\right)_{i \in [-2,2]}$ & $f$, $\apar$, $\aper$, $b_{1}$, $b_{2}$, $\sigma_{v}$, $A_{g}$, $\left(A_{i}\right)_{i \in [-2,2]}$ \\
NGC/SGC specific & $b_{1}$, $b_{2}$, $\sigma_{v}$, $A_{g}$ & $B_{\mathrm{nw}}$, $\left(A_{i}\right)_{i \in [-2,2]}$ & $b_{1}$, $b_{2}$, $\sigma_{v}$, $A_{g}$, $\left(A_{i}\right)_{i \in [-2,2]}$ \\
priors (MCMC) & $f \geq 0$, $b_{1} \geq 0$, $\sigma_{v} \geq 0$ & - &$f \geq 0$, $b_{1} \geq 0$, $\sigma_{v} \geq 0$ \\
\hline
\end{tabular}
\end{table*}

To sample the posterior distribution we run Markov Chain Monte Carlo (MCMC) with the package \texttt{emcee}~\citep{emcee}.
We run $8$ chains in parallel and check their convergence using the Gelman-Rubin criterion $R-1 < 0.02$~\citep{Gelman1992}.

\section{Mock challenge}
\label{sec:mock_challenge}

In this section we validate our implementation of the RSD TNS model and isotropic BAO template (presented in \Sec{model}) against mocks based on N-body simulations, which are expected to more faithfully reproduce the small-scale, non-linear galaxy clustering.
We estimate the potential modelling bias in the measurement of cosmological parameters. We refer the reader to~\citet{Alam2020} and~\citet{Avila2020} for a complete description of this mock challenge.

\subsection{MultiDark mocks}

A first set of mocks is based on the $z_{\mathrm{snap}} = 0.859$ snapshot of the MultiDark simulation MDPL2~\citep{Klypin2016:1411.4001v2} of volume $1 \Gpchc$ and $3840^{3}$ dark matter particles of mass $1.51 \times 10^{9} M_{\sun}/h$, run with the flat \LCDM\ cosmology\footnote{\url{https://www.cosmosim.org/cms/simulations/mdpl2/}}:
\begin{equation}
\begin{split}
h = 0.6777, \,\, \Omega_{m} = 0.307115, \,\, \Omega_{b} = 0.048206, \\
\sigma_{8} = 0.8228, \,\, n_{s} = 0.9611. \qquad
\end{split}
\label{eq:mdpl2_cosmology}
\end{equation}

Dark matter halos were populated with galaxies following two halo occupation distribution (HOD) models: a standard HOD (\emph{SHOD}), and a HOD quenched at high mass (\emph{HMQ}, see~\citealt{Alam2020} for details). Eleven types of mocks were produced for each HOD; in addition to the baseline (type~1), these include $50\%$ variations in the halo concentration in dark matter and the velocity dispersion of satellite galaxies (types~2, 3, 4, 5), a shift in the position of the central galaxy (type~6) and assembly bias prescriptions (types~7, 8 and~9). In the last two types of mocks (types~10 and~11), galaxy velocities were upscaled (downscaled) by $20 \%$, for which we thus expected a $20 \%$ increase (decrease) of the $\fsig$ measurement. The galaxy density reached $3 \times 10^{-3} \hMpcc$, about $10$ times the mean eBOSS ELG density, such that the shot noise is very low.

We derived a covariance matrix from a set of $500$ lognormal mocks produced with \texttt{nbodykit}, in the MDPL2 cosmology of \Eq{mdpl2_cosmology}, assuming a bias of $1.4$ and with the same density of $3 \times 10^{-3} \hMpcc$.  We checked that the agreement between N-body based and lognormal mocks was satisfactory on the whole $k$-range of the cosmological fit.

Both N-body based and lognormal mocks were analysed with the fiducial cosmology of \Eq{fiducial_cosmology}, as for the eBOSS ELG data. We thus accounted for the appropriate window function and global IC effect (\Sec{model_integral_constraints}) in the model, and we included the correction for the irregular $\mu$ distribution (\Sec{model_irregular_mu}) at large scales. As reported in~\citet{Alam2020} (Figure~4), the fitted cosmological parameters were found to be within $1 \sigma$ of the expected values, even for mocks with rescaled galaxy velocities, where the offset in the fitted $\fsig$ values corresponds to the $20 \%$ offset in velocity. However, the obtained uncertainties were only half of those expected with the eBOSS ELG sample, which was not sufficient to derive an accurate modelling systematic budget. We thus focused on larger mocks.

\subsection{OuterRim mocks}
\label{sec:mock_challenge_outerrim}

Two other sets of mocks were based on the $z_{\mathrm{snap}} = 0.865$ snapshot of the OuterRim~\citep{Heitmann2019:1904.11970v2} simulation of volume $27 \Gpchc$ and $10,240^{3}$ dark matter particles of mass $1.85 \times 10^{9} M_{\sun}/h$, run with the flat \LCDM\ cosmology:
\begin{equation}
\begin{split}
h = 0.71, \,\, \omega_{cdm} = 0.1109, \,\, \omega_{b} = 0.02258, \\
\sigma_{8} = 0.8, \,\, n_{s} = 0.963. \qquad
\end{split}
\label{eq:outerrim_cosmology}
\end{equation}
A first set of mocks using the \emph{SHOD} and \emph{HMQ} HODs was produced, with 6 types, corresponding to types 1, 4, 5, 6, 10 and~11 of the MultiDark-based mocks, with a density between $\simeq 3 \times 10^{-3} \hMpcc$ (\emph{SHOD}) and $\simeq 4 \times 10^{-3} \hMpcc$ (\emph{HMQ}).

A second set of OuterRim mocks was produced based on results from models of galaxy formation and evolution~\citep{Avila2020}. Three different HODs were considered: the mean number of satellite galaxies was fixed to a power-law (in the halo mass), but central galaxies followed either a smoothed step function ($\mathrm{erf}$, \emph{HOD-1}), a Gaussian (\emph{HOD-2}) or a Gaussian extended by a decaying power-law (baseline), based on results obtained by~\cite{Gonzalez-Perez2018:1708.07628v2}. The fraction of satellite galaxies $f_{\mathrm{sat}}$ was varied. Satellites were either directly assigned the positions and velocities of random particles in the dark matter halo (\emph{part.}) or they were sampled from a~\cite{Navarro1995:astro-ph/9508025v1} profile (\emph{NFW}); in the latter case the virial theorem~\citep{Bryan1998:astro-ph/9710107v1,Avila2018:1712.06232v2} was used to sample velocities.
The concentration was varied and the probability law for sampling satellites was also changed (Poisson, nearest integer, binomial, e.g. \cite{Jimenez2019:1906.04298v1}). These variations led to minor changes in the fitted cosmological parameters. %
Finally, velocities of satellite galaxies were biased with respect to dark matter by a factor $\alpha_{v}$ ($\alpha_{v} = 1$ in the baseline case), which is referred to as the satellite velocity bias \citep{Guo2015:1505.07861v3}, or were given an infall component following a Gaussian of mean $v_{t} = 500 \kmps$ and dispersion $200 \kmps$~\citep{Orsi2018:1708.00956v2}. The number density of the $24$ mocks we analysed ranges between $\simeq 2 \times 10^{-4} \hMpcc$ (\emph{part.} and some \emph{NFW}) and $\simeq 2 \times 10^{-3} \hMpcc$ (\emph{NFW}).

We analysed these mocks with the OuterRim cosmology, imposing periodic boundary conditions. Therefore, there is no window effect and we only included the correction for the irregular $\mu$ distribution (\Sec{model_integral_constraints}) at large scales. A Gaussian covariance matrix was calculated following~\citet{Grieb2016:1509.04293v2} for each of these mocks, taking their measured power spectrum as input.

No evidence for an overall systematic bias of the model was found when analysing these mocks, as reported in~\cite{Alam2020}.

\subsection{Blind OuterRim mocks}
\label{sec:mock_challenge_blind_outerrim}

We participated in a blind mock challenge dedicated to the ELG sample (see \citet{Alam2020}, Section~8). For simplicity, only velocities and HOD parameters were changed, while the background cosmology was kept fixed. Therefore, only the value of $\fsig$ was blind. $30$ to $40$ realisations for each of the $6$ types of mocks ($3$ for \emph{SHOD} and \emph{HMQ}) were produced with a density of the order of the eBOSS ELG mean density ($\simeq 2 \times 10^{-4} \hMpcc$). The OuterRim boxes were analysed the same way as in \Sec{mock_challenge_outerrim}. Though velocities were scaled by as much as $50\%$, no significant systematic shift in $\fsig$ could be seen, confirming the robustness of our RSD model.

The systematic uncertainties resulting from this blind mock challenge were derived in \citet{Alam2020} (Section~9): $1.6\%$ on $\fsig$, $0.8\%$ on $\apar$ and $0.7\%$ on $\aper$\footnote{This systematic budget was consistently updated using our prescription for $\sig$ discussed in \Sec{parameter_estimation} --- leading to a minor relative decrease of $4\%$ on the systematics for $\fsig$.}. We do not scale these errors by a factor of $2$ as in \cite{Alam2020}, since we further take into account the effect of the fiducial cosmology in \Sec{mock_challenge_fiducial_cosmology} in a conservative way.

\subsection{Fiducial cosmology}
\label{sec:mock_challenge_fiducial_cosmology}

As~\cite{Hou2020,Neveux2020,GilMarin2020,Bautista2020} we test the dependence of the measurement of cosmological parameters with respect to the cosmology --- dubbed as \emph{template cosmology} --- used to compute the linear power spectrum for the RSD model ($P_{\mathrm{m}}^{\mathrm{lin}}(k)$ in see \Sec{model_tns}). To this end, we reanalyse the first set of OuterRim mocks (type~1, 4, 5, 6, with \emph{SHOD} and \emph{HMQ} HODs) presented in \Sec{mock_challenge_outerrim}, but using different template cosmologies. We consider first the fiducial cosmology of the data analysis~\eqref{eq:fiducial_cosmology} and also scale each of the cosmological parameters ($h$, $\omega_{cdm}$, $\omega_{b}$ and $n_{s}$) of~\eqref{eq:outerrim_cosmology} by $\pm 10\%$ to $\pm 20\%$ (typically $30\sigma$ variations of \cite{Planck2018:1807.06209v2} CMB (TT, TE, EE, lowE, lensing) and BAO constraints).

Note that for simplicity we do not change the fiducial cosmology~\eqref{eq:outerrim_cosmology} used in the analysis (power spectrum estimation and Gaussian covariance matrix) and thus rescale the fitted $\apar$ and $\aper$ accordingly to determine $\sig$ as in \Sec{parameter_estimation}. 

\begin{figure}
\centering
\includegraphics[width=\columnwidth]{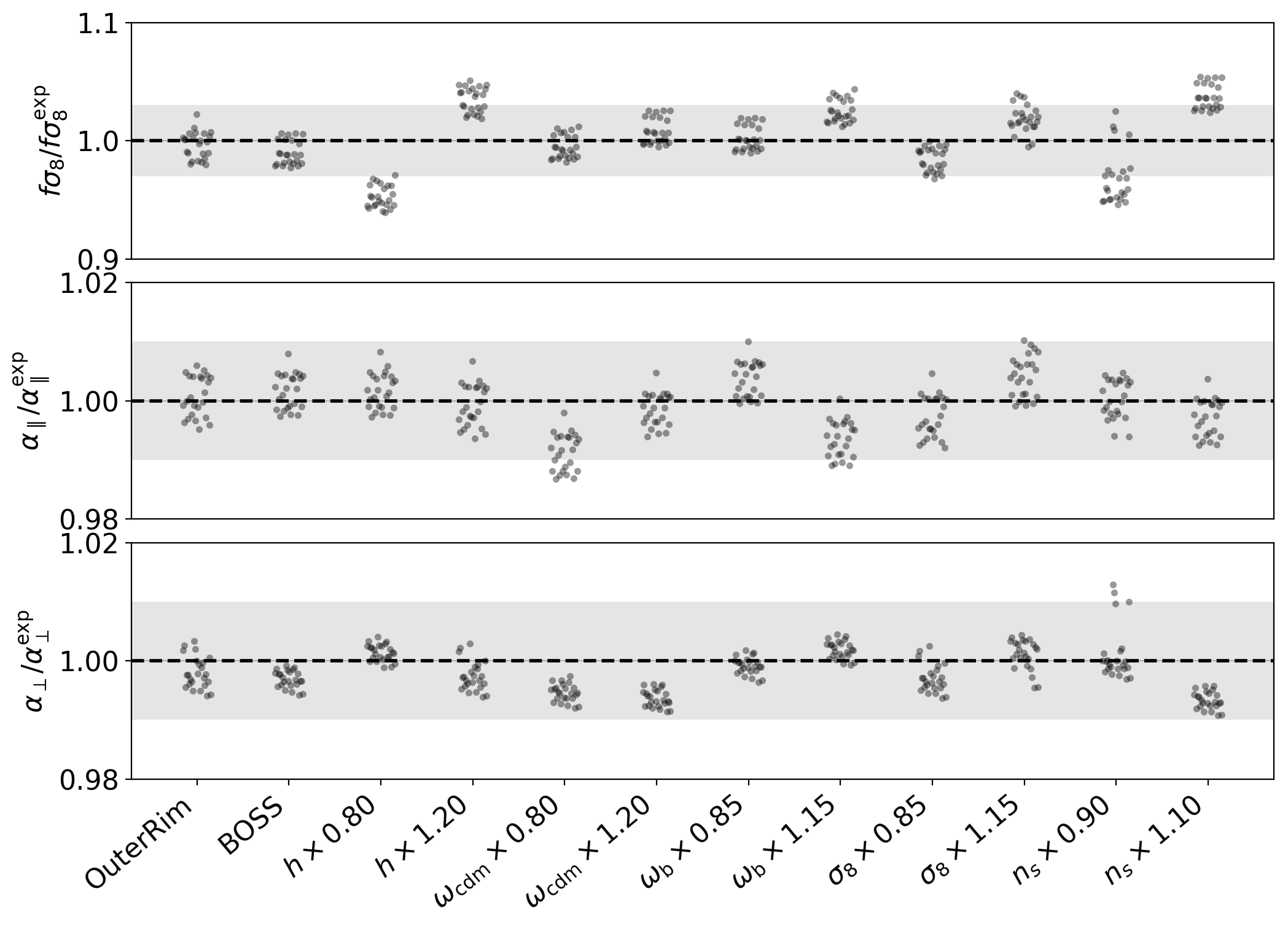}
\caption{Ratio of the RSD best fits to the OuterRim-based mocks (of type~1, 4, 5, 6 with \emph{SHOD} and \emph{HMQ} HODs, using three line-of-sight axes --- $x$, $y$, $z$) to their expected values, using different template cosmologies. The gray shaded area represents an error of $3\%$ on $\fsig$ and $1\%$ on the scaling parameters on either side of the reference values in the OuterRim cosmology.}
\label{fig:mock_challenge_oralam_cosmo}
\end{figure}

Results are shown in \Fig{mock_challenge_oralam_cosmo}. Scaling parameters are well recovered. The best fit $\fsig$ value is primarily sensitive to the template $h$ and $n_{s}$ values. Taking the root mean square of the difference (averaged over all types, HODs, and lines-of-sight) to the expected values gives the following systematic uncertainties: $2.6\%$ on $\fsig$ and $0.4\%$ on scaling parameters. Note that without the $\sig$ rescaling described in \Sec{parameter_estimation} the systematic uncertainties related to the choice of template cosmology would have been twice larger for $\fsig$.

We add the above uncertainties in quadrature to those derived in \Sec{mock_challenge_blind_outerrim} to obtain the final RSD modelling systematics: $3.0\%$ on $\fsig$ and $0.9\%$ on $\apar$ and $0.8\%$ on $\aper$.

\subsection{Isotropic BAO}
\label{sec:mock_challenge_isotropic_bao}

We also test the robustness of the isotropic BAO analysis with respect to variations in the HOD and template cosmology. Again, we consider the first set of OuterRim mocks (type~1, 4, 5, 6, with \emph{SHOD} and \emph{HMQ} HODs) presented in \Sec{mock_challenge_outerrim}, apply them reconstruction (with the parameters set in \Sec{reconstruction}), and measure their power spectrum using the OuterRim cosmology~\eqref{eq:outerrim_cosmology}.

Using the isotropic BAO model described in \Sec{model_isotropic_bao}, we first find (with the OuterRim cosmology as template cosmology) a damping parameter $\Sigma_{\mathrm{nl}}$ value of $8 \Mpch$ ($4 \Mpch$) to fit the pre-reconstruction (post-reconstruction) power spectrum. We use these values in the rest of the paper, unless stated otherwise.

\begin{figure}
\centering
\includegraphics[width=\columnwidth]{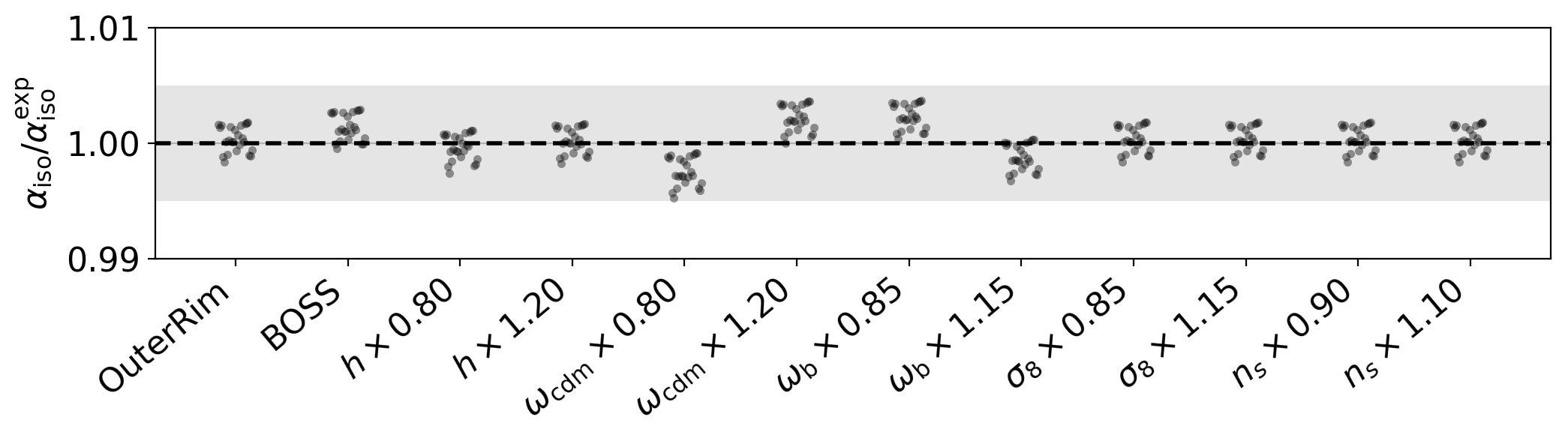}
\caption{Ratio of the isotropic BAO best fits to the OuterRim-based mocks (of type~1, 4, 5, 6 with \emph{SHOD} and \emph{HMQ} HODs, using three line-of-sight axes --- $x$, $y$, $z$) to their expected values, using different template cosmologies. The gray shaded area represents an error of $0.5\%$ on $\alpha$ on either side of the reference value in the OuterRim cosmology.}
\label{fig:mock_challenge_oralam_bao_cosmo}
\end{figure}

We then perform the post-reconstruction isotropic BAO fits with the different template cosmologies introduced in \Sec{mock_challenge_fiducial_cosmology}. Results are shown in \Fig{mock_challenge_oralam_bao_cosmo}. The measured $\alpha$ value shows very small dependence with the template cosmology, as also found in e.g. \cite{Carter2020:1906.03035v1}. The same is true for the HOD model. Taking the root mean square of differences between best fit and expected values gives a systematic uncertainty of $0.2\%$ on $\alpha$, which we take as BAO modelling systematics.

In addition, in order to quantify how typical the data BAO measurements are (see \Sec{results_bao_data}), we generate accurate mocks designed to match the ELG sample survey geometry. An OuterRim box (satellite fraction of $0.14$, no velocity bias) is trimmed to the eBOSS ELG footprint, including veto masks and radial selection function. We then cut $6$ nearly independent mocks for NGC and SGC with $3$ different orientations. The original box was replicated by $20\%$ to enclose the total SGC footprint. As the ELG density in the OuterRim box is much larger than the observed ELG density, we draw $4$ disjoint random subsamples for each sky-cut mock. The number of galaxies in the mock samples match that of the data to better than $1\%$. Then, we randomly generate $1000$ fake mock power spectra following a multivariate Gaussian. The Gaussian mean comes from the pre- and post-reconstruction power spectrum measurements of the above sky-cut OuterRim mocks, and its covariance matrix is given by the \emph{baseline} EZ mocks. These fake post-reconstruction power spectra will be used to quantify the probability of the BAO measurements of the data in \Sec{results_bao_data}.

\section{Data and approximate mocks}
\label{sec:data_mocks}

In this section we briefly describe the eBOSS ELG sample and the approximate EZ and GLAM-QPM mocks used to produce the covariance matrix of the measured power spectrum multipoles and to assess the impact of observational systematics on the final clustering measurements.

\subsection{Data}
\label{sec:data}

We use the ELG clustering catalogues from SDSS DR16 published in~\cite{Raichoor2020}. We refer the reader to that paper for a complete description of the different survey masks and weighting schemes adopted to account for variations of the survey selection function in the data.

Contrary to other BOSS and eBOSS surveys making use of the SDSS-I-II-III optical imaging data, ELG targets were selected from the data release 3 and 5 of the Dark Energy Camera Legacy Survey \citep[DECaLS][]{Dey2019:1804.08657v2} in the {\it grz} bands \citep[see][]{Raichoor2017:1704.00338v1}. DECaLS photometry, which is at least one magnitude deeper than the SDSS imaging in all bands, will be used by the next generation survey Dark Energy Spectroscopic Instrument \citep{DESI2016:1611.00036v2}.
However, the homogeneity of the imaging quality over the eBOSS ELG footprint was not fully under control in this early version of DECaLS, a point we will further discuss below.

\begin{figure*}
\centering
\includegraphics[width=0.27\textwidth]{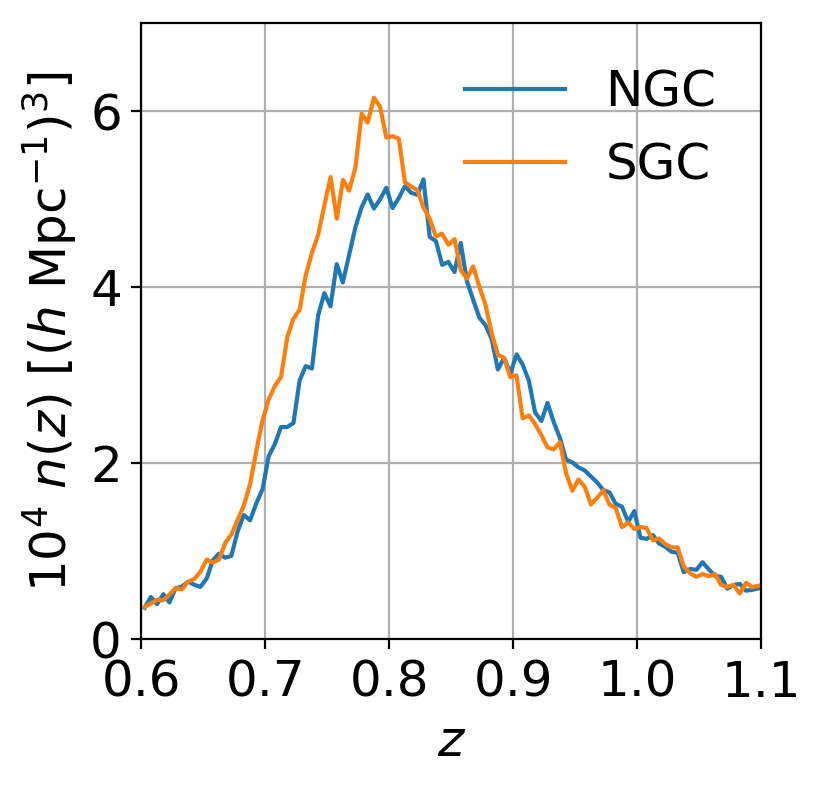}
\includegraphics[width=0.6\textwidth]{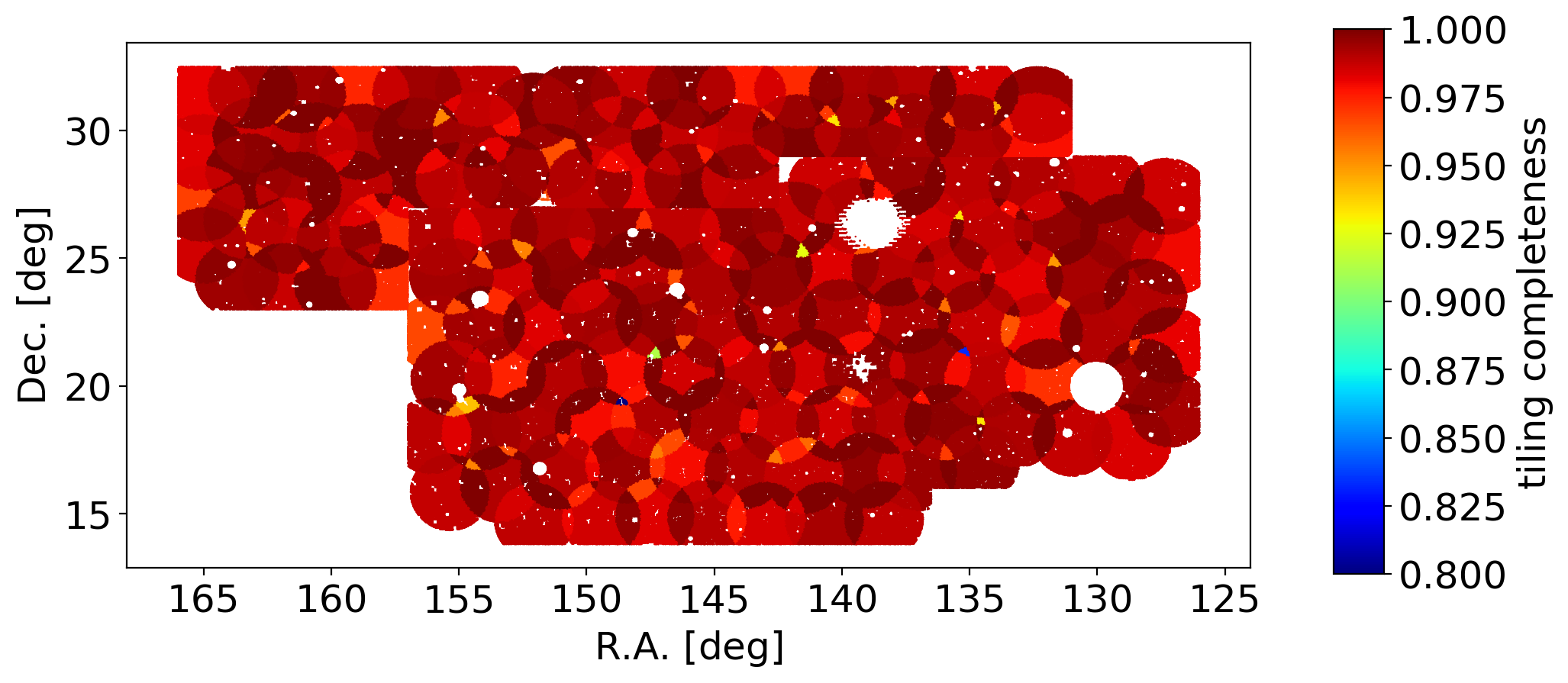}
\includegraphics[width=0.8\textwidth]{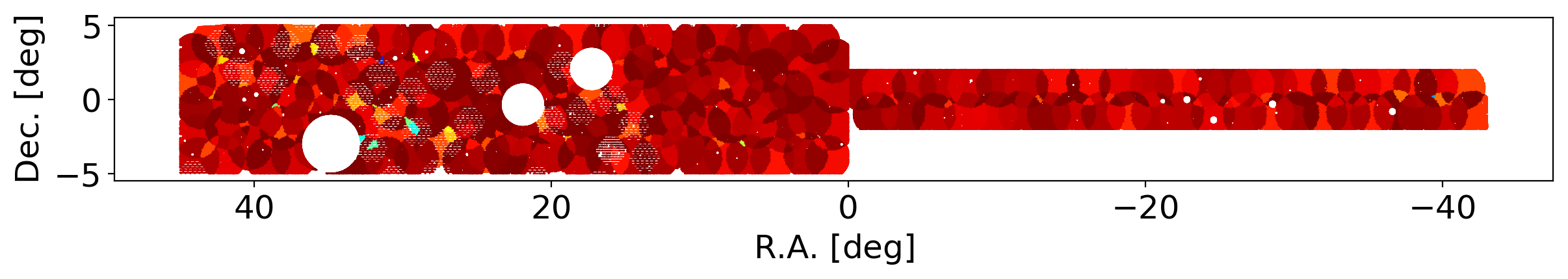}
\caption{eBOSS ELG footprint. Top left: comoving redshift density. Top right: tiling completeness in NGC. Bottom: tiling completeness in SGC. 
}
\label{fig:footprint_density}
\end{figure*}

The footprint and redshift density of the eBOSS ELG survey are shown in \Fig{footprint_density}. The eBOSS ELG sample selects star-forming ELGs in the redshift range $0.6 < z < 1.1$ and within tiling chunks\footnote{Regions in which the fibre assignment (determining the positions of the plates and fibres) is run independently.} \texttt{eboss21} and \texttt{eboss22} in the SGC and chunks \texttt{eboss23} and \texttt{eboss25} in the NGC. The number of targets ($N_{\mathrm{targ}}$) and ELGs ($N_{\mathrm{used}}$) in the final clustering sample is given in \Tab{data_stats}.

\begin{table}
\caption{Statistics of the eBOSS ELG sample. $N_{\mathrm{targ}}$ is the number of targets (after veto masks are applied). $N_{\mathrm{used}}$ is the number of objects in the final clustering catalogues, in $0.6 < z < 1.1$ (except otherwise stated). The effective area is the unvetoed area multiplied by the tiling completeness.}
\label{tab:data_stats}
\centering
\begin{tabular}{l|rrr}\hline
& NGC & SGC & ALL\\
\hline\hline
$N_{\mathrm{targ}}$ & $113,500$ & $116,194$ & $229,694$\\
$N_{\mathrm{used}}$ & $83,769$ & $89,967$ & $173,736$\\
$N_{\mathrm{used}}$ in $0.7 < z < 1.1$ & $79,106$ & $84,542$ & $163,648$\\
Effective area ($\deg^{2}$) & $369.5$ & $357.5$ & $727.0$\\
\hline
\end{tabular}
\end{table}

Three types of weights are introduced to correct for variations of the selection function in the data. The systematic weight $\wsys$ corrects for fluctuations of the ELG density with imaging quality. The close-pair weight $\wcp$ accounts for fibre collisions. Finally, $\wnoz$ corrects for redshift failures. 

A synthetic (\emph{random}) catalogue is built to sample the survey selection function of the weighted data. Angular coordinates of the synthetic catalogue are uniformly random, and random objects outside the footprint, including veto masks, are removed. 
Data redshifts are assigned to random objects, following the \emph{shuffled} scheme proposed in~\cite{Ross2012:1203.6499v3}. The previously mentioned anisotropies of the DECaLS imaging quality induce fluctuations of the eBOSS ELG redshift density which shall be introduced in the synthetic catalogue~\citep{Raichoor2020}. We found the main driver for these fluctuations to be imaging depth. We therefore assign data redshifts to randoms in $3$ separate sub-regions (dubbed \chunkz) of each tiling chunk defined according to their value of imaging depth. The depth-bins are chosen such that the redshift distribution is considered sufficiently (i.e. within shot noise and cosmic variance) constant within each \chunkz. In the synthetic catalogue, $\wsys$ accounts for the tiling completeness while $\wcp$ and $\wnoz$ are all set to $1$. $\wsys$ is then scaled such that the weighted number of random objects and data objects match in each \chunkz.

Each data and random object is weighted by the total weight $\wtot = \wfkp \wcomp$ with $\wcomp = \wsys \wcp \wnoz$ its completeness weight and $\wfkp$ the FKP weight:
\begin{equation}
\wfkp = \frac{1}{1+n_{g,i}P_{0}},
\end{equation}
where we take $P_{0} = 4000 \Mpchc$, close to the measured power spectrum monopole at $k \simeq 0.1 \hMpc$ (see \Fig{power_spectrum_data_ezmocks}). The redshift density $n_{g,i}$ is calculated in each chunk by binning data weighted by $\wcomp$ into redshift slices of size $\Delta z = 0.005$, starting at $z = 0$, and dividing the result by the comoving volume of each shell, assuming the fiducial cosmology of~\Eq{fiducial_cosmology}. The effective area used for the calculation is given by the number of randoms weighted by the tiling completeness in the final clustering sample divided by their original density (see~\Tab{data_stats} and~\citealt{Raichoor2020}).

In order to match the definition used for other eBOSS tracers and analyses, the effective redshift $\zeff$ of the ELG sample between $0.6 < z < 1.1$ is calculated as:
\begin{equation}
\zeff = \frac{\sum_{i,j} w_{\mathrm{tot},i} w_{\mathrm{tot},j} (z_{g,i} + z_{g,j})/2}{\sum_{i,j} w_{\mathrm{tot},i} w_{\mathrm{tot},j}},
\label{eq:effective_redshift}
\end{equation}
where the sum is performed over all galaxy pairs between $25 \Mpch$ and $120 \Mpch$. We measure $\zeff = 0.845$ for the combined NGC and SGC (NGC alone: $0.849$, SGC alone: $0.841$). We checked that this result varies by less than $0.4 \%$ when including pairs between $0 \Mpch$ and $200 \Mpch$. In \App{effective_redshift} we provide a definition of the effective redshift more specific to the power spectrum analysis, which quantitatively gives the same value as that adopted in \Eq{effective_redshift}. We also compute the effective redshift corresponding to the cuts $0.7 < z < 1.1$, which will be used in \Sec{results}: $\zeff = 0.857$ for the combined NGC and SGC (NGC alone: $0.860$, SGC alone: $0.853$). The typical variation (using \Eq{fiducial_cosmology}) corresponding to the $\simeq 0.8\%$ difference between the effective redshift of NGC and SGC is $0.2\%$ on $\fsig$, $0.4\%$ on $\DHu/\rdrag$ and $0.6\%$ on $\DM/\rdrag$, small compared to the statistical uncertainty (see \Sec{results}).

\Fig{power_spectrum_data_ezmocks} displays the power spectrum multipoles as measured on the data (blue curve). In the following sections we briefly recap the creation of EZ and GLAM-QPM mocks, as well as the implementation and correction of observational systematics. For more details we refer the reader to~\cite{Raichoor2020}.

\subsection{EZ mocks}
\label{sec:ezmocks}

The generation of EZ mocks is detailed in~\cite{Zhao2020}. The $1000$ EZ mocks (NGC, SGC) are built from EZ boxes of side $5 \Gpch$, with a galaxy number density of $6.4 \times 10^{-4} \hMpcc$ at different snapshots $z_{\mathrm{snap}} = 0.658, 0.725, 0.755, 0.825, 0.876, 0.950, 1.047$ used to cover the redshift ranges $0.6 - 0.7$, $0.7 - 0.75$, $0.75 - 0.8$, $0.8 - 0.85$, $0.85 - 0.9$, $0.9 - 1.0$, and $1.0 - 1.1$, respectively. The fiducial cosmology of these mocks is that of the MultiDark simulation (except for $\sigma_{8}$), i.e. flat \LCDM{} with:
\begin{equation}
\begin{split}
h = 0.6777, \,\, \Omega_{m} = 0.307115, \Omega_{b} = 0.048206,\\
\sigma_{8} = 0.8225, \,\, n_{s} = 0.9611.  \qquad
\end{split}
\label{eq:ez_cosmology}
\end{equation}

Mocks are trimmed to the tiling geometry and veto masks. We implement in the mocks the observational systematics seen in the data. The data redshift distribution is applied to the mocks in each \chunkz. We introduce angular systematics by trimming mock objects according to a map built from the data observed density with a Gaussian smoothing of radius $1\deg$. %
Contaminants, such as stars, or objects outside the redshift range $0.6 < z < 1.1$ are added to the catalogues, such that the target density matches in average that of the observed data. Fibre collisions are modelled using an extension of the~\cite{Guo2012:1111.6598v2} algorithm implemented in \texttt{nbodykit}, accounting for the plate overlaps and target priority. We include the TDSS~\citep{Ruan2016:1602.02752v1} targets ("FES" and "RQS1") which were tiled at the same time as eBOSS ELGs. Finally, some objects are declared as redshift failures following their nearest neighbour in the observed data.
All systematic corrections (weighting scheme and $n(z)$ dependence in the imaging depth) are applied the exact same way to the mocks as in the data clustering catalogues.

\begin{figure*}
\centering
\includegraphics[width=0.3\textwidth]{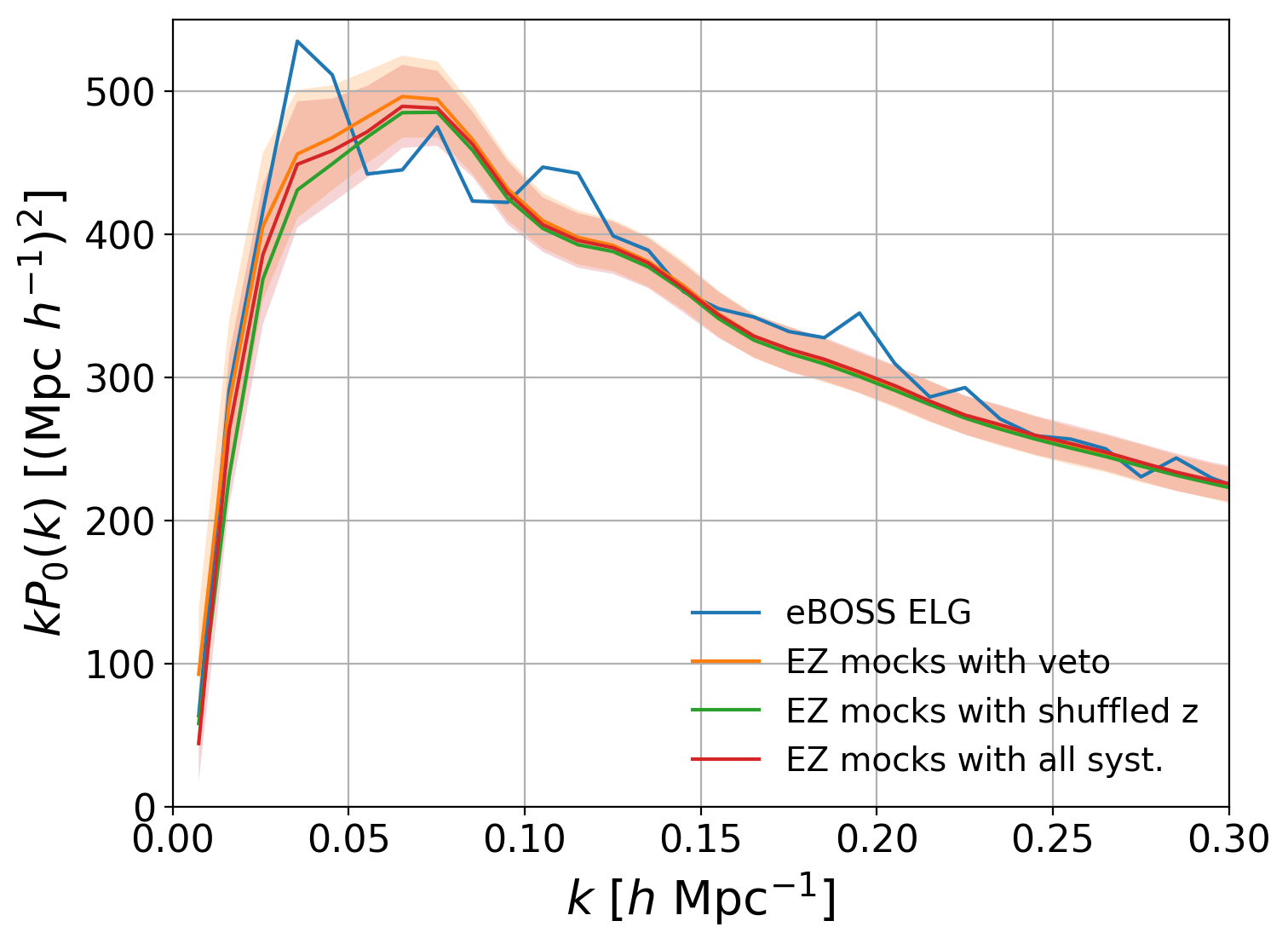}
\includegraphics[width=0.3\textwidth]{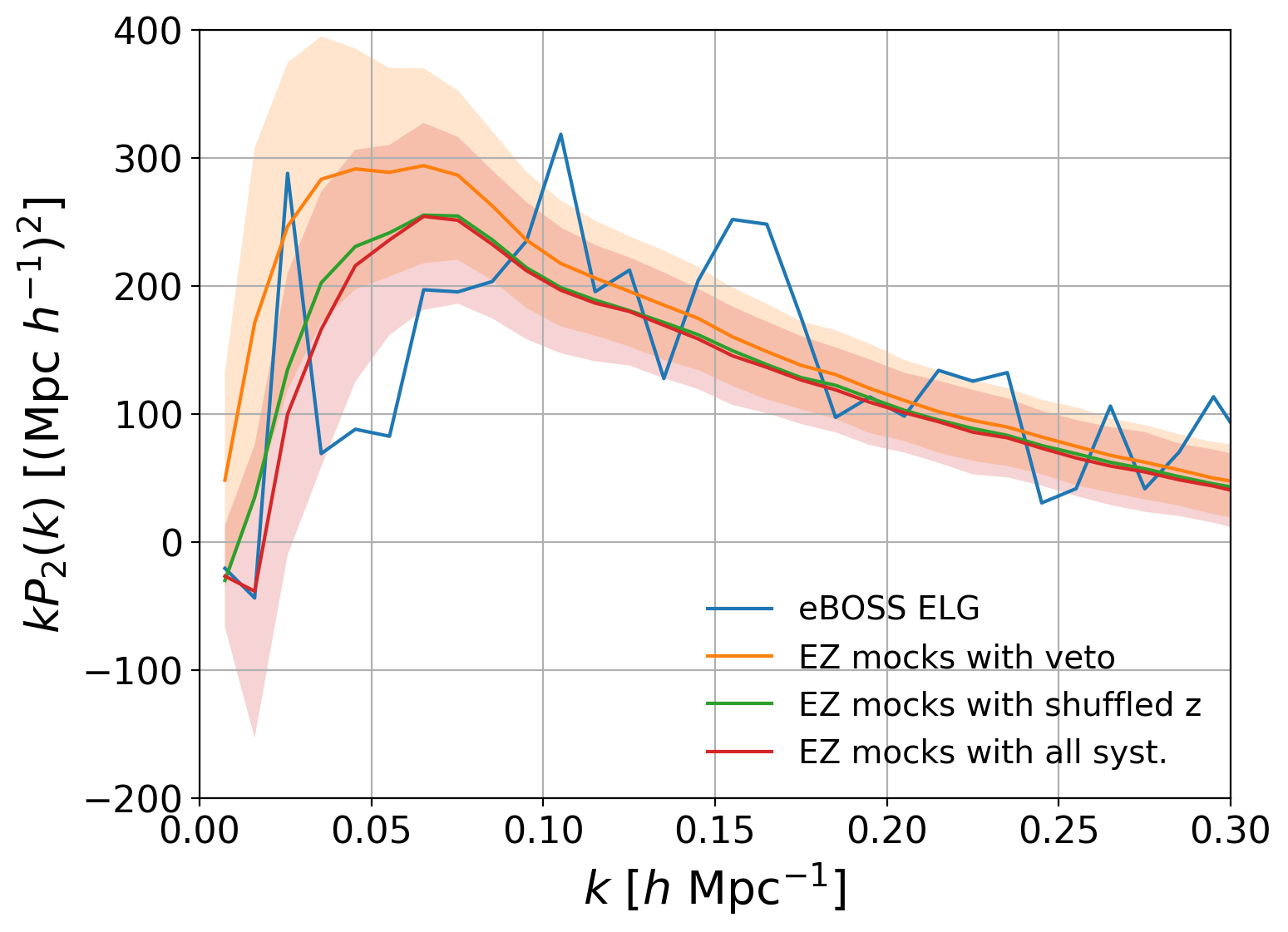}
\includegraphics[width=0.3\textwidth]{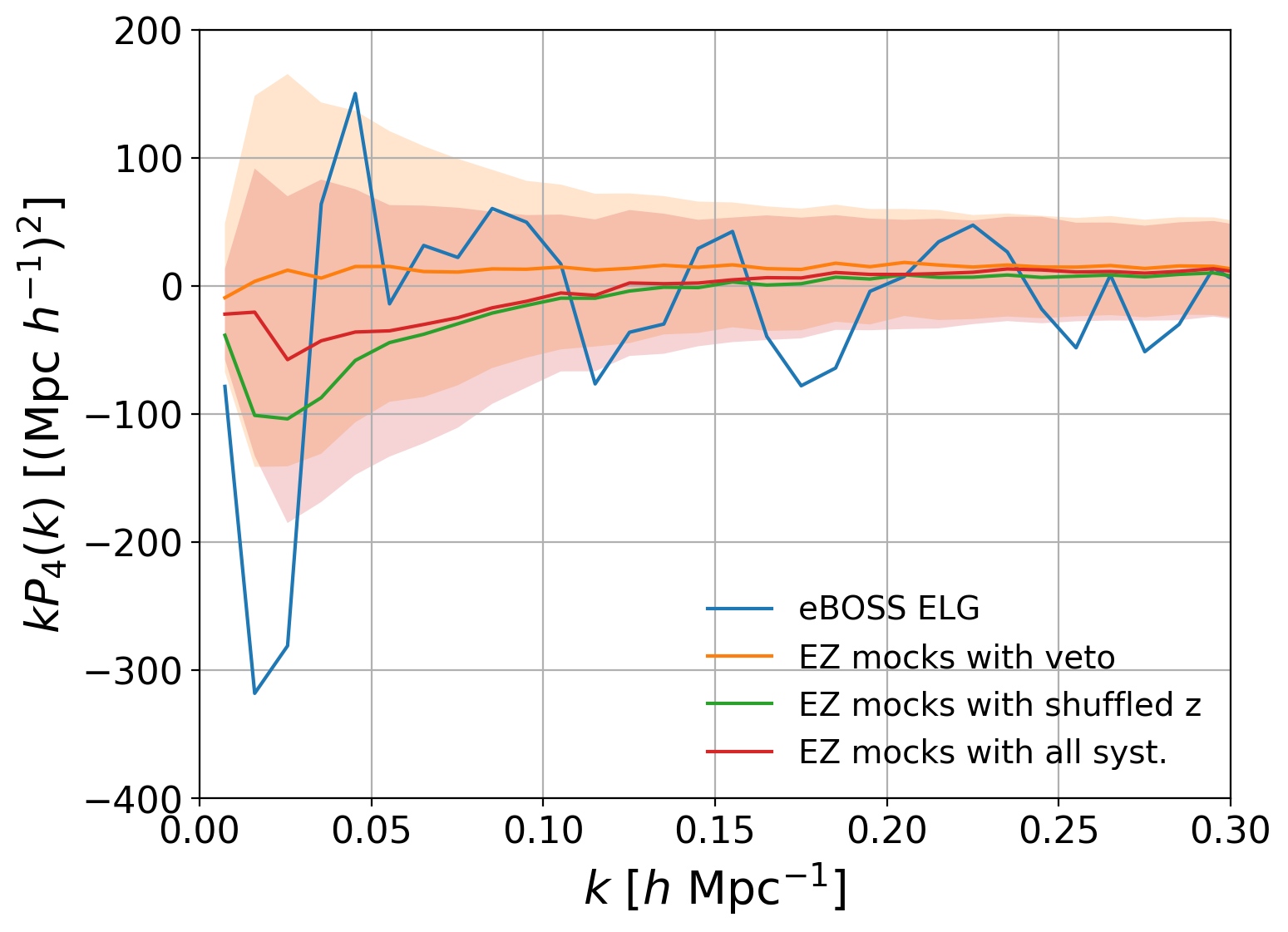}
\includegraphics[width=0.3\textwidth]{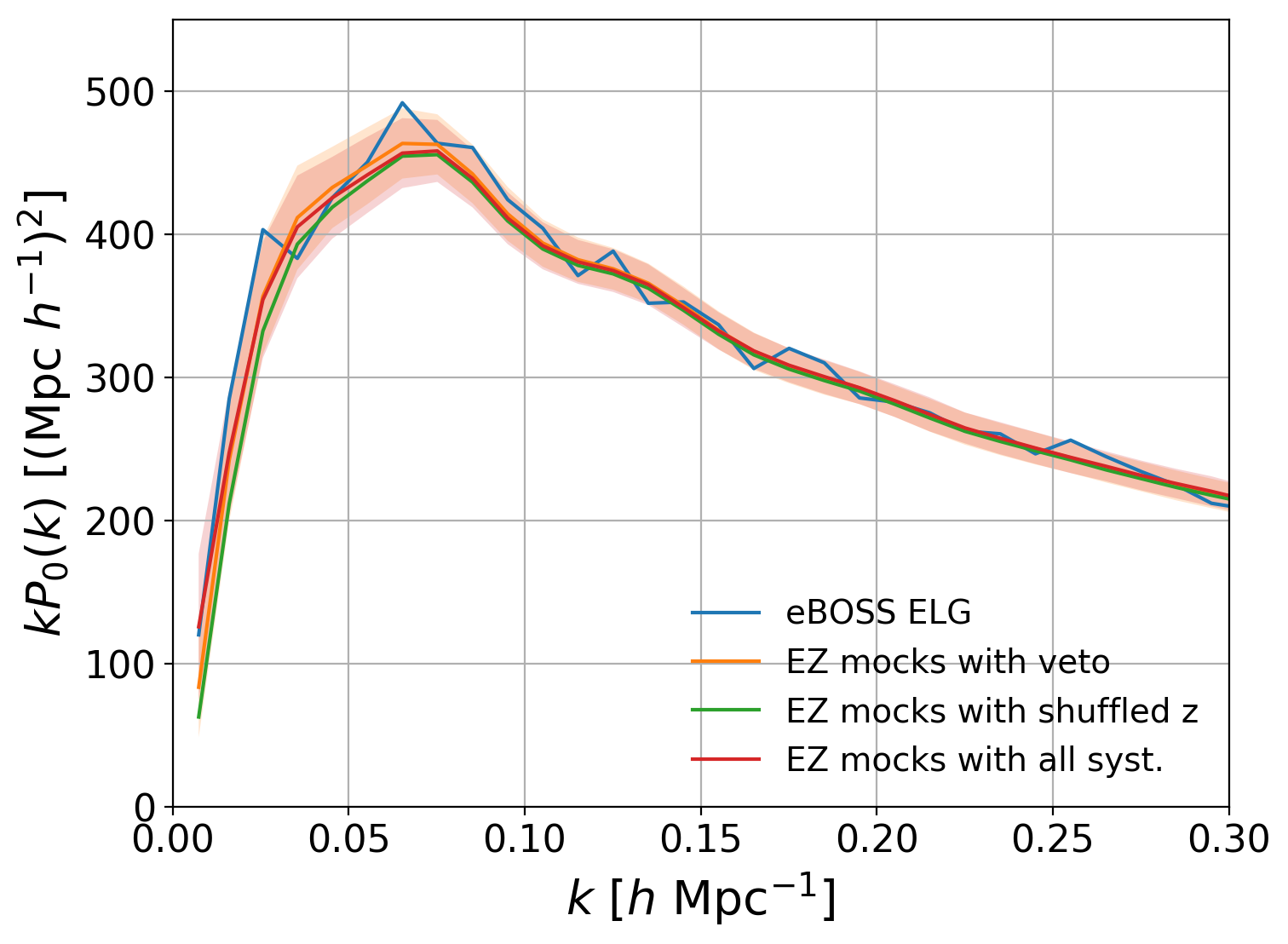}
\includegraphics[width=0.3\textwidth]{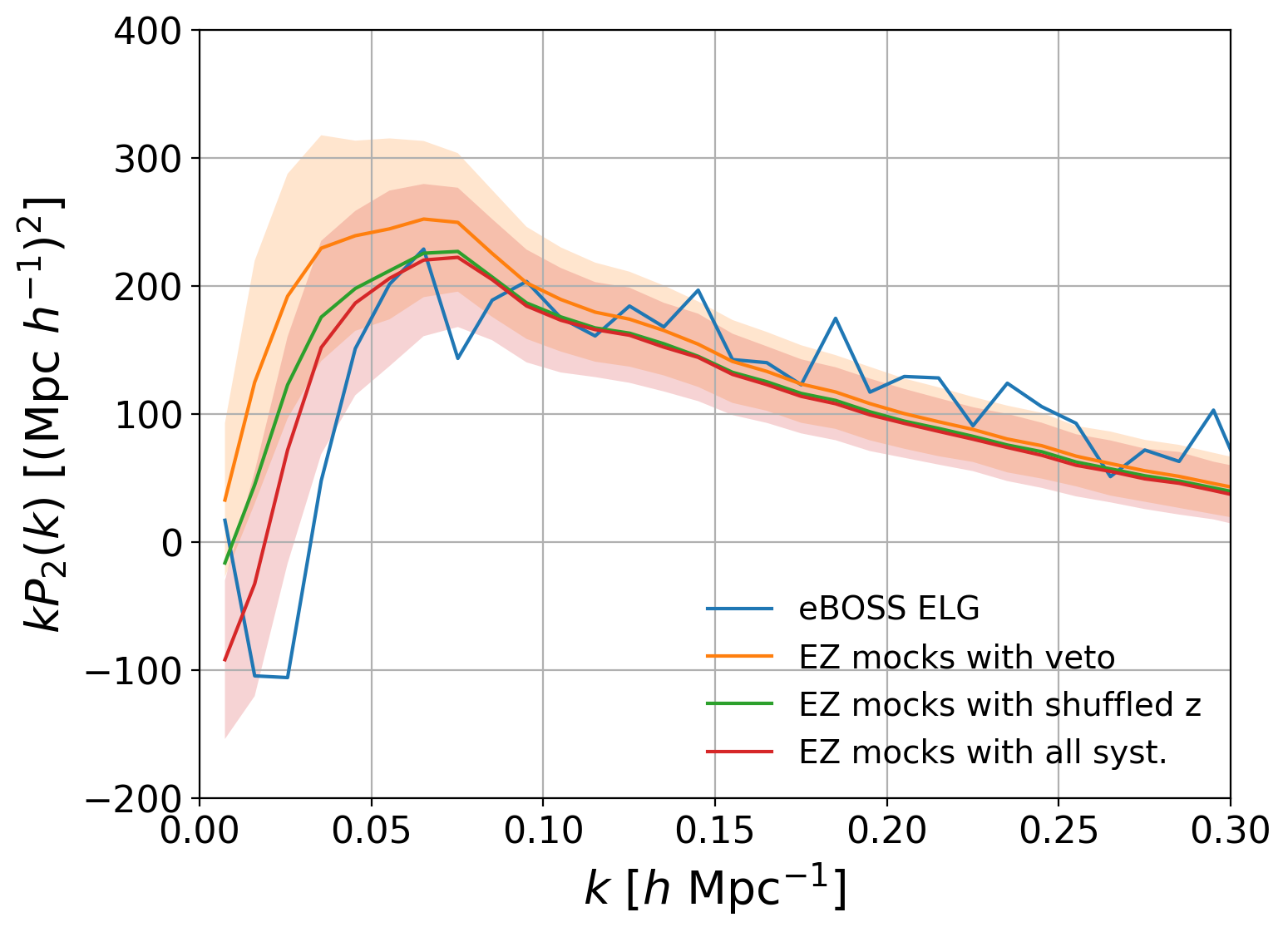}
\includegraphics[width=0.3\textwidth]{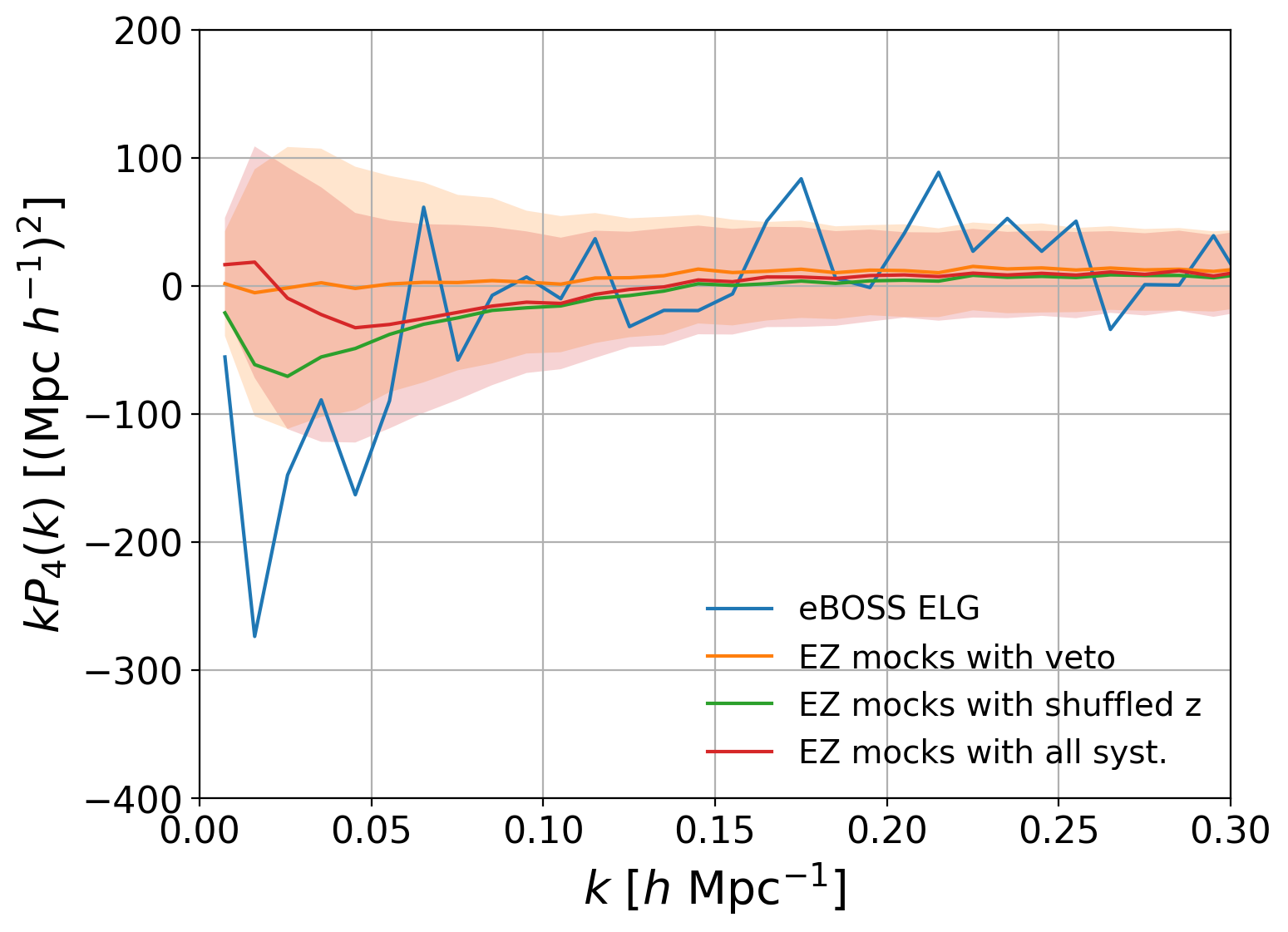}
\caption{Power spectrum measurements (left: monopole, middle: quadrupole, right: hexadecapole, top: NGC, bottom: SGC) of the eBOSS data (blue) and the mean and standard deviation (shaded region) of the EZ mocks without (orange) and with (red) all systematics. EZ mocks with the \emph{shuffled} scheme only (green) do not include observational systematics. The fitted $k$-range of the RSD measurement is $0.03 - 0.2 \hMpc$ for the monopole and quadrupole and $0.03 - 0.15 \hMpc$ for the hexadecapole.}
\label{fig:power_spectrum_data_ezmocks}
\end{figure*}

\begin{figure*}
\centering
\includegraphics[width=0.3\textwidth]{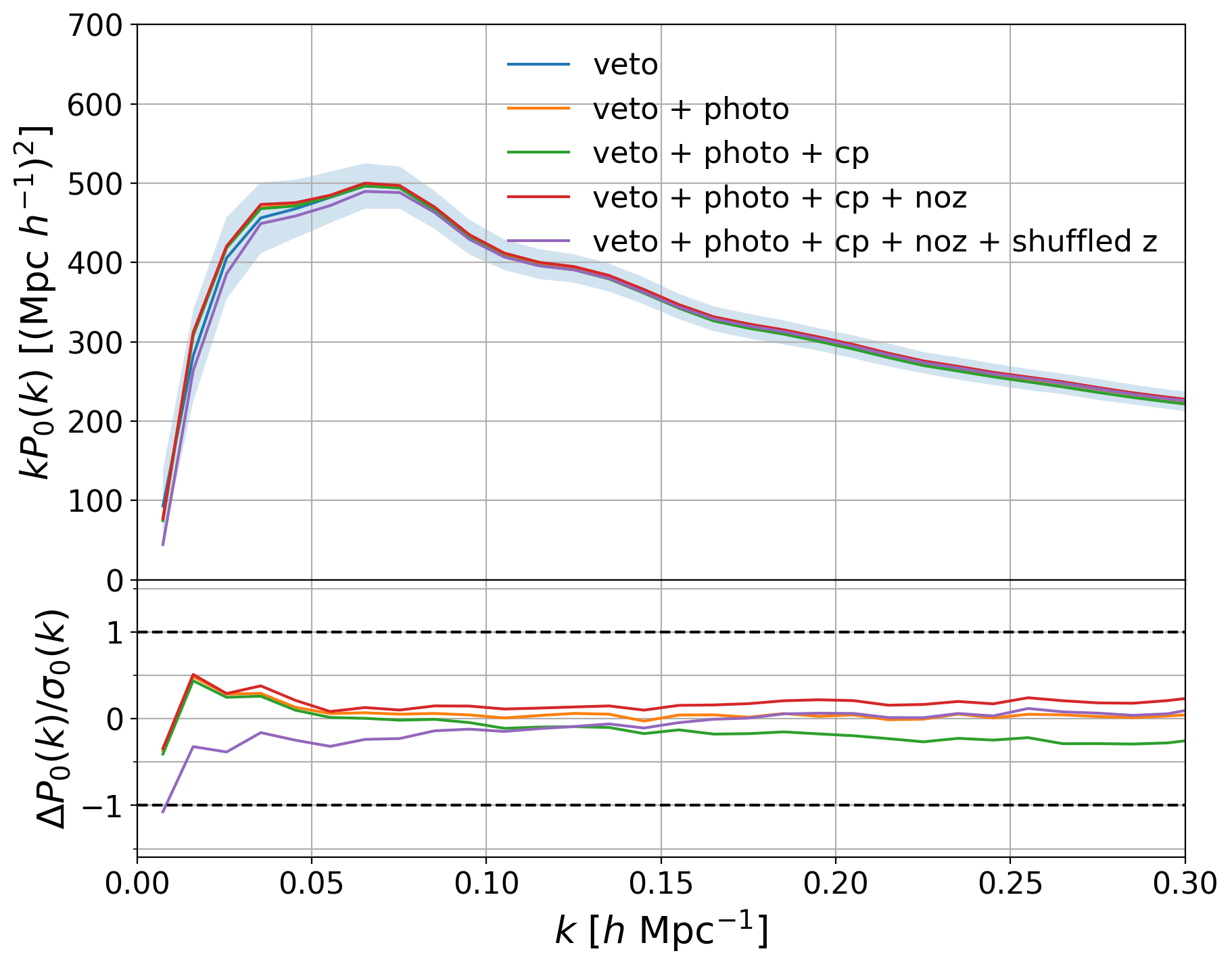}
\includegraphics[width=0.3\textwidth]{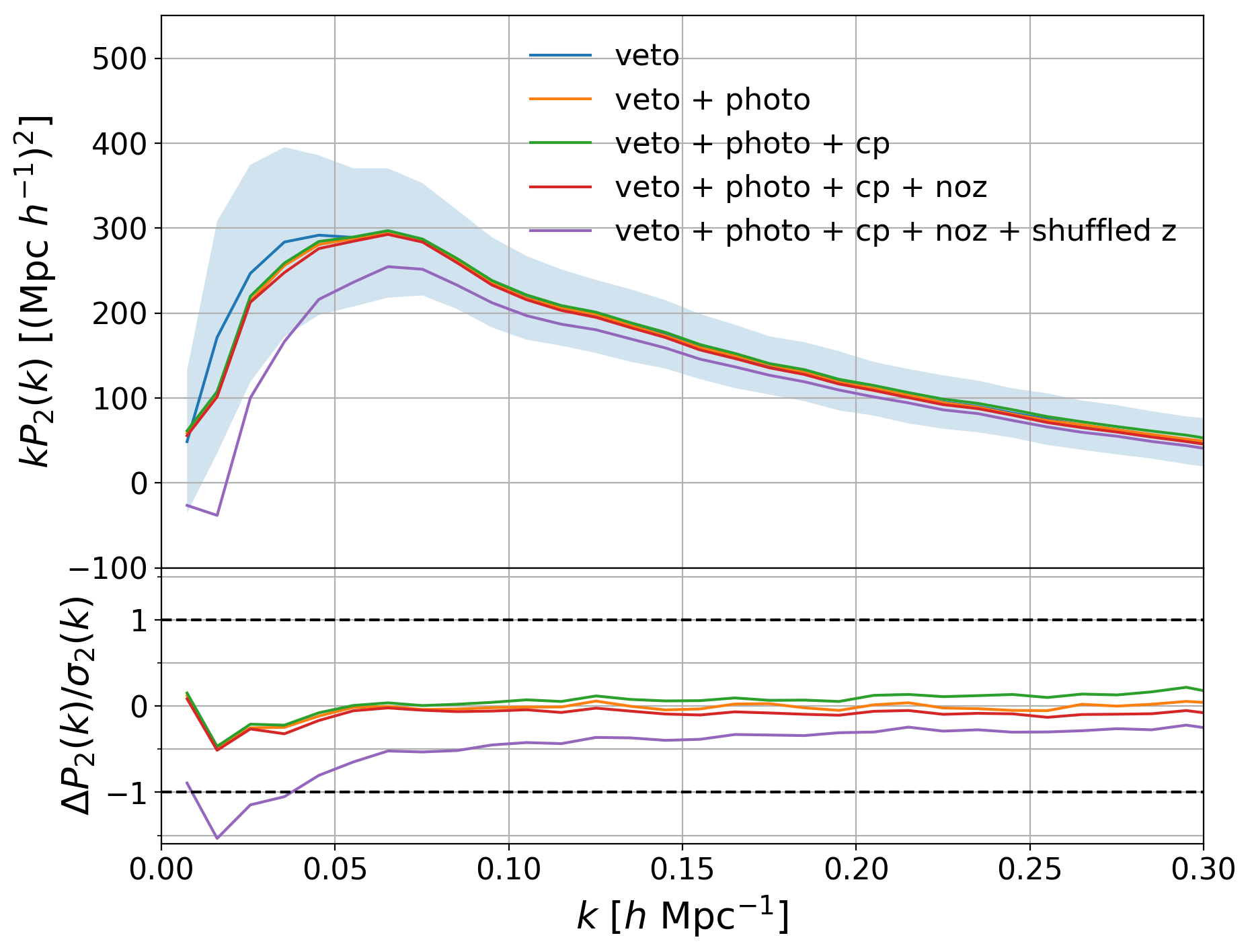}
\includegraphics[width=0.3\textwidth]{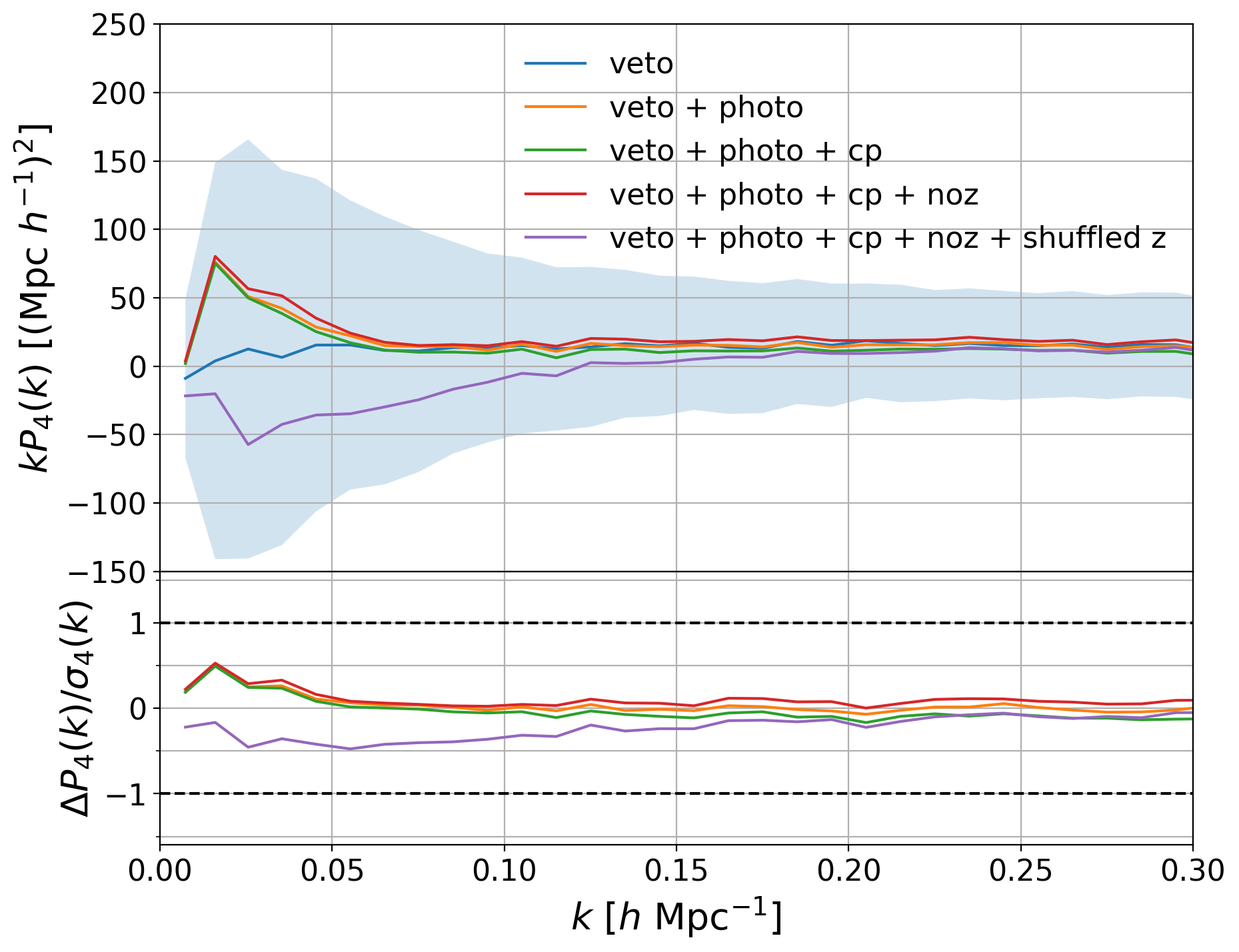}
\includegraphics[width=0.3\textwidth]{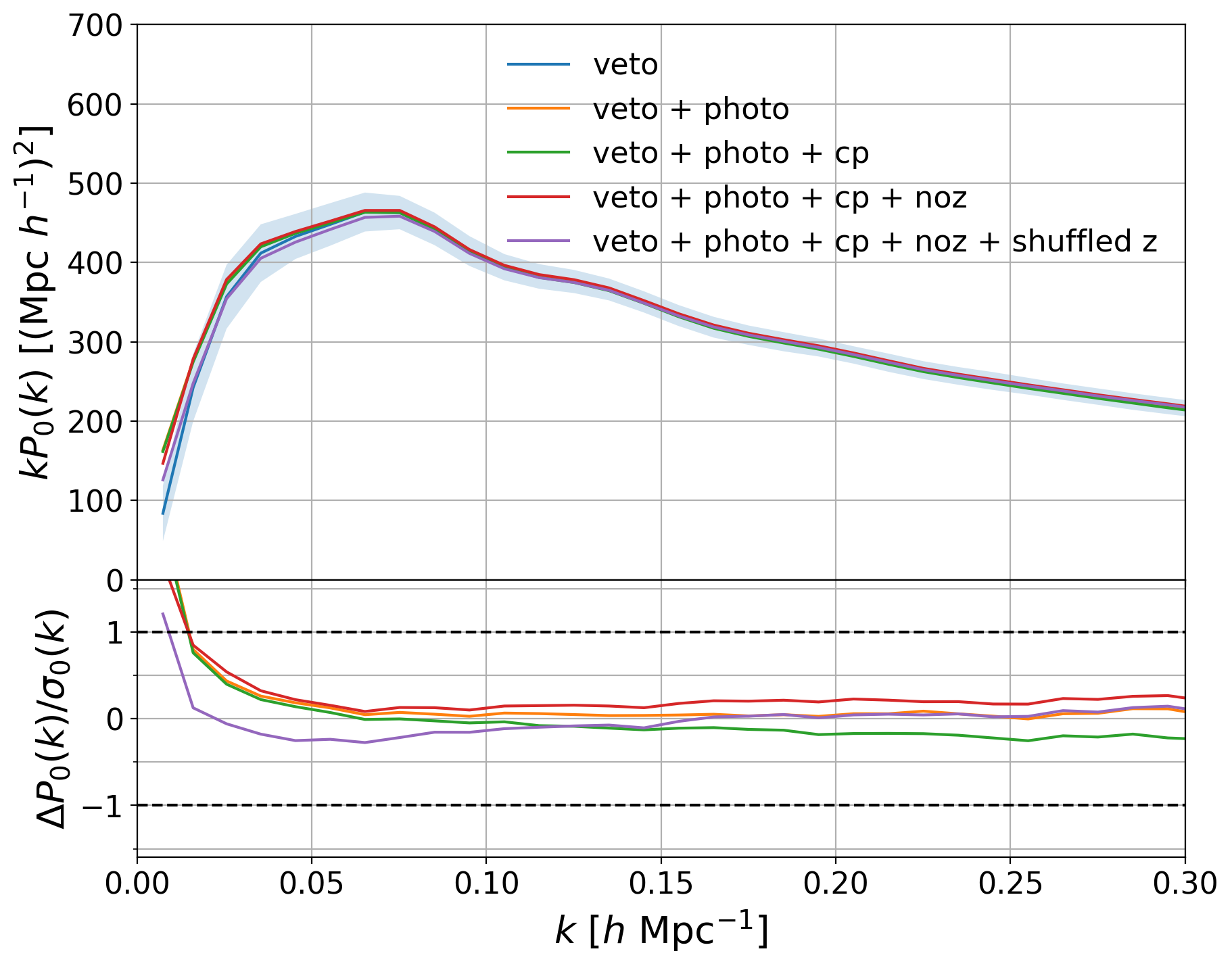}
\includegraphics[width=0.3\textwidth]{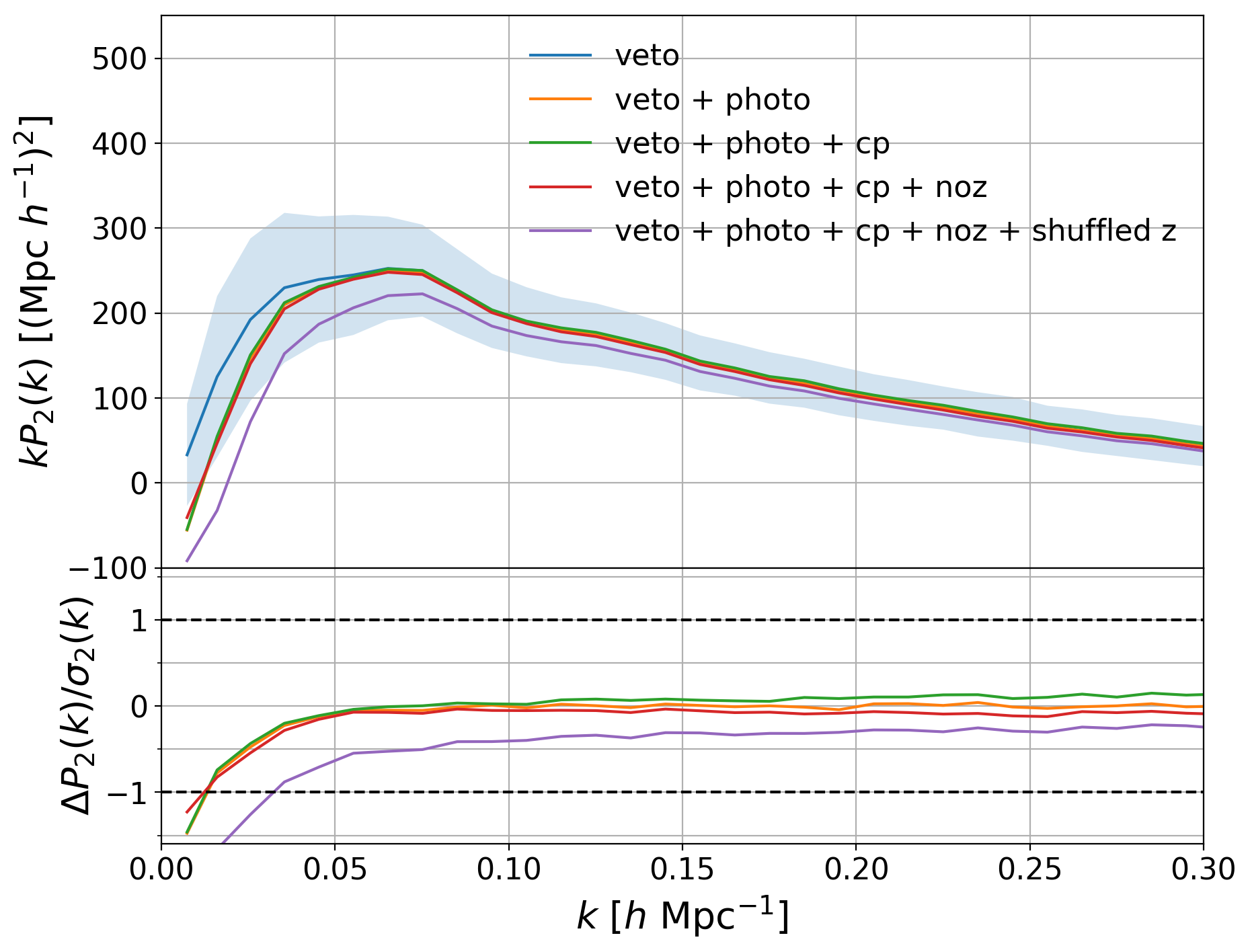}
\includegraphics[width=0.3\textwidth]{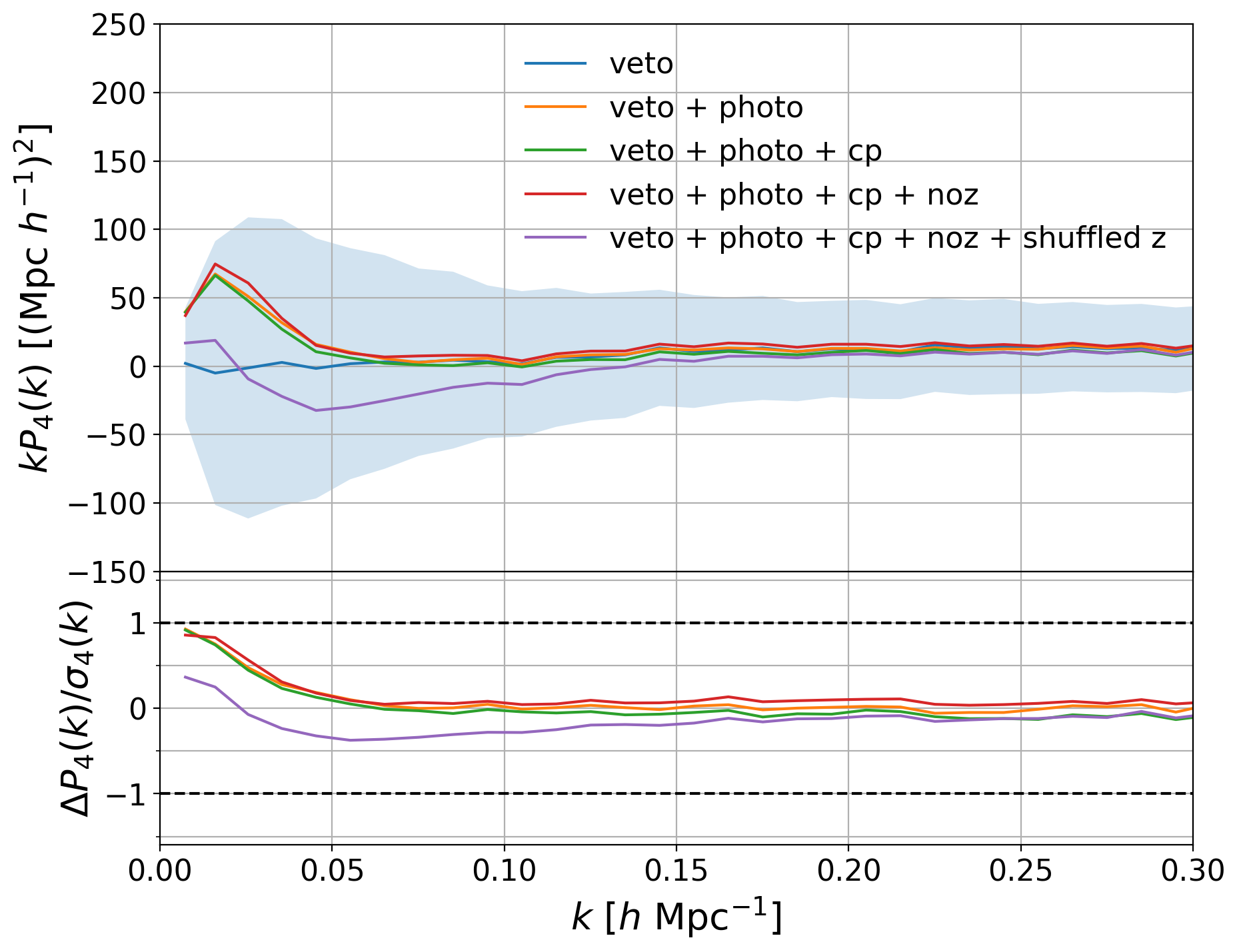}
\caption{Power spectrum measurements (left: monopole, middle: quadrupole, right: hexadecapole, top: NGC, bottom: SGC) of the EZ mocks, with different systematics and corrections applied successively. The blue shaded region represents the standard deviation of the mocks with veto masks only. %
Bottom panels: difference of the various schemes to the reference (with veto flag only), normalised by the standard deviation of the mocks. 
Note that redshift failures, as implemented in the EZ mocks, partially cancel the effect of fibre collisions.}
\label{fig:comparison_power_spectrum_ezmocks}
\end{figure*}

\Fig{comparison_power_spectrum_ezmocks} shows the different systematics and corrections applied successively to the EZ mocks. One can already see that angular photometric systematics (\emph{photo}) are the dominant ones. Another important effect is due to the \emph{shuffled} scheme used to assign redshifts to randoms from the mock data redshift distribution, which leads to the aforementioned radial integral constraint, clearly visible in the quadrupole and hexadecapole at large scale.

\Fig{power_spectrum_data_ezmocks} displays the power spectrum measurement of the eBOSS ELG sample (blue curve), together with the mean of the EZ mocks with veto masks only applied (\emph{baseline}, orange). Accounting for the \emph{shuffled} scheme in the mocks (green) resolves part of the difference between data and mocks in the quadrupole and hexadecapole on large scales. Including all systematics and corrections (red), the agreement with observed data is improved in the quadrupole.

\subsection{GLAM-QPM mocks}
\label{sec:qpmmocks}

The generation of GLAM-QPM mocks is detailed in~\cite{Lin2020}. The $2003$ GLAM-QPM mocks are built from boxes of side $3 \Gpch$, with cosmology:
\begin{equation}
\begin{split}
h = 0.678, \,\, \Omega_{m} = 0.307, \omega_{b} = 0.022,\\
\sigma_{8} = 0.828, \,\, n_{s} = 0.96. \qquad
\end{split}
\end{equation}

Mocks are trimmed to the tiling geometry and veto masks. Contrary to EZ mocks, we do not implement variations of the redshift distribution with imaging depth, nor imaging systematics. However, all other systematics (fibre collisions and redshift failures) are treated the same way as for EZ mocks. Again, all systematic corrections are applied the exact same way to the mocks as in the data catalogues.

\section{Testing the analysis pipeline using mock catalogues}
\label{sec:mocks_systematics}

In this section we first check our analysis pipeline and review how the observational systematics introduced in the approximate mocks impact BAO and RSD measurements. 
Although some systematic effects are difficult to model accurately in mocks, these can still be used to derive reliable estimates for part of the systematic uncertainties, a point we discuss also in this section. The other systematic uncertainties will be estimated from the data itself in \Sec{results}.

In all tests, to fit each type of mocks, we use the covariance matrix built from the same mocks, unless otherwise stated.

For reasons that will be justified in \Sec{results}, we will use NGC and SGC (NGC~+~SGC) or SGC only power spectrum measurements and vary the redshift range. The baseline result will use NGC~+~SGC, and the redshift ranges $0.7 < z < 1.1$ and $0.6 < z < 1.1$ for the RSD and BAO fits, respectively.

\subsection{Survey geometry effects}

The model presented in \Sec{model} neglects the evolution of the cosmological background within the redshift range of the eBOSS ELG sample. To test the impact of this assumption on clustering measurements, we first fit $300$ EZ periodic boxes at redshift $z_{\mathrm{snap}} = 0.876$ (see \Sec{ezmocks}), using a Gaussian covariance matrix, as in \Sec{mock_challenge_outerrim}. We compare these measurements to those obtained on the mean of the \emph{no veto} EZ mocks, that is including the (approximate) light-cone and global (tiling) footprint. In this case, we apply the corresponding window function treatment (\Sec{model_window_function}) and the global integral constraint (\Sec{model_integral_constraints}) in the model. To ease the comparison, which we present in \Tab{ezmocks_box_lightcone}, we extrapolate the best fits to the EZ boxes at redshift $z_{\mathrm{snap}} = 0.876$ to the effective redshift $\zeff = 0.845$ of the EZ mocks, using their input cosmology of \Eq{ez_cosmology}. The difference between the extrapolated mean of the best fits to the EZ boxes and the best fit to the mean of the EZ mocks is $0.2\%$ on $\fsig$, $0.4\%$ on $\DHu(z)/\rdrag$ and $0.2\%$ on $\DM(z)/\rdrag$, fully negligible compared to the dispersion of the mocks ($12\%$, $6\%$ and $5\%$ respectively, see \Tab{mocks_systematics}), validating our modelling approximation of the eBOSS ELG survey as a single snapshot at redshift $\zeff = 0.845$.
We finally apply veto masks to EZ mocks and in the window function calculation. In this case, again, the change in best fit parameters is $\simeq 0.1\%$, compatible with the error bars (\emph{baseline} versus \emph{no veto}). We also checked that increasing the sampling of the window function in the $s \rightarrow 0$ limit has virtually no impact ($0.01\%$) on the cosmological measurement. These tests validate our treatment of the window function with the fine-grained eBOSS ELG veto masks.

The total shifts between the \emph{baseline} sky-cut mocks and the EZ boxes are $0.1\%$, $0.5\%$ and $0.1\%$ for $\fsig$, $\apar$ and $\aper$. We take them as systematic shifts (under the generic denomination \emph{survey geometry}).

\begin{table*}
\caption{Comparison of the RSD measurements on EZ boxes at redshift $z_{\mathrm{snap}}=0.876$ and extrapolated at $\zeff=0.845$ (given their cosmology), with those from the sky-cut EZ mocks, with and without veto masks. For the EZ boxes we quote the mean and standard deviation of the best fit measurements, divided by the square root of the number of realisations ($300$). For the sky-cut mocks, error bars are given by the $\Delta \chi^{2} = 1$ level on the mean of the mocks.}
\label{tab:ezmocks_box_lightcone}
\centering
\begin{tabular}{lccc} 
\hline
& $\fsig$ & $\DHu/\rdrag$ & $\DM/\rdrag$\\
\hline
\hline
EZ boxes at $z_{\mathrm{snap}}=0.876$ & ${0.43088}_{-0.00017}^{+0.00017}$ & ${18.2186}_{-0.0034}^{+0.0034}$ & ${20.7175}_{-0.0027}^{+0.0027}$\\
EZ boxes at $\zeff=0.845$ & ${0.43391}_{-0.00017}^{+0.00017}$ & ${18.5590}_{-0.0034}^{+0.0034}$ & ${20.1526}_{-0.0026}^{+0.0026}$\\
Mean of EZ mocks no veto ($\zeff=0.845$) & ${0.4346}_{-0.0017}^{+0.0017}$ & ${18.477}_{-0.033}^{+0.033}$ & ${20.103}_{-0.028}^{+0.028}$\\
Mean of EZ mocks baseline ($\zeff=0.845$) & ${0.4341}_{-0.0017}^{+0.0017}$ & ${18.472}_{-0.034}^{+0.033}$ & ${20.127}_{-0.030}^{+0.031}$\\
\hline
\end{tabular}
\end{table*}

\subsection{Fibre collisions}
\label{sec:mocks_fibre_collisions}

Fibre collisions are shown to be the dominant observational systematics in the eBOSS QSO sample~\citep{Neveux2020,Hou2020}. The impact of fibre collisions can be seen on EZ mocks by comparing the green to the orange curves in \Fig{comparison_power_spectrum_ezmocks}. Here we test their effect on cosmological fits to GLAM-QPM mocks, as these mocks are not further impacted by photometric systematics.

We report in \Tab{mocks_systematics} the best fits to $2003$ GLAM-QPM \emph{baseline} mocks with only geometry and veto masks applied (\emph{baseline}) and to the mocks where fibre collisions are simulated (\emph{fibre collisions}). We find a systematic shift of $2.5\%$ on $\fsig$ ($22\%$ of the dispersion of the mocks), $0.6\%$ on $\apar$ ($9\%$) and $0.5\%$ on $\aper$ ($10\%$).

The impact of fibre collisions can be mitigated following~\cite{Hahn2017:1609.01714v1}, if the fraction of collided pairs $f_{s}$ and the fibre collision angular scale $D_{fc}$ are known. In the \cite{Hahn2017:1609.01714v1} correction, $f_{s} = 1$ corresponds to all galaxy pairs closer than the fibre collision angular scale being unobserved. Because of tile overlaps, this fraction is reduced. The fraction of collided pairs $f_{s}$ can then be estimated in several ways:
\begin{itemize}
\setlength\itemsep{-1em}
\item \emph{tile overlap}: the fraction of the survey area without tile overlap, estimated using the synthetic catalogue. This assumes all collisions are resolved in tile overlaps;\\
\item \emph{collision fraction}: the number of targets which were collided with another one (including the relevant TDSS targets), divided by the number of targets that would be assigned a fibre without tile overlap. This number is simulated with the same algorithm as that used for the EZ and GLAM-QPM mocks to implement fibre collisions, except the effect of tile overlaps (see \Sec{ezmocks} and~\ref{sec:qpmmocks});\\
\item \emph{simulated collision fraction}: same as \emph{collision fraction}, but also simulating the number of data targets which were collided with another one (including the relevant TDSS targets), taking into account tile overlaps;\\
\item \emph{EZ simulated collision fraction}: same as \emph{simulated collision fraction}, in the EZ mocks;\\
\item \emph{GLAM-QPM simulated collision fraction}: same as \emph{simulated collision fraction}, in the GLAM-QPM mocks.

\end{itemize}
\begin{table}
\caption{Different estimates of the fibre collisions fraction $f_{s}$. See text for details.}
\label{tab:fibre_collisions}
\centering
\begin{tabular}{lcc}
\hline
& NGC & SGC\\
\hline\hline
tile overlap & $0.44$ & $0.35$\\
collision fraction & $0.47$ & $0.39$\\
simulated collision fraction & $0.46$ & $0.38$\\
EZ simulated collision fraction & $0.46 \pm 0.005$ & $0.38 \pm 0.004$\\
GLAM-QPM simulated collision fraction & $0.46 \pm 0.005 $ & $0.39 \pm 0.005$\\
\hline
\end{tabular}
\end{table}

\begin{figure}
\centering
\includegraphics[width=0.45\columnwidth]{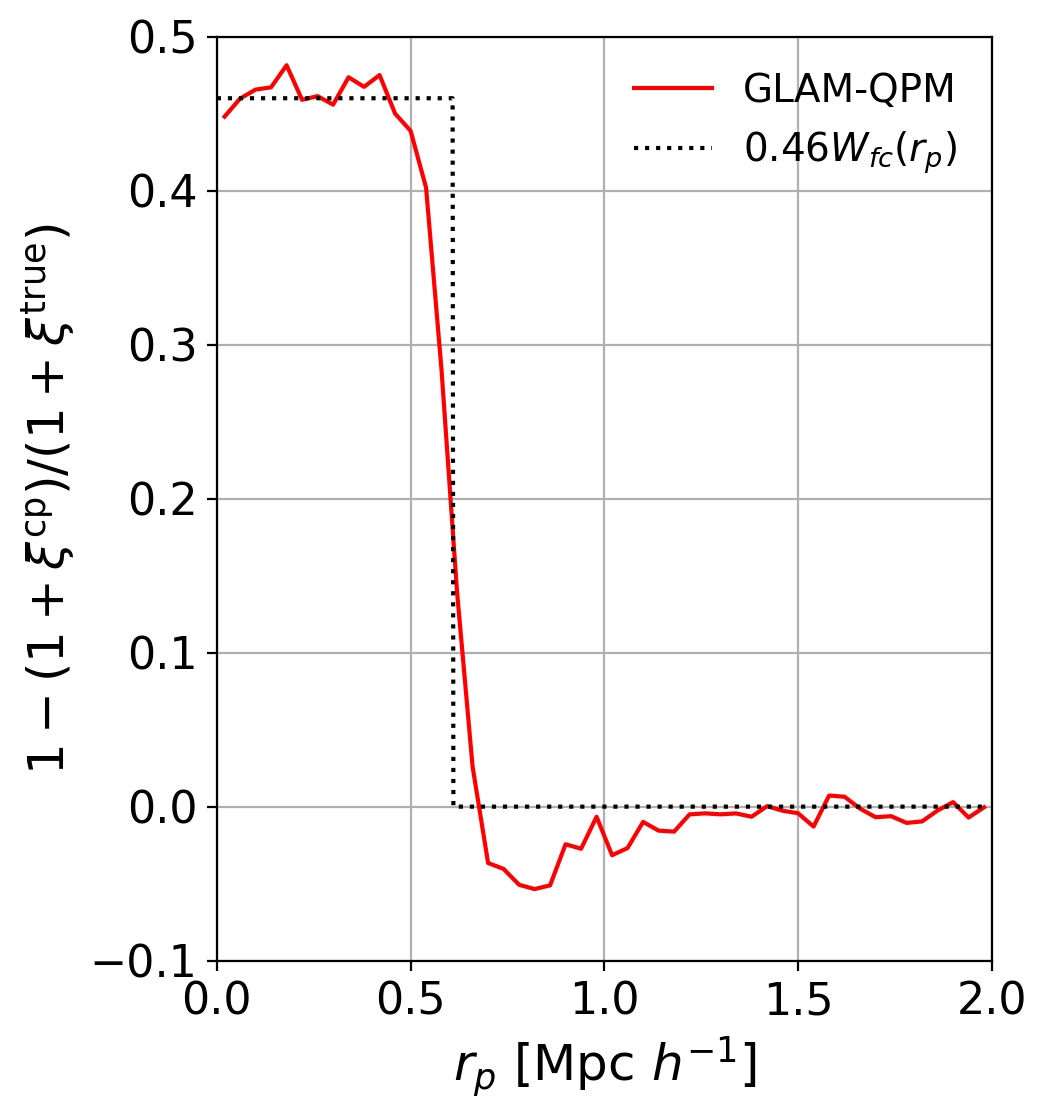}
\includegraphics[width=0.45\columnwidth]{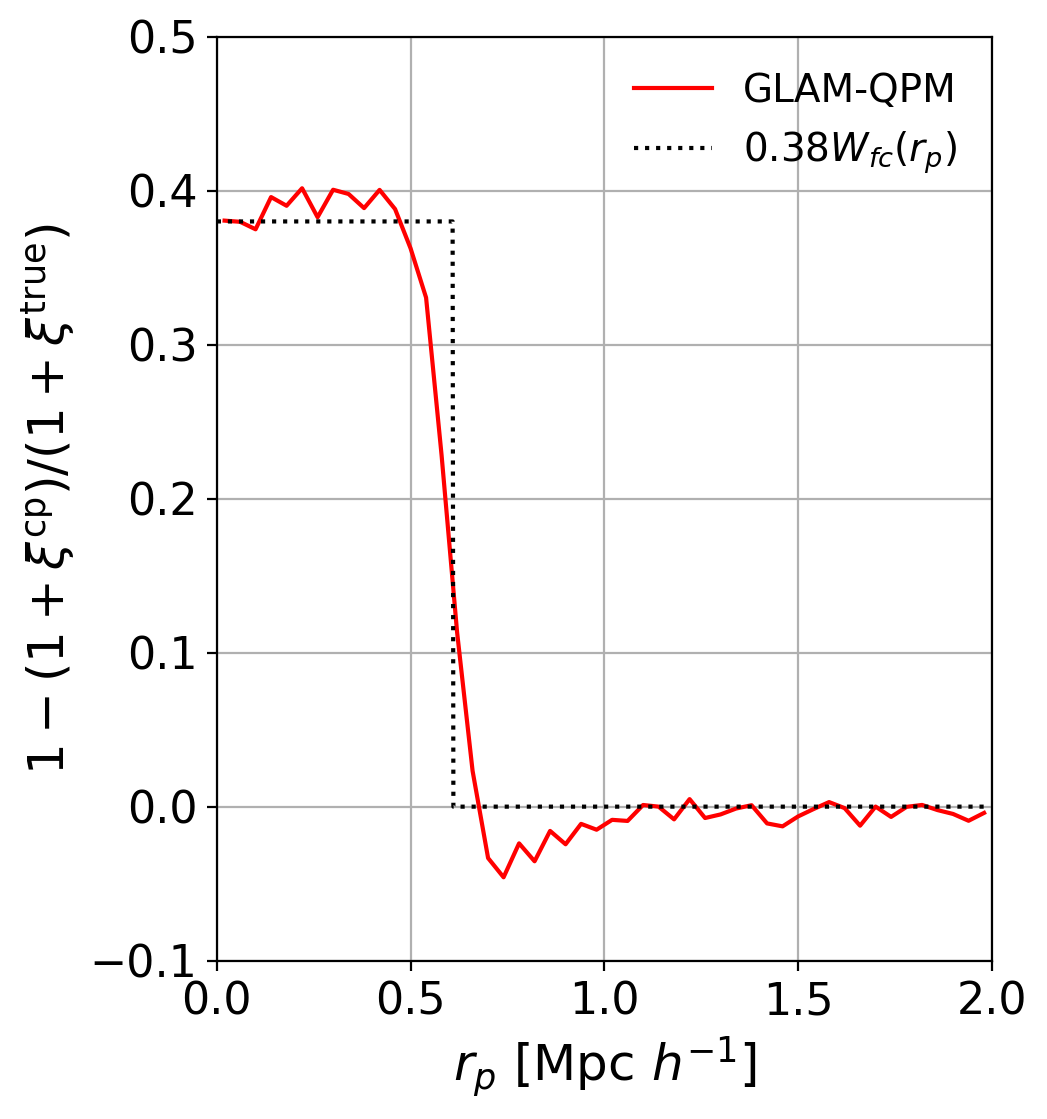}
\caption{Ratio of the $\wcp$-corrected projected correlation function to the true projected correlation function, presented in the form $1 - (1 + \xi^{\mathrm{cp}})/(1 + \xi^{\mathrm{true}})$, as obtained in $379$ GLAM-QPM mocks and in the model of~\citet{Hahn2017:1609.01714v1} (left: NGC, right: SGC). See text for details, and Figure~8 of~\citet{Hahn2017:1609.01714v1} for comparison.}
\label{fig:projected_correlation_qpmmocks}
\end{figure}

All these estimates are calculated with veto masks applied and are reported in \Tab{fibre_collisions}, using $50$ mocks (for \emph{EZ} and \emph{GLAM-QPM simulated collision fraction}). They all agree within $2\%$. 
The modelling of fibre collisions in~\citet{Hahn2017:1609.01714v1} is actually based on their impact on the projected correlation function. \Fig{projected_correlation_qpmmocks} displays the ratio of the projected correlation function of the GLAM-QPM mocks with fibre collisions corrected by $\wcp$ to the true one (without fibre collisions): $f_{s}$, given by the height of the step function (see~\citealt{Hahn2017:1609.01714v1}), is in very good agreement with the above estimates provided in \Tab{fibre_collisions}. We therefore choose the corresponding values $f_{s} = 0.46$ for NGC and $f_{s} = 0.38$ for SGC.

For the fibre collision angular scale $D_{fc}$, we take the comoving distance corresponding to the fibre collision radius $62''$ at the effective redshift of the eBOSS ELG sample $\zeff = 0.845$. The obtained value, $0.61 \Mpch$,
provides good modelling of the effect as can be seen in \Fig{projected_correlation_qpmmocks}. For the redshift cut $0.7 < z < 1.1$, a similar calculation provides $D_{fc} = 0.62 \Mpch$.

The parameters $f_{s}$ and $D_{fc}$ being determined, the~\cite{Hahn2017:1609.01714v1} correction can be included in the RSD model. Best fits to the GLAM-QPM mocks with fibre collisions are in very good agreement with the \emph{baseline} mocks %
once the correction is included: the potential remaining systematic bias is $0.3\%$ on $\fsig$, $0.1\%$ on $\apar$ and $0.0\%$ on $\apar$ --- $3\%$, $2\%$ and $0\%$ of the dispersion of the mocks, respectively (\emph{fibre collisions + Hahn et al.} versus \emph{baseline} in \Tab{mocks_systematics}). We therefore include this correction as a baseline in the following.

Note that~\cite{Bianchi2017:1703.02070v2,Percival2017:1703.02071v3} developed a method to correct for such missing observations in the $n$-point (configuration space) correlation function using $n$-tuple upweighting; for an application to the eBOSS samples (including ELG), we refer the reader to~\cite{Mohammad2020}. This method has been very recently extended to the Fourier space analysis by~\cite{Bianchi2019:1912.08803v1}. We do not apply this technique to the eBOSS ELG sample, since most of this analysis was completed before this publication and because the effect of fibre collisions appears subdominant, especially after the~\cite{Hahn2017:1609.01714v1} correction.

\subsection{Radial integral constraint}
\label{sec:mocks_radial_integral_constraint}

As mentioned in \Sec{ezmocks}, the \emph{shuffled} scheme, used to assign data redshifts to randoms is responsible for a major shift of the power spectrum multipoles (purple versus red curves in \Fig{comparison_power_spectrum_ezmocks}). As discussed in~\cite{deMattia2019:1904.08851v3}, this damping of the power spectrum multipoles on large scales is not specific to the \emph{shuffled} scheme, but to any method measuring the radial selection function on the observed data itself. We report in \Tab{mocks_systematics} the cosmological measurements from RSD fits without (\emph{baseline}, GIC) and with the \emph{shuffled} scheme (\emph{shuffled}, GIC), while keeping the global integral constraint (GIC) in the model: the induced systematic shift is $0.4\%$ on $\fsig$ ($4\%$ of the dispersion of the mocks), $4.4\%$ on $\apar$ ($66\%$) and $3.9\%$ on $\aper$ ($78\%$). Modelling the radial integral constraint (RIC) removes most of this bias: the remaining shift is $0.2\%$ on $\fsig$ ($2\%$ of the dispersion of the mocks), $0.2\%$ on $\apar$ ($3\%$) and $0.3\%$ on $\aper$ ($5\%$).

\subsection{Remaining angular systematics}
\label{sec:mocks_angular_systematics}

In \Sec{data_mocks} we mentioned the large angular photometric systematics of the eBOSS ELG sample, which we attempted to introduce in the EZ mocks (orange versus blue curves in \Fig{comparison_power_spectrum_ezmocks}). These systematics bias cosmological measurements from RSD fits, as can be seen in \Tab{mocks_systematics}: comparing the fits on contaminated mocks, including the fibre collision correction of \Sec{mocks_fibre_collisions} (\emph{all syst., fc}) to uncontaminated mocks (\emph{baseline}, GIC), one notices a bias of $8.8\%$ on $\fsig$, $2.4\%$ on $\apar$ and $1.6\%$ on $\aper$, corresponding to a significant shift of respectively $75\%$, $35\%$ and $32\%$ of the dispersion of the best fits to the mocks.

We propose to mitigate these residual systematics by rescaling weighted randoms in \texttt{HEALPix}\footnote{\url{http://healpix.jpl.nasa.gov/}}~\citep{Gorski2005:astro-ph/0409513v1} pixels such that the density fluctuations $F(\vr)$ of \Eq{fkp_field} are forced to $0$ in each pixel (a scheme which will be referred to as the \emph{pixelated} scheme in the following). This leads to an angular integral constraint (AIC), which we model and combine with the radial IC following~\cite{deMattia2019:1904.08851v3}. In \Tab{mocks_systematics} we report the RSD measurements without (\emph{baseline}, GIC) and with the full angular and radial integral constraints (ARIC) modelled, applying the \emph{shuffled} and \emph{pixelated} schemes to the  uncontaminated mock data, for two pixel sizes: $\mathrm{nside} = 64$ ($\simeq 0.84 \deg^{2}$) and $\mathrm{nside} = 128$ ($\simeq 0.21 \deg^{2}$). The combined radial and angular integral constraint is correctly modelled, generating only a small potential bias of $1.1\%$ on $\fsig$, $0.5\%$ and $0.4\%$ on scaling parameters (which amounts to $10\%$, $8\%$ and $7\%$ of the dispersion of the mocks, respectively) for $\mathrm{nside} = 64$. A similar shift is seen with $\mathrm{nside} = 128$. The \emph{pixelated} scheme increases statistical uncertainties by a reasonable fraction of $\simeq 10\%$.

\begin{figure}
\centering
\includegraphics[width=\columnwidth]{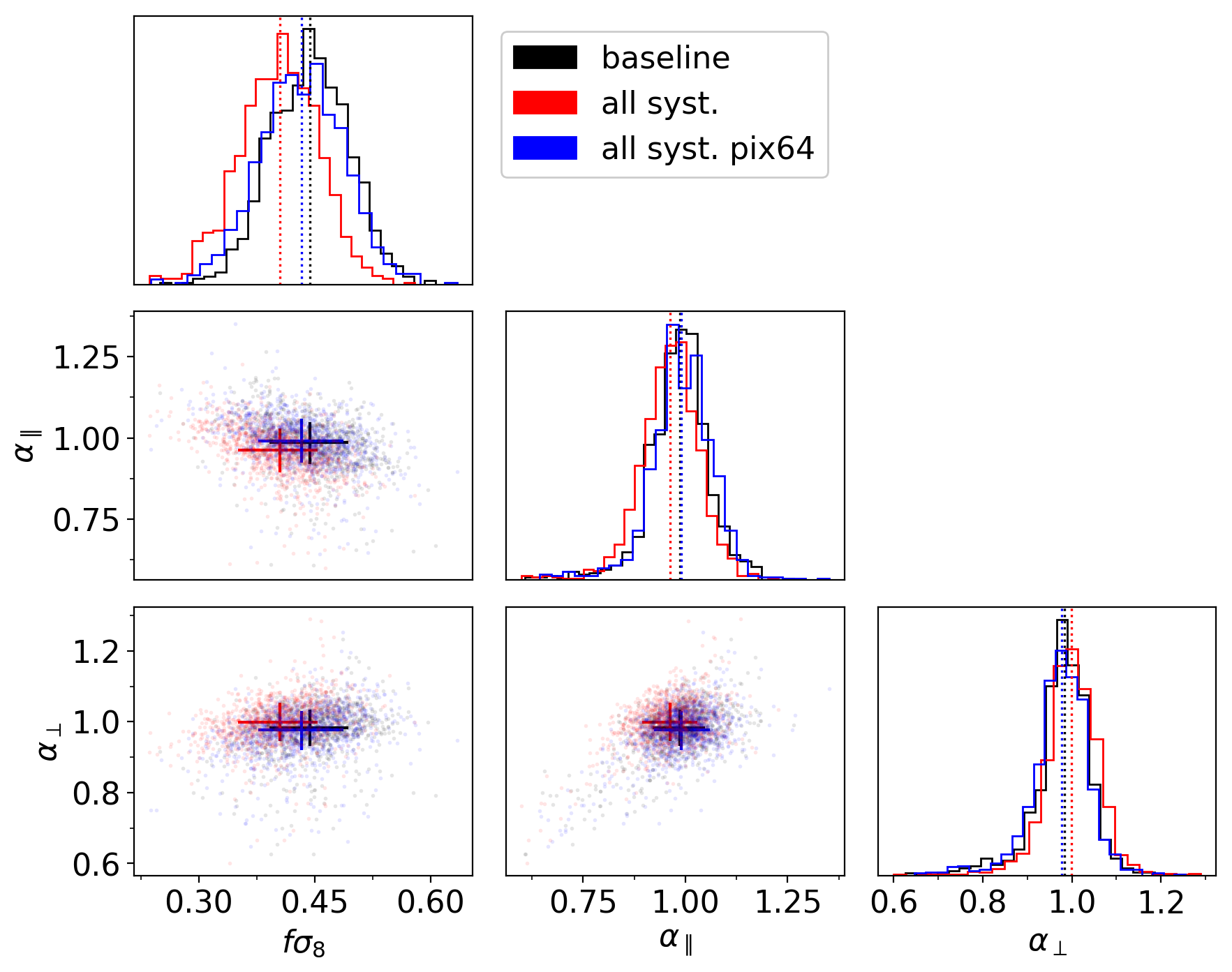}
\caption{Best fits to the baseline (black) and fully contaminated (red) EZ mocks. In blue, the \emph{pixelated} scheme is applied on contaminated mocks to mitigate angular systematics. Dotted vertical lines in the histograms and crosses in the scatter plots point to the median of the best fit values, while the size of the crosses is given by the $16\%$ and $84\%$ percentiles.
}
\label{fig:mocks_angular_systematics}
\end{figure}

Finally, \Fig{mocks_angular_systematics} shows the best fits to the \emph{baseline} (blue) and contaminated (red) EZ mocks. Measurements obtained when applying the \emph{pixelated} scheme ($\mathrm{nside} = 64$) to the contaminated mocks and modelling the ARIC are shown in blue. The systematic bias quoted at the beginning of the section is clearly reduced and becomes $2.5\%$ on $\fsig$ ($21\%$ of the dispersion of the mocks), $0.4\%$ on $\apar$ ($6\%$) and $0.5\%$ on $\aper$ ($12\%$) with $\mathrm{nside} = 64$, slightly less with $\mathrm{nside} = 128$ (see \Tab{mocks_systematics}, \emph{all syst \& pix64, fc} with respect to \emph{baseline, GIC}).

\subsection{Likelihood Gaussianity}
\label{sec:mocks_gaussianity}

In \Sec{fitting_methodology} we assumed that we could use a Gaussian likelihood to compare data and model. While this may be accurate enough by virtue of the central limit theorem when the number of modes is high enough, it may break down on large scales where statistics is lower and mode coupling due to the survey geometry, and, in our specific case, RIC and ARIC, occurs (see e.g.~\citealt{Hahn2019:1803.06348v1}).

Comparing the median of the fits to each individual \emph{baseline} EZ mocks to the fit to the mean of the mocks (see first two rows of second series of results in \Tab{mocks_systematics}, \emph{baseline, GIC} versus \emph{mean of mocks baseline, GIC}), we observe shifts of $2.2\%$ on $\fsig$ ($19\%$ of the dispersion of the mocks), $1.3\%$ on $\apar$ ($21\%$) and $0.9\%$ on $\aper$ ($18\%$). This bias could be due to either non-Gaussianity of the power spectrum likelihood or model non-linearity.

To test a potential bias coming from the breakdown of such a Gaussian assumption, we produce $1000$ fake power spectra following a Gaussian distribution around the mean of the EZ mocks, with the covariance of the mocks, and fit them with our model (using the same covariance matrix). Results are reported in \Tab{mocks_systematics} (\emph{fake all syst. \& pix64, fc}). Shifts with respect to the true mocks (\emph{all syst. \& pix64, fc}) are $0.2\%$ on $\fsig$, $0.1\%$ on $\apar$ and $0.2\%$ on $\aper$.

We therefore conclude that one can safely use the Gaussian likelihood to compare data and model power spectra. We also attribute the shifts between the fit to the mean of the EZ mocks and the median of the fits to each mock to model non-linearity.

\begin{table*}
\caption{Impact of systematics on RSD and RSD~+~BAO measurements on the GLAM-QPM and EZ mocks. We quote the median and the $16\%$ and $84\%$ percentiles as a metric of the centre and dispersion of the measurements. 
}
\label{tab:mocks_systematics}
\centering
\begin{tabular}{lccc}
\hline
& $\fsig$ & $\apar$ & $\aper$ \\
\hline
\hline
RSD only & GLAM-QPM mocks & &\\
\hline
baseline & ${0.442}_{-0.050}^{+0.050}$ & ${0.997}_{-0.066}^{+0.064}$ & ${0.990}_{-0.054}^{+0.048}$\\
fibre collisions & ${0.453}_{-0.049}^{+0.050}$ & ${0.990}_{-0.066}^{+0.063}$ & ${0.995}_{-0.054}^{+0.048}$\\
fibre collisions + Hahn et al. & ${0.443}_{-0.050}^{+0.049}$ & ${0.998}_{-0.066}^{+0.063}$ & ${0.990}_{-0.055}^{+0.048}$\\
\hline

RSD only & EZ mocks & tests of IC &\\
\hline
mean of mocks baseline, GIC & ${0.4341}_{-0.0017}^{+0.0017}$ & ${0.9997}_{-0.0019}^{+0.0018}$ & ${0.9926}_{-0.0015}^{+0.0015}$\\
baseline, GIC & ${0.444}_{-0.052}^{+0.050}$ & ${0.987}_{-0.066}^{+0.061}$ & ${0.984}_{-0.053}^{+0.050}$\\
shuffled, GIC & ${0.446}_{-0.054}^{+0.049}$ & ${0.943}_{-0.069}^{+0.060}$ & ${1.022}_{-0.052}^{+0.053}$\\
shuffled, RIC & ${0.443}_{-0.054}^{+0.051}$ & ${0.985}_{-0.064}^{+0.061}$ & ${0.986}_{-0.052}^{+0.049}$\\
shuffled \& pix64, ARIC & ${0.449}_{-0.057}^{+0.053}$ & ${0.982}_{-0.064}^{+0.063}$ & ${0.987}_{-0.053}^{+0.055}$\\
shuffled \& pix128, ARIC & ${0.450}_{-0.061}^{+0.054}$ & ${0.983}_{-0.066}^{+0.064}$ & ${0.987}_{-0.053}^{+0.055}$\\
\hline

RSD only & EZ mocks & mitigation &\\
\hline
all syst. fc & ${0.405}_{-0.054}^{+0.049}$ & ${0.964}_{-0.069}^{+0.065}$ & ${0.999}_{-0.054}^{+0.055}$\\
all syst. \& pix64 fc & ${0.433}_{-0.056}^{+0.054}$ & ${0.990}_{-0.067}^{+0.069}$ & ${0.978}_{-0.059}^{+0.052}$\\
all syst. \& pix128 fc & ${0.438}_{-0.057}^{+0.056}$ & ${0.987}_{-0.068}^{+0.070}$ & ${0.979}_{-0.058}^{+0.054}$\\
fake all syst. \& pix64 fc & ${0.434}_{-0.059}^{+0.054}$ & ${0.991}_{-0.058}^{+0.060}$ & ${0.979}_{-0.055}^{+0.053}$\\
\hline

RSD~+~BAO & EZ mocks & &\\
\hline
mean of mocks baseline, GIC & ${0.4384}_{-0.0017}^{+0.0016}$ & ${1.0031}_{-0.0019}^{+0.0017}$ & ${0.9979}_{-0.0016}^{+0.0012}$\\
baseline, GIC & ${0.445}_{-0.048}^{+0.048}$ & ${0.994}_{-0.051}^{+0.057}$ & ${0.994}_{-0.040}^{+0.037}$\\
shuffled \& pix64, ARIC & ${0.449}_{-0.051}^{+0.050}$ & ${0.987}_{-0.055}^{+0.057}$ & ${0.995}_{-0.044}^{+0.041}$\\
all syst., fc & ${0.404}_{-0.049}^{+0.049}$ & ${0.973}_{-0.054}^{+0.055}$ & ${1.010}_{-0.048}^{+0.043}$\\
all syst. \& pix64, fc & ${0.434}_{-0.051}^{+0.053}$ & ${1.000}_{-0.056}^{+0.058}$ & ${0.989}_{-0.048}^{+0.046}$\\
all syst. \& pix64, fc, $B_{\mathrm{nw}}$ free & ${0.434}_{-0.052}^{+0.053}$ & ${1.002}_{-0.056}^{+0.057}$ & ${0.989}_{-0.048}^{+0.045}$\\
all syst. \& pix128, fc & ${0.439}_{-0.051}^{+0.054}$ & ${0.998}_{-0.057}^{+0.059}$ & ${0.990}_{-0.047}^{+0.046}$\\
\hline

RSD~+~BAO & EZ mocks & $0.7 < z < 1.1$ &\\
\hline
all syst. \& pix64, fc & ${0.436}_{-0.058}^{+0.054}$ & ${0.998}_{-0.055}^{+0.061}$ & ${0.991}_{-0.051}^{+0.045}$\\
photo syst. \& pix64 & ${0.451}_{-0.054}^{+0.055}$ & ${0.987}_{-0.057}^{+0.061}$ & ${0.996}_{-0.046}^{+0.043}$\\
photo + cp syst. \& pix64, no fc & ${0.457}_{-0.055}^{+0.051}$ & ${0.986}_{-0.064}^{+0.056}$ & ${1.001}_{-0.047}^{+0.042}$\\
photo + cp syst. \& pix64, fc & ${0.446}_{-0.055}^{+0.051}$ & ${0.993}_{-0.063}^{+0.059}$ & ${0.996}_{-0.046}^{+0.042}$\\
all syst. \& pix64, fc, no $\wnoz$ & ${0.435}_{-0.057}^{+0.056}$ & ${1.002}_{-0.056}^{+0.063}$ & ${0.992}_{-0.050}^{+0.043}$\\
all syst. randnoz \& pix64, fc & ${0.446}_{-0.051}^{+0.054}$ & ${0.989}_{-0.061}^{+0.058}$ & ${0.994}_{-0.051}^{+0.046}$\\
all syst. randnoz \& pix64, fc, no $\wnoz$ & ${0.445}_{-0.056}^{+0.056}$ & ${0.992}_{-0.060}^{+0.059}$ & ${0.995}_{-0.050}^{+0.045}$\\
all syst. \& pix64, fc, GLAM-QPM cov & ${0.435}_{-0.061}^{+0.058}$ & ${1.000}_{-0.061}^{+0.065}$ & ${0.988}_{-0.053}^{+0.045}$\\
all syst. \& pix64, fc, no syst. cov & ${0.435}_{-0.059}^{+0.056}$ & ${1.002}_{-0.058}^{+0.061}$ & ${0.992}_{-0.050}^{+0.044}$\\
all syst. \& pix64, $\Sigma_{\mathrm{nl}} = 6 \Mpch$ & ${0.435}_{-0.060}^{+0.056}$ & ${0.997}_{-0.058}^{+0.062}$ & ${0.987}_{-0.049}^{+0.046}$\\
all syst. \& pix64, fc +1/2 $k$-bin & ${0.439}_{-0.058}^{+0.055}$ & ${0.996}_{-0.053}^{+0.064}$ & ${0.993}_{-0.052}^{+0.046}$\\
fake all syst. \& pix64, fc & ${0.437}_{-0.066}^{+0.058}$ & ${0.999}_{-0.062}^{+0.061}$ & ${0.992}_{-0.059}^{+0.048}$\\
\hline
\end{tabular}
\end{table*}

\begin{table*}
\caption{Isotropic BAO measurements on EZ and GLAM-QPM OuterRim mocks in different conditions. We quote statistics for the $N_{\mathrm{det}}$ mocks with BAO \emph{detection}, i.e. mocks for which $\alpha - \sigma_{\mathrm{low}} > 0.8\alpha^{\mathrm{exp}}$ and $\alpha + \sigma_{\mathrm{up}} < 1.2\alpha^{\mathrm{exp}}$. $\aver{\alpha}$ is the mean $\alpha$, $\aver{\sigma}$ the mean $\Delta \chi^{2} = 1$ error ($=\left(\sigma_{\mathrm{low}}+\sigma_{\mathrm{up}}\right)/2$). $S$ is the standard deviation of $\alpha$, rescaled by $\sqrt{m_{2}}$, with $m_{2}$ given by \Eq{covariance_m_2} (the uncorrected value is provided in brackets)}. Expected values $\alpha^{\mathrm{exp}}$ are given at the top of each sub-table %
(in the $\aver{\alpha}$ column).

\label{tab:mocks_bao_systematics}
\centering
\begin{tabular}{lccccc}
\hline
& $\aver{\alpha}$ & $\aver{\sigma}$ & $S$ (uncorrected) & $N_{\mathrm{det}}/N_{\mathrm{tot}}$ & $\aver{\chi^{2}}/dof$\\
\hline
\hline
EZ mocks & 1.0003 & & & &\\
\hline
baseline pre-reconstruction & $1.004$ & $0.047$ & $0.051$ ($0.049$) & $942/1000$ & $40.7/(54-13)=0.992$\\
mean of mocks baseline & $1.0017$ & $0.0011$ & $-$ & $-$ & $671/(54-13)=16.4$
\\
baseline & $1.000$ & $0.033$ & $0.043$ ($0.042$) & $981/1000$ & $41.4/(54-13)=1.01$\\
shuffled & $1.000$ & $0.033$ & $0.043$ ($0.042$) & $979/1000$ & $41.4/(54-13)=1.01$\\
all syst. & $1.002$ & $0.033$ & $0.042$ ($0.042$) & $979/1000$ & $41.5/(54-13)=1.01$\\
photo syst. & $1.000$ & $0.034$ & $0.043$ ($0.042$) & $978/1000$ & $41.4/(54-13)=1.01$\\
photo + cp syst. & $1.001$ & $0.034$ & $0.043$ ($0.042$) & $983/1000$ & $41.6/(54-13)=1.01$\\
all syst., no $\wnoz$ & $1.001$ & $0.034$ & $0.043$ ($0.042$) & $985/1000$ & $42.3/(54-13)=1.03$\\
all syst. rand noz & $0.999$ & $0.034$ & $0.043$ ($0.042$) & $982/1000$ & $41.5/(54-13)=1.01$\\
all syst. rand noz, no $\wnoz$ & $1.000$ & $0.034$ & $0.044$ ($0.043$) & $986/1000$ & $42.1/(54-13)=1.03$\\
all syst., GLAM-QPM cov & $1.000$ & $0.036$ & $0.044$ ($0.043$) & $981/1000$ & $39.8/(54-13)=0.970$\\
all syst., no syst. cov & $1.000$ & $0.034$ & $0.043$ ($0.042$) & $986/1000$ & $42.0/(54-13)=1.03$\\
all syst. + 1/2 $k$-bin & $1.002$ & $0.034$ & $0.044$ ($0.043$) & $977/1000$ & $41.4/(54-13)=1.01$\\
all syst. $\Sigma_{\mathrm{nl}} = 6 \Mpch$ & $1.002$ & $0.038$ & $0.042$ ($0.041$) & $974/1000$ & $41.5/(54-13)=1.01$\\
fake all syst. & $1.001$ & $0.034$ & $0.044$ ($0.043$) & $982/1000$ & $41.8/(54-13)=1.02$\\
\hline
GLAM-QPM mocks & 0.9992 & & & &\\
\hline
baseline pre-reconstruction & $1.002$ & $0.046$ & $0.047$ ($0.047$) & $1907/2003$ & $41.2/(54-13)=1.01$\\
baseline & $0.998$ & $0.031$ & $0.040$ ($0.040$) & $1969/2003$ & $42.5/(54-13)=1.04$\\
all syst. & $0.997$ & $0.032$ & $0.043$ ($0.042$) & $1973/2003$ & $42.4/(54-13)=1.03$\\
\hline
\end{tabular}
\end{table*}

\subsection{Isotropic BAO}
\label{sec:mocks_isotropic_bao}

In \Tab{mocks_bao_systematics} and hereafter, as in~\cite{Ata2017:1705.06373v2,Raichoor2020}, we qualify BAO \emph{detections} as $\alpha$ measurements for which the best fit value and its error bar (determined by the $\Delta \chi^{2}=1$ level) are within the range $\left[0.8\alpha^{\mathrm{exp}},1.2\alpha^{\mathrm{exp}}\right]$ ($\alpha^{\mathrm{exp}}$ being the expected $\alpha$ value, given the fiducial and mock cosmologies). Statistics are provided for the $N_{\mathrm{det}}$ mocks with BAO \emph{detections}. As we include covariance matrix corrections (Hartlap factor $D$, given by \Eq{covariance_hartlap}) and correction to the parameter covariance matrix ($m_{1}$ factor, see \Eq{covariance_m_1}) in the $\alpha$ measurement on each mock, we follow \cite{Percival2014:1312.4841v1} and provide the standard deviation $S$ of the $\alpha$ measurement corrected by $\sqrt{m_{2}}$, with $m_{2}$ given by \Eq{covariance_m_2}.

As stated in \Sec{mock_challenge_isotropic_bao}, we fix $\Sigma_{\mathrm{nl}}$ to $8 \Mpch$ (respectively $4 \Mpch$) when fitting pre-reconstruction (respectively post-reconstruction) power spectra. Pre-reconstruction $\alpha$ measurements on both EZ and GLAM-QPM mocks are biased slightly high, as can be seen from \Tab{mocks_bao_systematics} (\emph{baseline pre-reconstruction} versus \emph{baseline}). This is in line with the expected shift of the BAO peak caused by the non-linearity of structure formation~\citep{Padmanabhan2009:0812.2905v3,Ding2018:1708.01297v2}. On the contrary, post-reconstruction $\alpha$ measurements do not show any bias, at the $0.43/\sqrt{1000} \simeq 0.1\%$ level.

The radial integral constraint effect was noticed to have a significant impact on RSD cosmological measurements (\Sec{mocks_radial_integral_constraint}). We find its impact to be negligible on the post-reconstruction isotropic BAO measurements (\emph{shuffled} versus \emph{baseline}). We thus do not model any RIC correction for the isotropic BAO fits, as it would have required an increased computation time.

Adding all observational systematics and their correction scheme (\emph{all syst.}), the isotropic BAO fits to EZ mocks shift by a negligible $0.1\%$, while no change is seen with GLAM-QPM mocks (which do not include angular photometric systematics).

As in \Sec{mocks_gaussianity} we again generate and fit (\emph{fake all syst.}) $1000$ fake power spectra following a Gaussian distribution with mean and covariance matrix inferred from the contaminated EZ mocks. A negligible shift of $0.1\%$ of $\alpha$ is seen with respect to the true mocks (\emph{all syst.}), showing that one can safely use a Gaussian likelihood to compare data and model power spectra. A small shift of $0.1\%$ is seen between the fit to the mean of the mocks and the mean of the fits to each individual mock, which we label as \emph{model non-linearity} in the following.

To support the data robustness tests presented in \Sec{results}, we apply systematics successively to the EZ mocks.

Fibre collisions lead to a negligible $\alpha$ shift of $0.1\%$ (\emph{photo + cp syst.} versus \emph{photo syst.}).

Redshift failures do not impact the $\alpha$ measurement (\emph{all syst.} versus \emph{photo + cp syst.}). Ignoring the correction weight $\wnoz$ and removing redshift failures from the mocks used to build the covariance matrix is equally harmless (\emph{all syst. no $\wnoz$} versus \emph{all syst.}). A negligible shift is seen as well when the correction weight $\wnoz$ is not used, and redshift failures are removed from the mocks used to build the covariance matrix (\emph{all syst. rand noz \& pix64, fc, no $\wnoz$}). Note however that the modelling of redshift failures in the mocks is complex since we have no perfect knowledge of the corresponding systematics in the observed data. In the above, redshift failures are implemented in the EZ mocks following a deterministic process: a mock object is declared a redshift failure if the redshift of its nearest neighbour in the data could not be reliably measured. Such a scheme overestimates the angular impact of redshift failures. We therefore produce and analyse a second set of mocks, where redshift failures are applied to the EZ mocks with a probability following the model fitted on the data catalogue. In this case (\emph{rand noz}), shifts in the measured $\alpha$ due to redshift failures are equally small.

The recovered $\alpha$ does not change when $k$-bin centres are shifted by half a bin ($0.005 \hMpc$, \emph{all syst. + 1/2 $k$-bin}). Fitting the EZ mocks with the covariance matrix estimated from GLAM-QPM mocks results in a small $0.1\%$ shift of $\alpha$ measurements. The same behaviour is seen when using the covariance matrix from EZ mocks without systematics (with the \emph{shuffled} scheme only, \emph{no syst. cov}).

Based on the previous tests, we determine two systematic effects to be included in the final systematic budget: the model non-linearity, since it can only be measured on mocks, and fibre collisions, as we believe our modelling of the effect in the EZ mocks (see \Sec{ezmocks}) to be quite representative of the actual data fibre collisions. We directly take the shift in $\alpha$ attributed to model non-linearity as a systematic bias. The $0.1\%$ $\alpha$ shift attributed to fibre collisions is below twice the mock-to-mock dispersion divided by the square root of the number of mocks (common \emph{detections}), i.e. $2 \times 3.8\%/\sqrt{962} \simeq 0.2\%$, a value which we take as a systematic uncertainty, following the same procedure as in \cite{Neveux2020,GilMarin2020}. 

\begin{figure}
\centering
\includegraphics[width=0.6\columnwidth]{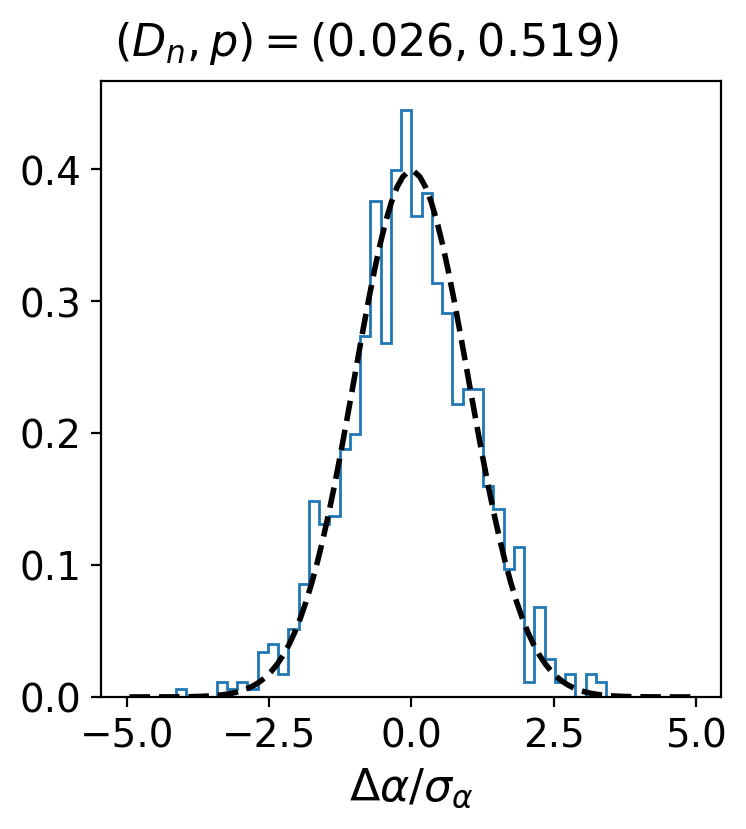}
\caption{Kolmogorov-Smirnov test on the residuals of EZ mocks, with $\Sigma_{\mathrm{nl}} = 6 \Mpch$, for the BAO fits.}
\label{fig:residuals_ezmocks_baoiso_rec}
\end{figure}

One would notice that the mean error on $\alpha$ measurements on EZ and GLAM-QPM mocks (defined by the $\Delta \chi^{2} = 1$ level, see \Sec{fitting_methodology_likelihood}) is systematically and significantly lower than the dispersion of the best fit values. The value of the damping parameter $\Sigma_{\mathrm{nl}} = 4 \Mpch$ is chosen to match the BAO amplitude seen in the reconstructed OuterRim mocks (see \Sec{mock_challenge_isotropic_bao}). However, the BAO amplitude is significantly less pronounced in the EZ mocks (see e.g.~\citealt{Raichoor2020}), hence favouring a larger $\Sigma_{\mathrm{nl}} = 6 \Mpch$. When this value is used, the distribution of the residuals $(\alpha - \aver{\alpha})/\sigma$ of mocks with BAO \emph{detection} is consistent with a standard normal distribution, as shown by the Kolmogorov-Smirnov test\footnote{Non-parametric statistical test to determine the consistency between a sample and a probability law or another sample, based on the supremum of the difference of their cumulative distribution function.} of \Fig{residuals_ezmocks_baoiso_rec}. Using a lower $\Sigma_{\mathrm{nl}}$ artificially decreases the error on the BAO fits to EZ or GLAM-QPM mocks. Since we determined $\Sigma_{\mathrm{nl}}$ on the more accurate OuterRim-based mocks, we conclude that statistical errors quoted on the data measurement are fairly estimated.

\subsection{Combination of RSD and BAO measurements}
\label{sec:mocks_combination_RSD_BAO}

As already mentioned in \Sec{fitting_methodology}, we combine the RSD and BAO likelihoods, taking into account the cross-covariance between pre- and post-reconstruction power spectrum measurements.

In \Tab{mocks_systematics}, one can notice a small shift between the RSD and the RSD~+~BAO fit to the mean of the \emph{baseline} EZ mocks: $1.0\%$ on $\fsig$, $0.3\%$ on $\apar$ and $0.5\%$ on $\aper$, which we quote as systematic error related to the technique of combining RSD and BAO likelihoods.
These shifts may come from residual systematic differences between BAO template and mocks which contaminate the RSD part of the likelihood through its cross-covariance with the BAO part. We do not investigate this effect further since these biases remain small ($< 10\%$) compared to the dispersion of the mocks (and thus to the data measurement errors).

Again, RSD~+~BAO measurements on contaminated (\emph{all syst, fc}) mocks are strongly biased: $9.2\%$ on $\fsig$ ($86\%$ of the dispersion of the mocks), $2.1\%$ on $\apar$ ($41\%$) and $1.6\%$ on $\aper$ ($42\%$). When applying the \emph{pixelated} scheme ($\mathrm{nside} = 64$), one recovers reasonable systematic shifts 
with respect to (\emph{baseline, GIC}) of $2.5\%$ on $\fsig$, $0.6\%$ on $\apar$ and $0.5\%$ on $\aper$. 
These shifts reduce further when using  $\mathrm{nside} = 128$, but we choose the \emph{pixelated} scheme with $\mathrm{nside} = 64$ as it induces a bias which we estimate small enough for our analysis since it represents $24\%$ of the dispersion of the mocks on $\fsig$, $11\%$ on $\apar$ and $13\%$ on $\aper$. In addition, the \emph{pixelated} scheme (which involves integrating over all scales of the model correlation function) has only been tested up to $\mathrm{nside} = 64$ with N-body based mocks in \cite{deMattia2019:1904.08851v3}. Moreover, the data clustering measurement is also plagued by the complex dependence of $n(z)$ with imaging quality, which we only partly removed through the \chunkz{} splitting of the radial selection function in \Sec{data}. This will require estimating the potential residual systematics from the data itself (see \Sec{results_rsd_bao_data}), which will prove to be large so that the previously mentioned shifts become subdominant.

We note the potential systematic bias induced by applying the radial and angular integral constraints (\emph{shuffled \& pix64, ARIC} versus \emph{baseline, GIC}): $0.8\%$ on $\fsig$, $0.6\%$ on $\apar$ and $0.1\%$ on $\aper$. These shifts are more than twice the mock-to-mock dispersion, divided by the square root of the number of mocks ($0.4\%$ on $\fsig$, $0.1\%$ on $\apar$ and $\aper$). We therefore account for the ARIC modelling in our systematic budget by taking an error of $0.8\%$ on $\fsig$, $0.6\%$ on $\apar$ and $0.1\%$ on $\aper$.

The isotropic BAO template of \Eq{template_bao} contains a bias term $B_{\mathrm{nw}}$, which we so far forced to be equal to the linear bias $b_{1}$ of the RSD model (see Eq.~\ref{eq:power_galaxy_galaxy}). We try to let it free (\emph{$B_{\mathrm{nw}}$ free}) and see no shift on cosmological parameters. We thus keep $B_{\mathrm{nw}} = b_{1}$ in the following.

As in \Sec{mocks_gaussianity} we generate and fit (\emph{fake all syst.}) $1000$ fake power spectra following a Gaussian distribution with mean and covariance matrix inferred from the contaminated EZ mocks. Negligible shifts of $0.3\%$ on $\fsig$, $0.1\%$ on $\apar$ and $0.1\%$ on $\aper$ are seen with respect to the true mocks (\emph{all syst. \& pix64, fc}), showing that using a Gaussian likelihood to compare data and model power spectra is accurate enough. However, we find small systematic shifts of $1.6\%$ on $\fsig$ ($15\%$ of the dispersion of the mocks), $1.0\%$ on $\apar$ ($17\%$) and $0.4\%$ on $\aper$ ($11\%$) between the median of the best fits to each individual mock and the fit to the mean of the mocks (\emph{baseline, GIC} versus \emph{mean of mocks baseline, GIC}). As in \Sec{mocks_gaussianity}, we attribute this bias to the model non-linearity, which one would note is slightly reduced compared to the RSD only analysis.

\begin{figure}
\centering
\includegraphics[width=\columnwidth]{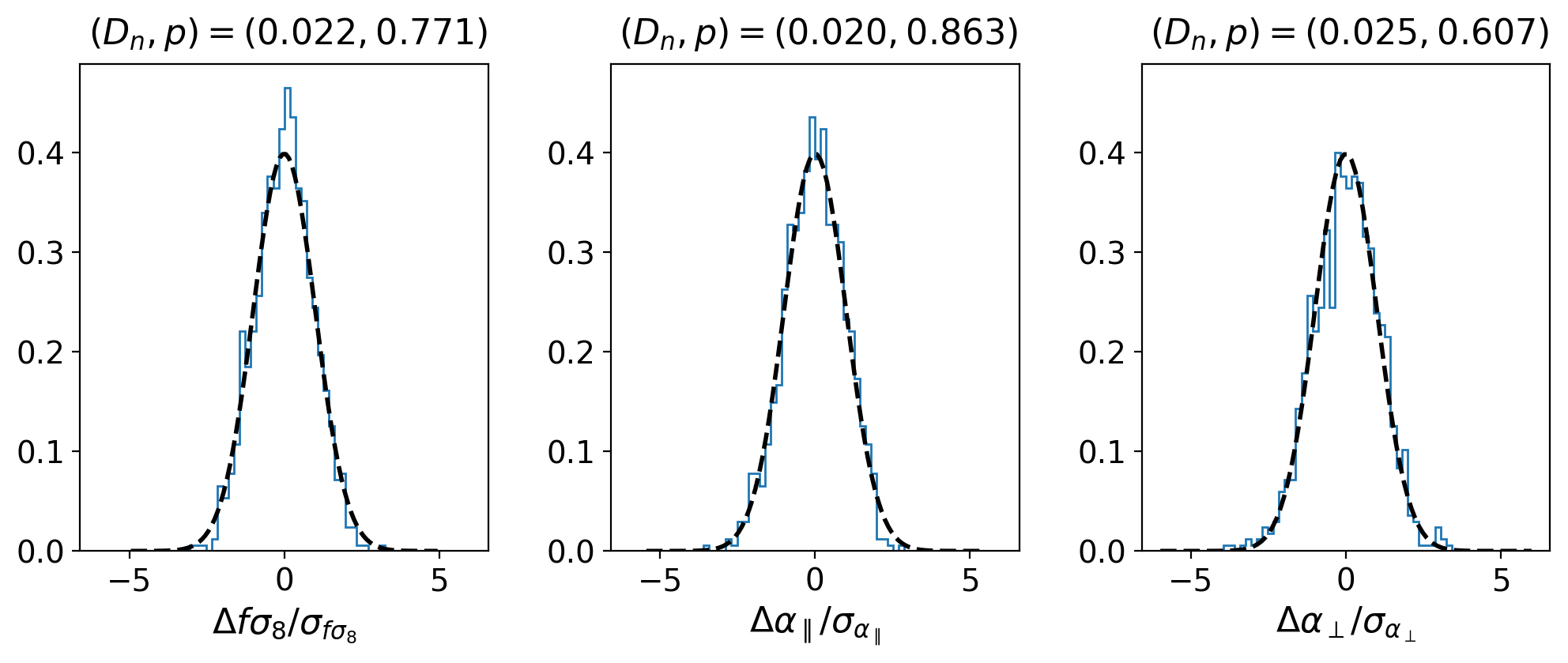}
\caption{Kolmogorov-Smirnov test on the residuals of EZ mocks, with $\Sigma_{\mathrm{nl}} = 6 \Mpch$, for the RSD~+~BAO fits.%
}
\label{fig:residuals_ezmocks_rsdbao}
\end{figure}

We check that the error bars measured on each individual mock (defined by the $\Delta \chi^{2} = 1$ level, see \Sec{fitting_methodology_likelihood}) are correctly estimated by performing a similar test as done on the post-reconstruction isotropic BAO fits in \Sec{mocks_isotropic_bao}. For this Kolmogorov-Smirnov test shown in \Fig{residuals_ezmocks_rsdbao}, we use $\Sigma_{\mathrm{nl}} = 6 \Mpch$, and keep only mocks for which the best fit $\apar$ and $\aper$ and their error bars (divided by the $\apar$ and $\aper$ expected values) are within the range $\left[0.8,1.2\right]$. The residuals seem to be in correct agreement with a standard normal distribution, as expected. We checked that the $\fsig$ residuals remain very compatible with a standard normal distribution when considering all mocks (i.e. without cut on $\apar$ and $\aper$ best fits and error bars).

\subsection{Further tests}
\label{sec:mocks_further_tests}

In \Sec{results}, we will justify our choice to fit the data with the redshift cut $0.7 < z < 1.1$. The expected shift on $\fsig$ due to the change in effective redshift is $0.3\%$. As can be seen in \Tab{mocks_systematics} (\emph{all syst. pix64, fc}), the effect of this redshift cut on the EZ mocks is negligible.

As in \Sec{mocks_isotropic_bao}, to support the data robustness tests presented in \Sec{results}, we apply systematics successively to the EZ mocks.

Fibre collisions shift $\fsig$, $\apar$ and $\aper$ by $1.1\%$, $0.6\%$ and $0.0\%$, respectively (\emph{photo + cp syst. \& pix64, fc} versus \emph{photo syst. \& pix64}). The $\aper$ shift lies below twice the mock-to-mock dispersion divided by the square root of the number of mocks, $0.2\%$, which we therefore take as a systematic uncertainty for this parameter. This is not the case for $\fsig$ ($0.5\%$) and $\apar$ ($0.2\%$), for which we take the measured shifts $1.1\%$, $0.6\%$ as systematic uncertainty, following the same procedure as in e.g.~\citet{Neveux2020}. In \Sec{mocks_fibre_collisions} our several estimates of the $f_{s}$ parameter required for the \citet{Hahn2017:1609.01714v1} fibre collision correction differed by $2\%$ at most. To assess the impact of this additional uncertainty, we compare best fits with and without the \citet{Hahn2017:1609.01714v1} fibre collision correction: we find shifts on $\fsig$, $\apar$ and $\aper$ of $1.3\%$, $0.1\%$ and $0.5\%$, respectively. Multiplying these variations by the uncertainty of $2\%$ leads to a very small additional uncertainty, which we thus neglect.

Despite the weights $\wnoz$ and the \emph{pixelated} scheme, redshift failures (red versus green curves in \Fig{comparison_power_spectrum_ezmocks}) produce shifts of $2.3\%$, $0.5\%$ and $0.5\%$ on $\fsig$, $\apar$ and $\aper$ (\emph{all syst. \& pix64, fc} versus \emph{photo + cp syst. \& pix64, fc}). These shifts become $2.6\%$, $0.9\%$ and $0.4\%$ on $\fsig$, $\apar$ and $\aper$ when the correction weight $\wnoz$ is not used, and redshift failures are removed from the mocks used to build the covariance matrix (\emph{all syst. \& pix64, fc, no $\wnoz$}). A much smaller systematic shift is seen with respect to angular photometric systematics and fibre collisions only (\emph{photo + cp syst. \& pix64, fc}) for EZ mocks with the stochastic implementation of redshift failures (\emph{all syst. rand noz \& pix64, fc}): $0.1\%$ on $\fsig$, $0.4\%$ on $\apar$ and $0.1\%$ on $\aper$.

We finally test the robustness of our analysis when using a covariance matrix measured from the GLAM-QPM mocks (without angular photometric systematics), from EZ mocks without systematics (\emph{no syst. cov}), and when shifting the $k$-bin centres by half a bin ($0.005 \hMpc$, \emph{all syst. + 1/2 $k$-bin}). In all these cases, best fits to the EZ mocks remain stable.

We therefore conclude that our analysis pipeline is robust enough to perform the BAO and RSD~+~BAO measurements on the eBOSS ELG data. Based on the previous tests, four systematic effects estimated on mocks (survey geometry, model non-linearity, ARIC modelling, fibre collisions) will be included in the final systematic budgets presented in the next section.

\section{Results}
\label{sec:results}

In this section we present isotropic BAO, RSD, and combined RSD~+~BAO measurements on the eBOSS DR16 ELG data,
discuss robustness tests of those results and provide the final error budget, including statistical and systematic contributions. In particular, systematic uncertainties are estimated from data itself where we consider mocks cannot give a reliable estimate.

\subsection{Isotropic BAO measurements}
\label{sec:results_bao_data}

\begin{figure}
\centering
\includegraphics[width=\columnwidth]{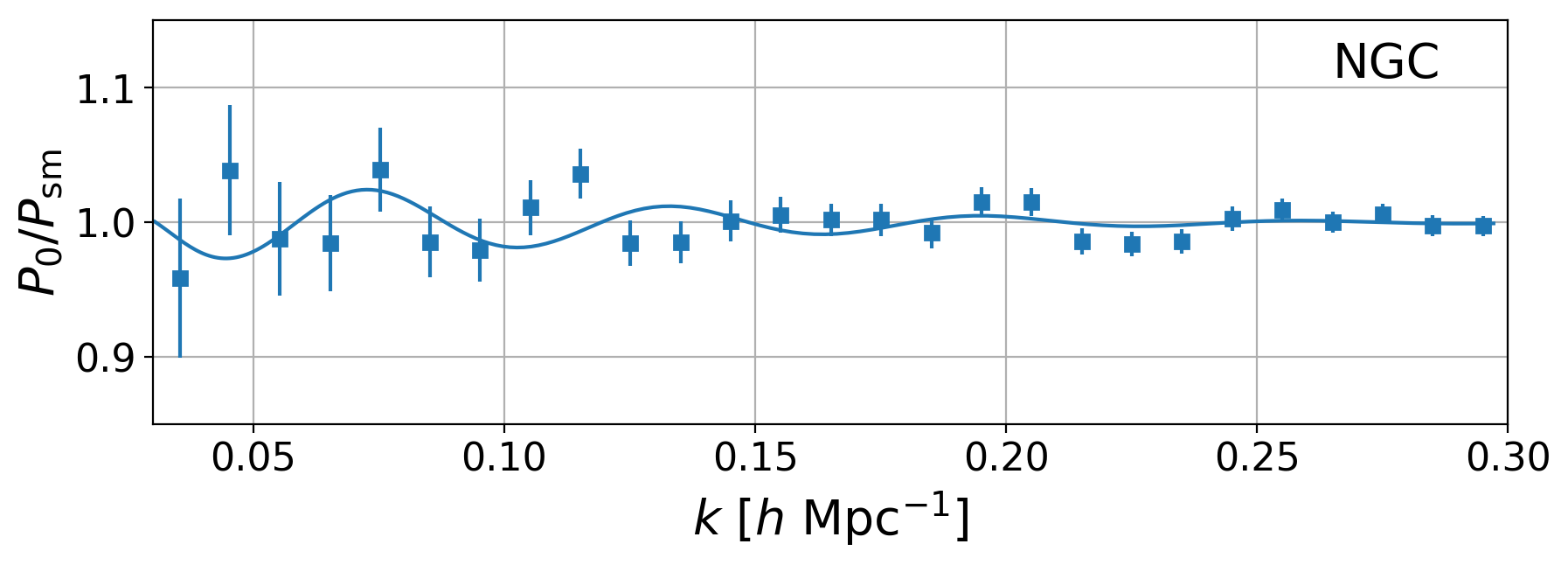}
\includegraphics[width=\columnwidth]{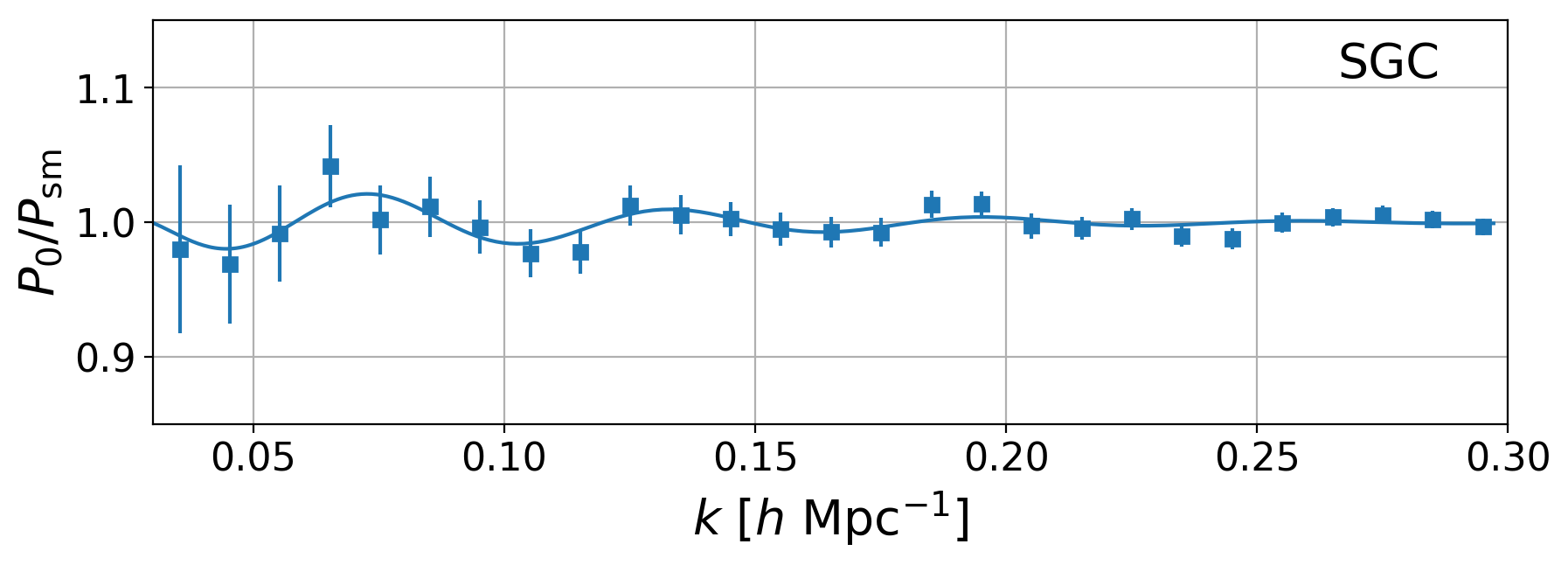}
\caption{Isotropic BAO fit (top: NGC, bottom: SGC), in the baseline case: NGC~+~SGC, $0.6 < z < 1.1$, $\Sigma_{\mathrm{nl}} = 4 \Mpch$. Both data (points with error bars from EZ mocks) and model (continuous line) are divided by the \emph{no-wiggle} power spectrum.}
\label{fig:fit_baoiso}
\end{figure}

\begin{table}
\caption{Isotropic BAO best fits on the eBOSS DR16 ELG sample. Error bars are defined by the $\Delta \chi^{2} = 1$ level. 
}
\label{tab:results_bao_data}
\centering
\begin{tabular}{lcc}
\hline
& $\alpha$ & $\chi^{2}/dof$\\
\hline
\hline
SGC only & $z$ cuts &\\
\hline
$0.6 < z < 1.1$ & ${0.997}_{-0.035}^{+0.032}$ & $12.4/(27-7)=0.619$\\
$0.65 < z < 1.1$ & ${0.989}_{-0.034}^{+0.034}$ & $13.1/(27-7)=0.655$\\
$0.7 < z < 1.1$ & ${0.995}_{-0.038}^{+0.039}$ & $13.1/(27-7)=0.654$\\
$0.75 < z < 1.1$ & ${0.993}_{-0.038}^{+0.040}$ & $20.1/(27-7)=1.01$\\
\hline
SGC only & $0.6 < z < 1.1$ &\\
\hline
no \chunkz{} & ${0.991}_{-0.036}^{+0.035}$ & $13.8/(27-7)=0.692$\\
no \chunkz{} G-Q cov & ${0.988}_{-0.036}^{+0.034}$ & $15.2/(27-7)=0.762$\\
$\Sigma_{\mathrm{nl}} = 6 \Mpch$ & ${0.993}_{-0.040}^{+0.040}$ & $13.1/(27-7)=0.653$\\
no $\wnoz$ & ${0.997}_{-0.033}^{+0.032}$ & $11.7/(27-7)=0.584$\\
GLAM-QPM cov & ${0.997}_{-0.036}^{+0.034}$ & $12.3/(27-7)=0.613$\\
no syst. cov & ${0.997}_{-0.033}^{+0.033}$ & $13.1/(27-7)=0.654$\\
500 mocks in cov & ${1.000}_{-0.033}^{+0.032}$ & $12.6/(27-7)=0.632$\\
+ 1/2 $k$-bin & ${1.003}_{-0.031}^{+0.030}$ & $16.2/(27-7)=0.810$\\
OR cosmo (rescaled) & ${0.999}_{-0.026}^{+0.025}$ & $13.3/(27-7)=0.664$\\
\hline
\hline
NGC~+~SGC & $z$ cuts &\\
\hline
$0.6 < z < 1.1$ (baseline) & ${0.986}_{-0.028}^{+0.025}$ & $42.8/(54-13)=1.04$\\
$0.65 < z < 1.1$ & ${0.984}_{-0.027}^{+0.026}$ & $42.5/(54-13)=1.04$\\
$0.7 < z < 1.1$ & ${0.982}_{-0.032}^{+0.028}$ & $44.8/(54-13)=1.09$\\
$0.75 < z < 1.1$ & ${0.961}_{-0.041}^{+0.035}$ & $49.0/(54-13)=1.19$\\
\hline
NGC~+~SGC & $0.6 < z < 1.1$ &\\
\hline
no \chunkz{} & ${0.973}_{-0.036}^{+0.031}$ & $42.1/(54-13)=1.03$\\
no \chunkz{} G-Q cov & ${0.970}_{-0.031}^{+0.029}$ & $40.2/(54-13)=0.980$\\
$\Sigma_{\mathrm{nl}} = 6 \Mpch$ & ${0.979}_{-0.038}^{+0.033}$ & $43.4/(54-13)=1.06$\\
no $\wnoz$ & ${0.984}_{-0.026}^{+0.025}$ & $46.1/(54-13)=1.12$\\
GLAM-QPM cov & ${0.988}_{-0.027}^{+0.026}$ & $38.8/(54-13)=0.946$\\
no syst. cov & ${0.984}_{-0.026}^{+0.026}$ & $40.3/(54-13)=0.984$\\
500 mocks in cov & ${0.991}_{-0.027}^{+0.025}$ & $38.8/(54-13)=0.946$\\
+ 1/2 $k$-bin & ${0.991}_{-0.027}^{+0.025}$ & $52.1/(54-13)=1.27$\\
OR cosmo (rescaled) & ${0.992}_{-0.023}^{+0.022}$ & $40.2/(54-13)=0.980$\\
OR cosmo (rescaled), & ${0.988}_{-0.029}^{+0.026}$ & $40.4/(54-13)=0.986$\\
$\Sigma_{\mathrm{nl}} = 6 \Mpch$ & & \\
\hline
\end{tabular}
\end{table}

As decided in \Sec{mock_challenge_isotropic_bao}, we take $\Sigma_{\mathrm{nl}} = 4 \Mpch$ as fiducial value for the BAO damping and BAO templates are computed within the fiducial cosmology~\eqref{eq:fiducial_cosmology}, except otherwise stated. \Fig{fit_baoiso} shows the BAO oscillation pattern fitted to the observed ELG NGC~+~SGC data. One would note that NGC does not show a clear BAO feature, contrary to SGC. The best fit $\alpha$ and its $1\sigma$ error are provided for both SGC and NGC~+~SGC fits in \Tab{results_bao_data}. For both fits, $\alpha \pm \sigma$ lies well in $\left[0.8,1.2\right]$\footnote{Here we assume that fiducial cosmology~\eqref{eq:fiducial_cosmology} agrees with the true one such that the BAO peak positions differ by much less than $20\%$.}, the criterion used in \Sec{mocks_isotropic_bao} to qualify \emph{detections} in the mocks. However, in the NGC alone, the best fit $\alpha$ value is $0.79$, such that the aforementioned criterion is not met. The same test, using the same BAO template in fiducial cosmology~\eqref{eq:fiducial_cosmology}, is applied to EZ mocks (with $\Sigma_{\mathrm{nl}} = 6 \Mpch$, to match their BAO signal amplitude) and to sky-cut OuterRim mocks of \Sec{mock_challenge_isotropic_bao} (with $\Sigma_{\mathrm{nl}} = 2.4 \Mpch$, to match their BAO signal amplitude), as reported in \Tab{results_bao_detection} ($\alpha\pm\sigma \notin [0.8,1.2]$ line). Sky-cut OuterRim mocks (hereafter OR mocks), based on N-body simulations, provide the expected BAO detections in the absence of non-Gaussian contributions due to systematics\footnote{We note however that in the OuterRim cosmology~\eqref{eq:outerrim_cosmology} the BAO amplitude (and hence signal-to-noise of BAO fits) is slightly larger than the~\cite{Planck2018:1807.06209v2} best fit model.}. In contrast, EZ mocks include known data systematics, but their BAO amplitude is lower than expected given their cosmology. Altogether, we expect the correct BAO \emph{detection} rate to lie between values derived from EZ mocks and OR mocks. One notices that $9\%$ of the EZ mocks and $2.5\%$ of the OR mocks fail to meet the $\alpha\pm\sigma \in [0.8,1.2]$ criterion in both the NGC and SGC. Therefore, the probability that $\alpha$ does not lie in $[0.8,1.2]$ within errors, for either the NGC or the SGC, ranges from $5\%$ (OR mocks) to $17\%$ (EZ mocks), such that, with this criterion, the behaviour of the data is not very unexpected. This is in line with conclusions drawn in the configuration space BAO analysis~\citep{Raichoor2020}.

\begin{figure}
\centering
\includegraphics[width=0.6\columnwidth]{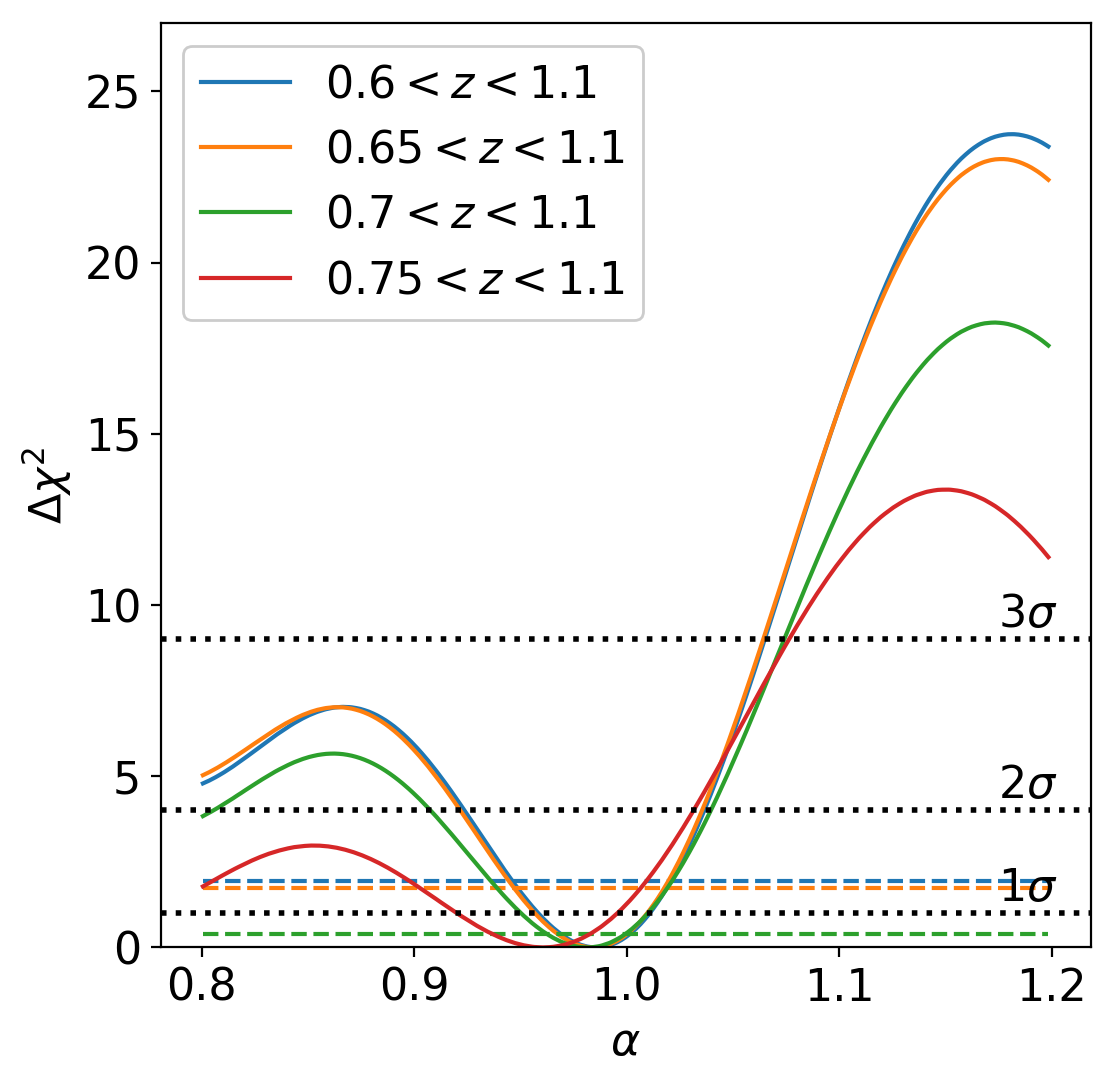}
\caption{$\chi^{2}$ profiles of the isotropic BAO fits for different lower redshift cuts, fitting NGC~+~SGC, relative to the minimum value obtained with the \emph{wiggle} power spectrum template. Continuous (respectively dashed) lines show the $\chi^{2}$ profile using the \emph{wiggle} (respectively \emph{no-wiggle}) power spectrum template. Systematic uncertainties of \Tab{error_budget_bao} are not included.}
\label{fig:profile_baoiso}
\end{figure}

To further quantify the BAO signal we compute the $\chi^{2}$ difference between the best fits obtained with the \emph{wiggle} and \emph{no-wiggle} power spectrum templates (see \Sec{model_isotropic_bao}). The $\chi^{2}$ profiles using the \emph{wiggle} and \emph{no-wiggle} power spectrum templates are shown in \Fig{profile_baoiso}. Combining NGC and SGC we find $\Delta \chi^{2} = -1.95$ ($1.4\sigma$) at a best fit value denoted $\alpha_{\mathrm{NSGC}}$ in the following. Note however that the best fit $\alpha$ value may not be relevant to compute the $\Delta \chi^{2}$ criterion %
when too far from the true one if the data (or mock) vector is too noisy. Therefore, for data or mock fits performed on each cap (NGC and SGC) separately, we quote in \Tab{results_bao_detection} the $\Delta \chi^{2}$ value evaluated at the corresponding NGC~+~SGC best fit value rather than at the respective NGC or SGC best fits, which are more subject to noise. We also provide the $\Delta \chi^{2}$ taken at the expected $\alpha$ value, given our fiducial cosmology, $\alpha^{\mathrm{exp}}$ ($\alpha^{\mathrm{exp}}=1$ for %
data). We find that for NGC~+~SGC, the mean $\Delta \chi^{2}(\alpha=\alpha_{\mathrm{NSGC}})$ is lower in the mocks, meaning a better BAO \emph{detection}. However, $18\%$ EZ mocks and $7\%$ OR mocks have larger $\Delta \chi^{2}$ values, i.e. worse BAO \emph{detection}, than the data (see $N\left(>\Delta \chi^{2}(\alpha=\alpha_{\mathrm{NSGC}})\right)$ line in \Tab{results_bao_detection}). So according to this criterion, the behaviour of the data is not very unexpected. A similar conclusion holds when taking $\Delta \chi^{2}$ at $\alpha = \alpha^{\mathrm{exp}}$ (see $N\left(>\Delta \chi^{2}(\alpha=\alpha^{\mathrm{exp}})\right)$ in \Tab{results_bao_detection}). Focusing on the SGC, $\Delta \chi^{2}(\alpha=\alpha_{\mathrm{NSGC}})$ is smaller (better BAO \emph{detection}) in data than in $85\%$ of the EZ mocks and $60\%$ of the OR mocks. However, only $1.8\%$ of EZ mocks and $0.6\%$ of OR mocks show a larger $\Delta \chi^{2}(\alpha=\alpha_{\mathrm{NSGC}})$ (worse BAO \emph{detection}) than the NGC data. Therefore, the probability for such a poor BAO \emph{detection} to happen in either NGC or SGC is approximately twice higher, of the order of a few percents. Again similar conclusions hold when taking $\Delta \chi^{2}$ at $\alpha = \alpha^{\mathrm{exp}}$. We emphasise however that the above figures are tied to the statistics used to qualify the BAO \emph{detection}.

\begin{table}
\caption{Data versus mock BAO \emph{detection}, according to different criteria. The BAO signal is noticeable in eBOSS SGC post-reconstruction data, not in NGC data, where $\alpha$ and its error bar are not within $[0.8,1.2]$ ($\alpha \pm \sigma \notin [0.8,1.2]$). $N\left(\alpha\pm\sigma \notin [0.8,1.2]\right)$ is the number of mocks not satisfying this criterion, for EZ mocks (using $\Sigma_{\mathrm{nl}} = 6 \Mpch$) with all systematics implemented and for OuterRim mocks (using $\Sigma_{\mathrm{nl}} = 2.4 \Mpch$).
The $\chi^{2}$ difference $\Delta \chi^{2}$ between the \emph{wiggle} and \emph{no-wiggle} template best fits for data and mocks is provided at the $\alpha$ value measured in the NGC~+~SGC combination ($\Delta \chi^{2}(\alpha=\alpha_{\mathrm{NSGC}})$) and at the fiducial $\alpha$ value ($\Delta \chi^{2}(\alpha=\alpha^{\mathrm{exp}})$). $N\left(>\Delta \chi^{2}(\alpha=\alpha_{\mathrm{NSGC}})\right)$ and $N\left(>\Delta \chi^{2}(\alpha=\alpha^{\mathrm{exp}})\right)$ are the number of mocks which show a larger $\Delta \chi^{2}$ than the data (and hence weaker BAO \emph{detection}) at $\alpha=\alpha_{\mathrm{NSGC}}$ and $\alpha=\alpha^{\mathrm{exp}}$, respectively.}
\label{tab:results_bao_detection}
\centering
\begin{tabular}{lccc}
\hline
& NGC & SGC & NGC~+~SGC \\
\hline
\hline
data & & & \\
\hline
\hline
$\Delta \chi^{2}(\alpha=\alpha_{\mathrm{NSGC}})$ & $3.58$ & $-5.53$ & $-1.95$ \\
$\Delta \chi^{2}(\alpha=\alpha^{\mathrm{exp}})$ & $3.97$ & $-5.61$ & $-1.63$ \\
\hline
\hline
all syst. EZ mocks & & & \\
\hline
$N\left(\alpha\pm\sigma \notin [0.8,1.2]\right)$ & $91$ & $90$ & $26$ \\
$\Delta \chi^{2}(\alpha=\alpha_{\mathrm{NSGC}})$ & $-3.22$ & $-2.43$ & $-5.65$ \\
$N\left(>\Delta \chi^{2}(\alpha=\alpha_{\mathrm{NSGC}})\right)$ & $18$ & $846$ & $184$ \\
$\Delta \chi^{2}(\alpha=\alpha^{\mathrm{exp}})$ & $-2.62$ & $-1.92$ & $-4.54$ \\
$N\left(>\Delta \chi^{2}(\alpha=\alpha^{\mathrm{exp}})\right)$ & $21$ & $881$ & $269$ \\
\hline
\hline
sky-cut OuterRim mocks & & & \\
\hline
$N(\alpha\pm\sigma \notin [0.8,1.2])$ & $25$ & $23$ & $6$ \\
$\Delta \chi^{2}(\alpha=\alpha_{\mathrm{NSGC}})$ & $-5.38$ & $-4.48$ & $-9.85$ \\
$N\left(>\Delta \chi^{2}(\alpha=\alpha_{\mathrm{NSGC}})\right)$ & $6$ & $601$ & $68$ \\
$\Delta \chi^{2}(\alpha=\alpha^{\mathrm{exp}})$ & $-4.72$ & $-4.07$ & $-8.79$ \\
$N\left(>\Delta \chi^{2}(\alpha=\alpha^{\mathrm{exp}})\right)$ & $17$ & $654$ & $102$ \\
\hline
\end{tabular}
\end{table}

We have seen that contrary to SGC the poor BAO detection in the NGC is statistically unlikely (even given the known observational systematics implemented in the EZ mocks). Let us now discuss whether one can combine the two caps.
We note that NGC photometry is shallower than SGC and thus more prone to (potentially unknown) photometric systematics~\citep{Raichoor2020}. SGC only and combined NGC~+~SGC fits are similar, to the $1.1\%$ level ($0.6 < z < 1.1$ in \Tab{results_bao_data}), which is statistically expected, as seen in $755/1000$ EZ mocks (considering both tails). Hence, there is no hint of a strong, unexpected systematic shift in the combined fit, due the addition of potentially contaminated NGC data. We also note that this $1.1\%$ shift is smaller than $1.3\%$, the uncertainty related to photometric systematics included in our systematic budget (see \Tab{error_budget_bao}), and hence is already accounted for if the NGC was the major source of photometric systematics. As a conclusion, we do not see any reason to reject NGC data in the fit. %
Moreover, combining NGC~+~SGC turns out to be more optimal than considering SGC alone, even given the poor BAO detection in NGC. To check this, we select EZ mocks with $\Delta \chi^{2}(\alpha=\alpha_{\mathrm{NSGC}}) > 0$ in the NGC but $\Delta \chi^{2}(\alpha=\alpha_{\mathrm{NSGC}}) < 0$ in the SGC among mocks for which the NGC~+~SGC combination fulfils $\alpha\pm\sigma \in [0.8,1.2]$. The dispersion of $\alpha$ measurements in the obtained sample of $130$ mocks is $0.048$ for NGC~+~SGC, less than $0.058$ for SGC alone. In addition, a Kolmogorov-Smirnov test (similar to that of \Fig{residuals_ezmocks_baoiso_rec}) does not show any misestimation of error bars ($p$-value of $0.533$ for the residuals to be consistent with a standard normal distribution). A similar reduction of error bars is seen in data in \Tab{results_bao_data} (${}_{-0.028}^{+0.025}$ versus ${}_{-0.035}^{+0.032}$). Hence, we find it legitimate to combine NGC and SGC measurements.

To further check that error bars are still correctly estimated in the low signal-to-noise regime, we select NGC~+~SGC EZ mocks for which $\Delta \chi^{2}(\alpha=\alpha_{\mathrm{NSGC}})$ is larger than that observed in the data (while still restricting to $\alpha\pm\sigma \in [0.8,1.2]$). Again, in this sample of $165$ EZ mocks, a Kolmogorov-Smirnov test shows no hint for a misestimation of error bars (the $p$-value for the residuals to be consistent with a standard normal distribution is $0.459$). Therefore, the method to estimate statistical uncertainties in data appears to be correct. We note however that data statistical error bars (${}_{-0.028}^{+0.025}$, see \Tab{results_bao_data}) are smaller than those seen in mocks on average ($\sigma \simeq 0.033$, see \Tab{mocks_bao_systematics}); $246/1000$ EZ mocks have smaller statistical error bars than data (using $\Sigma_{\mathrm{nl}} = 4 \Mpch$ for both data and mocks). In a sample of EZ mocks for which $\Delta \chi^{2}(\alpha=\alpha_{\mathrm{NSGC}})$ is within $\pm 1$ of the data $\Delta \chi^{2}(\alpha=\alpha_{\mathrm{NSGC}})$, which may be considered as representative of the data affinity for BAO, we find $6/157$ mocks to have smaller statistical error bars than the data\footnote{This fraction increases to $39/157$ when comparing data total error bars (including systematics, see \Tab{error_budget_bao}) to mock statistical error bars.}. Due to the low average BAO amplitude in the EZ mocks and the additional residual systematics in the data, this fraction slightly underestimates the probability to obtain smaller error bars than in data. Altogether, though small, data statistical error bars are not too unlikely.

We now turn to stability tests performed on data. Beforehand, we emphasise that despite the low significance of the BAO signal, a robust measurement of the BAO position is possible because the relative amplitude of oscillations is imposed as a prior in the BAO model (\Sec{model_isotropic_bao}). In other words, though a model without BAO is not disfavoured by the data, a model with BAO far from the maximum of the likelihood is a significantly worse fit to the data.

In~\cite{Raichoor2020} (Fig.~10), variations of the redshift density with photometric depth were noted to be relatively higher in the low redshift end, $0.6 < z \lesssim 0.7$. We therefore test the robustness of our measurement with the lower $z$-limit. Best fits do not move significantly with the lower $z$ cut (given the change in the sample statistics), as can also be seen in \Fig{profile_baoiso}. For example, between $0.6 < z < 1.1$ and $0.7 < z < 1.1$ the best fit $\alpha$ moves by $0.4\%$. This is not significant as a larger shifts happen in $378/1000$ (considering both tails: $741/1000$) EZ mocks. We thus use the full redshift range $0.6 < z < 1.1$ for our isotropic BAO measurement.

The $\alpha$ measurement remains very stable with the assumed $\Sigma_{\mathrm{nl}}$ value in SGC only. For NGC~+~SGC, some $0.8\%$ shift ($27\%$ of the statistical uncertainty) is obtained between $\Sigma_{\mathrm{nl}} = 4 \Mpch$ and $\Sigma_{\mathrm{nl}} = 6 \Mpch$. A larger shift happens for $97/1000$ (considering both tails: $214/1000$) EZ mocks. To account for the uncertainty in the expected amplitude of the BAO signal, we include this shift as an additional uncertainty.

We estimate the potential residual systematics due to the imperfect modelling of the variations of the survey selection function with imaging quality (see \Sec{data}) by comparing the measurement obtained with the baseline correction to that without any mitigation. Namely, random redshifts are taken separately in each \texttt{chunk} (instead of \chunkz{}) to measure the data power spectrum. The covariance matrix is built from mocks where we do not introduce photometric systematics nor variations of the redshift density with \chunkz{}. We take care to change the model window functions (see \Sec{model}) accordingly. A shift in $\alpha$ of $1.3\%$, which we take as systematic uncertainty, is noticed between these two configurations. One would notice that a variation of $0.6\%$ is seen in the SGC alone. Hence, the shift of the NGC~+~SGC best fit appears to be mainly driven by the NGC, as can be expected since photometry is shallower in the NGC than in the SGC. %

The residual systematics remaining after the redshift failure correction (see \Sec{data}) are similarly estimated by comparing the measurement obtained with the baseline correction to that without any mitigation (\emph{no $\wnoz$}). Namely, no $\wnoz$ are applied to measure the data power spectrum, while we do not introduce redshift failures in the mocks to construct the covariance matrix. This leads to a shift in $\alpha$ of $0.2\%$; a larger variation is seen in $438/1000$ (considering both tails: $851/1000$) EZ mocks, making it quite likely. Note however that data and EZ mock shifts are expected to match since a mock object is deterministically declared a redshift failure if the redshift of its nearest neighbour in the data could not be reliably measured. As mentioned in \Sec{mocks_further_tests}, a second set of mocks was produced with redshift failures being randomly drawn from the model fitted to the data; in this case, we find a larger variation in $378/1000$ (considering both tails: $826/1000$) EZ mocks. Hence, the shift seen in the data is well explained by the mocks. We conservatively include it in the systematic budget to account for the uncertainty in the redshift failure correction.

Changing the covariance matrix for that based on GLAM-QPM mocks, which does not include angular nor radial photometric systematics leads to a small $\alpha$ shift of $0.4\%$ in the \emph{no \chunkz{}} case (\emph{no \chunkz{} G-Q cov} versus \emph{no \chunkz{}}). Similarly, in the baseline case, using a covariance matrix built from GLAM-QPM or EZ mocks without systematics (only the \emph{shuffled} scheme) induces small shifts of $0.2\%$ (\emph{GLAM-QPM cov} versus \emph{baseline}). A larger shift is observed in the former case for $283/1000$ (considering both tails: $732/1000$) EZ mocks, in the latter case for $377/1000$ (considering both tails: $713/1000$) EZ mocks. The parameter covariance matrix correction \Eq{covariance_m_1} already leads to an increase of $0.7\%$ of the statistical error, which is higher than obtained by summing the shifts above in quadrature ($0.3\%$). We therefore do not include any systematic uncertainty related to the choice of the covariance matrix. Dividing the number of mocks used to build the covariance matrix by $2$, we find a shift in $\alpha$ of $0.5\%$ (\emph{500 mocks in cov} versus \emph{baseline}). This would lead to a $1.9\%$ increase of the error if added in quadrature, while the increase of error bars required to account for the change in covariance matrix between these two configurations is $1.5\%$. Hence, the shift seen in the fit is compatible with a statistical fluctuation and we conclude that the estimation of the covariance matrix is robust enough for this measurement.

Moving the centre of the $k$-bin by half a bin ($0.005 \hMpc$, \emph{+ 1/2 $k$-bin}) leads to a $\alpha$ shift of $0.5\%$, which is compatible with a statistical fluctuation since a larger shift is observed for $307/1000$ (considering both tails: $595/1000$) EZ mocks. We therefore do not include any additional uncertainty related to the choice of $k$-bins in our systematic budget.

Finally, we change the fiducial cosmology \Eq{fiducial_cosmology} for the OuterRim cosmology \Eq{outerrim_cosmology}, and for comparison purposes we report the $\alpha$ value rescaled to our fiducial cosmology. Some shift can be seen (\emph{OR cosmo} versus \emph{baseline}), which we relate to the change of damping term. Indeed, we checked that analysing a power spectrum in our fiducial cosmology with an OuterRim template leads to a preferred $\Sigma_{\mathrm{nl}} \simeq 5.4 \Mpch$. The measurement obtained with $\Sigma_{\mathrm{nl}} = 6 \Mpch$ and OuterRim cosmology (\emph{OR cosmo $\Sigma_{\mathrm{nl}} = 6 \Mpch$}) is indeed close (within $0.2 \%$) to the measurement using $\Sigma_{\mathrm{nl}} = 4 \Mpch$ and our fiducial cosmology (\emph{$0.6 < z < 1.1$ (fiducial)}). This shift can be seen in $428/1000$ (considering both tails: $840/1000$) EZ mocks and is within the $0.2\%$ modelling uncertainty derived in \Sec{mock_challenge_isotropic_bao}. We therefore do not include any additional uncertainty due to the assumed fiducial cosmology.

\begin{table}
\caption{Error budget for post-reconstruction isotropic BAO measurements on the eBOSS DR16 ELG sample. Percentages are provided with respect to the $\alpha$ value. The last three lines (statistics, systematics and total) recap the absolute statistical error bar, the systematic contribution (total minus statistics), and the total error bar, respectively.}
\label{tab:error_budget_bao}
\centering
\begin{tabular}{lc} 
\hline
source & $\alpha$ \\
\hline
\hline
linear & \\
\hline
model non-linearity (from EZ mocks) & $0.1\%$ \\
\hline
\hline
quadrature & \\
\hline
modelling systematics (from mock challenge) & $0.2\%$ \\
damping term $\Sigma_{\mathrm{nl}}$ & $0.8\%$ \\
photometric systematics & $1.3\%$ \\
fibre collisions (from EZ mocks) & $0.2\%$ \\
redshift failures & $0.2\%$ \\
\hline
\hline
statistics & ${}_{-0.028}^{+0.025}$ \\
systematics & ${}_{-0.005}^{+0.006}$ \\
total & ${}_{-0.033}^{+0.031}$ \\
\hline
\end{tabular}
\end{table}

Overall, despite the mild preference for BAO in the eBOSS ELG sample, the BAO measurement appears relatively robust.
The final error budget is reported in \Tab{error_budget_bao}. Since the $0.1\%$ $\alpha$ shift attributed to model non-linearity is closer to a bias than a systematic uncertainty, we decided to be conservative and to add it linearly to the statistical uncertainty. All other contributions to the systematic budget are uncertainties due to the model accuracy (modelling systematics) or our limited knowledge of the survey selection function (photometric systematics, fibre collisions, redshift failures) or are analysis choices (damping term). We thus add them in quadrature to the statistical uncertainty.

Our final post-reconstruction isotropic BAO measurement is 
$\alpha = 0.986_{-0.033}^{+0.031}$, including statistical and systematic uncertainties.

In terms of the volume-averaged distance $\DV(z)$, we find: 
\begin{equation}
\DV(\zeff=0.845)/\rdrag = 18.33_{-0.62}^{+0.57},
\end{equation}
independently of the assumed fiducial cosmology.

In order to generate the BAO likelihood profile for further cosmological inference, we first rescale $\Delta \chi^{2}(\alpha) = \chi^{2}(\alpha) - \chi^{2}(\alpha_{0})$ (with $\alpha_{0}$ the best fit value) by the inverse of the parameter covariance rescaling~\eqref{eq:covariance_m_1}. To include the systematic error budget we further rescale $\Delta \chi^{2}(\alpha)$ by the ratio $\left(\sigma_{\alpha,\mathrm{stat}}/\sigma_{\alpha,\mathrm{tot}}\right)^{2}$, with $\sigma_{\alpha,\mathrm{tot}}$ and $\sigma_{\alpha,\mathrm{stat}}$ the total and statistical upper (lower) error bars when $\alpha>\alpha_{0}$ ($\alpha<\alpha_{0}$). We finally provide the BAO likelihood $e^{-\Delta\chi^{2}(\DV/\rdrag)/2}$, with $\DV = \alpha \DV^{\fid}$.

\subsection{Combined RSD and BAO measurements}
\label{sec:results_rsd_bao_data}

RSD and combined RSD~+~BAO measurements are reported under different fitting conditions in \Tab{results_rsdbao_data}. As in \Sec{results_bao_data}, we take $\Sigma_{\mathrm{nl}} = 4 \Mpch$ as fiducial value for the BAO damping and RSD and BAO templates are computed within the fiducial cosmology~\eqref{eq:fiducial_cosmology}, except otherwise stated. We quote results for the combined caps NGC~+~SGC and SGC alone.

\begin{table*}
\caption{RSD and RSD~+~BAO best fits on the eBOSS DR16 ELG sample. Error bars are defined by the $\Delta \chi^{2} = 1$ level, except in the \emph{Bayesian} case, where we consider the minimum interval which contains $68\%$ of the MCMC samples.}
\label{tab:results_rsdbao_data}
\centering
\begin{tabular}{lcccc}
\hline
& $\fsig$ & $\apar$ & $\aper$ & $\chi^{2}/dof$\\
\hline
\hline
RSD only & SGC only & $z$ cuts & &\\
\hline
$0.6 < z < 1.1$ & ${0.348}_{-0.079}^{+0.078}$ & ${0.974}_{-0.084}^{+0.100}$ & ${0.952}_{-0.062}^{+0.071}$ & $31.7/(46-7)=0.812$\\
$0.65 < z < 1.1$ & ${0.367}_{-0.085}^{+0.086}$ & ${0.982}_{-0.093}^{+0.097}$ & ${0.968}_{-0.073}^{+0.090}$ & $29.2/(46-7)=0.749$\\
$0.7 < z < 1.1$ & ${0.38}_{-0.11}^{+0.10}$ & ${1.013}_{-0.098}^{+0.087}$ & ${1.04}_{-0.13}^{+0.13}$ & $29.8/(46-7)=0.763$\\
$0.75 < z < 1.1$ & ${0.33}_{-0.10}^{+0.23}$ & ${0.98}_{-0.16}^{+0.11}$ & ${0.936}_{-0.082}^{+0.328}$ & $31.3/(46-7)=0.802$\\
\hline
RSD only & SGC only & $0.7 < z < 1.1$ & &\\
\hline
GLAM-QPM cov & ${0.429}_{-0.082}^{+0.081}$ & ${1.022}_{-0.079}^{+0.082}$ & ${1.120}_{-0.100}^{+0.098}$ & $28.6/(46-7)=0.733$\\
no syst. cov & ${0.409}_{-0.082}^{+0.077}$ & ${1.027}_{-0.074}^{+0.076}$ & ${1.110}_{-0.098}^{+0.087}$ & $29.3/(46-7)=0.752$\\
+ 1/2 $k$-bin & ${0.374}_{-0.099}^{+0.093}$ & ${1.023}_{-0.079}^{+0.083}$ & ${1.052}_{-0.098}^{+0.104}$ & $25.0/(46-7)=0.640$\\
OR cosmo (rescaled) & ${0.371}_{-0.106}^{+0.090}$ & ${1.029}_{-0.082}^{+0.075}$ & ${1.07}_{-0.12}^{+0.10}$ & $24.9/(46-7)=0.638$\\
\hline
\hline
RSD~+~BAO & SGC only & $z$ cuts & &\\
\hline
$0.6 < z < 1.1$ & ${0.327}_{-0.105}^{+0.084}$ & ${1.017}_{-0.087}^{+0.133}$ & ${0.967}_{-0.061}^{+0.059}$ & $51.0/(73-12)=0.837$\\
$0.65 < z < 1.1$ & ${0.348}_{-0.107}^{+0.090}$ & ${1.013}_{-0.087}^{+0.120}$ & ${0.976}_{-0.066}^{+0.064}$ & $46.1/(73-12)=0.756$\\
$0.7 < z < 1.1$ & ${0.335}_{-0.124}^{+0.099}$ & ${1.004}_{-0.097}^{+0.122}$ & ${0.986}_{-0.083}^{+0.071}$ & $50.0/(73-12)=0.819$\\
$0.75 < z < 1.1$ & ${0.33}_{-0.16}^{+0.12}$ & ${1.01}_{-0.12}^{+0.16}$ & ${0.964}_{-0.096}^{+0.087}$ & $52.4/(73-12)=0.859$\\
\hline
RSD + BA0 & SGC only & $0.7 < z < 1.1$ & &\\
\hline
GLAM-QPM cov & ${0.425}_{-0.092}^{+0.088}$ & ${0.950}_{-0.103}^{+0.095}$ & ${1.033}_{-0.066}^{+0.058}$ & $53.5/(73-12)=0.877$\\
no syst. cov & ${0.386}_{-0.085}^{+0.079}$ & ${0.979}_{-0.073}^{+0.079}$ & ${1.028}_{-0.062}^{+0.057}$ & $47.6/(73-12)=0.780$\\
+ 1/2 $k$-bin & ${0.31}_{-0.12}^{+0.11}$ & ${1.04}_{-0.10}^{+0.12}$ & ${0.973}_{-0.078}^{+0.074}$ & $47.6/(73-12)=0.780$\\
OR cosmo (rescaled) & ${0.31}_{-0.11}^{+0.10}$ & ${1.01}_{-0.11}^{+0.11}$ & ${0.983}_{-0.070}^{+0.069}$ & $45.4/(73-12)=0.743$\\
\hline
\hline
RSD only & NGC~+~SGC & $z$ cuts & &\\
\hline
$0.6 < z < 1.1$ & ${0.250}_{-0.067}^{+0.124}$ & ${1.15}_{-0.28}^{+0.11}$ & ${0.919}_{-0.039}^{+0.038}$ & $87.6/(92-11)=1.08$\\
$0.65 < z < 1.1$ & ${0.259}_{-0.068}^{+0.116}$ & ${1.15}_{-0.27}^{+0.10}$ & ${0.922}_{-0.041}^{+0.040}$ & $83.9/(92-11)=1.04$\\
$0.7 < z < 1.1$ & ${0.382}_{-0.056}^{+0.053}$ & ${0.871}_{-0.061}^{+0.109}$ & ${0.901}_{-0.050}^{+0.043}$ & $81.3/(92-11)=1.00$\\
$0.75 < z < 1.1$ & ${0.365}_{-0.073}^{+0.062}$ & ${0.905}_{-0.080}^{+0.146}$ & ${0.887}_{-0.053}^{+0.045}$ & $66.6/(92-11)=0.822$\\
\hline
RSD only & NGC~+~SGC & $0.7 < z < 1.1$ & &\\
\hline
GLAM-QPM cov & ${0.308}_{-0.074}^{+0.104}$ & ${1.08}_{-0.27}^{+0.11}$ & ${0.942}_{-0.083}^{+0.047}$ & $80.3/(92-11)=0.991$\\
no syst. cov & ${0.388}_{-0.054}^{+0.052}$ & ${0.865}_{-0.061}^{+0.106}$ & ${0.910}_{-0.052}^{+0.046}$ & $83.2/(92-11)=1.03$\\
+ 1/2 $k$-bin & ${0.376}_{-0.134}^{+0.055}$ & ${0.887}_{-0.071}^{+0.272}$ & ${0.910}_{-0.048}^{+0.047}$ & $68.5/(92-11)=0.846$\\
OR cosmo (rescaled) & ${0.289}_{-0.075}^{+0.128}$ & ${1.08}_{-0.27}^{+0.11}$ & ${0.933}_{-0.075}^{+0.043}$ & $73.4/(92-11)=0.906$\\
\hline

RSD~+~BAO & NGC~+~SGC & $z$ cuts & &\\
\hline
$0.6 < z < 1.1$ & ${0.271}_{-0.057}^{+0.059}$ & ${1.129}_{-0.090}^{+0.078}$ & ${0.938}_{-0.030}^{+0.030}$ & $140/(146-21)=1.12$\\
$0.65 < z < 1.1$ & ${0.281}_{-0.057}^{+0.059}$ & ${1.122}_{-0.085}^{+0.076}$ & ${0.938}_{-0.031}^{+0.031}$ & $137/(146-21)=1.10$\\
$0.7 < z < 1.1$ (baseline) & ${0.289}_{-0.066}^{+0.068}$ & ${1.085}_{-0.107}^{+0.087}$ & ${0.941}_{-0.034}^{+0.035}$ & $141/(146-21)=1.13$\\
$0.75 < z < 1.1$ & ${0.319}_{-0.069}^{+0.068}$ & ${1.062}_{-0.092}^{+0.088}$ & ${0.937}_{-0.037}^{+0.037}$ & $123/(146-21)=0.984$\\
\hline\hline
RSD~+~BAO & NGC~+~SGC & $0.7 < z < 1.1$ & &\\
\hline
no \chunkz{} & ${0.261}_{-0.058}^{+0.059}$ & ${1.109}_{-0.074}^{+0.070}$ & ${0.928}_{-0.036}^{+0.036}$ & $153/(146-21)=1.23$\\
no \chunkz{} GLAM-QPM cov & ${0.262}_{-0.061}^{+0.061}$ & ${1.137}_{-0.078}^{+0.077}$ & ${0.926}_{-0.034}^{+0.035}$ & $136/(146-21)=1.09$\\
$\Sigma_{\mathrm{nl}} = 6 \Mpch$ & ${0.283}_{-0.069}^{+0.101}$ & ${1.086}_{-0.284}^{+0.096}$ & ${0.934}_{-0.036}^{+0.036}$ & $141/(146-21)=1.13$\\
no $\wnoz$ & ${0.306}_{-0.068}^{+0.071}$ & ${1.088}_{-0.112}^{+0.089}$ & ${0.939}_{-0.034}^{+0.034}$ & $144/(146-21)=1.15$\\
GLAM-QPM cov & ${0.305}_{-0.065}^{+0.066}$ & ${1.101}_{-0.081}^{+0.078}$ & ${0.956}_{-0.035}^{+0.037}$ & $132/(146-21)=1.06$\\
no syst. cov & ${0.326}_{-0.078}^{+0.072}$ & ${1.03}_{-0.11}^{+0.10}$ & ${0.952}_{-0.038}^{+0.037}$ & $125/(146-21)=1.00$\\
500 mocks in cov & ${0.301}_{-0.070}^{+0.071}$ & ${1.084}_{-0.092}^{+0.082}$ & ${0.941}_{-0.036}^{+0.037}$ & $127/(146-21)=1.02$\\
+ 1/2 $k$-bin & ${0.287}_{-0.064}^{+0.067}$ & ${1.082}_{-0.087}^{+0.081}$ & ${0.943}_{-0.033}^{+0.035}$ & $140./(146-21)=1.12$\\
OR cosmo (rescaled) & ${0.271}_{-0.063}^{+0.066}$ & ${1.098}_{-0.086}^{+0.075}$ & ${0.938}_{-0.032}^{+0.034}$ & $129/(146-21)=1.03$\\
\hline
\hline
\hline
Bayesian & $0.289_{-0.075}^{+0.060}$ & $1.085_{-0.090}^{+0.104}$ & $0.941_{-0.037}^{+0.036}$ & $141/(146-21) = 1.13$ \\
\hline
\end{tabular}
\end{table*}

As already noted in \Sec{results_bao_data}, variations of the redshift density with photometric depth were observed to be relatively higher in the low redshift end. We therefore test the robustness of our measurement with respect to the lower redshift cut.
For the RSD~+~BAO measurement, shifts of $6.4\%$, $4.1\%$ and $0.4\%$ are seen for $\fsig$, $\apar$ and $\aper$, respectively, when changing the redshift range $0.6 < z < 1.1$ to $0.7 < z < 1.1$ for the pre-reconstruction power spectrum (RSD part of the likelihood). Larger shifts are observed for $159/1000$ (considering both tails: $296/1000$) EZ mocks for $\fsig$, $19/1000$ ($32/1000$) for $\apar$ and $374/1000$ ($732/1000$) for $\aper$. The somewhat low probability of the change in $\apar$ may point towards some unaccounted systematics in the lower redshift end. These potential systematics were without effect on the BAO measurement, as noted in \Sec{results_bao_data}.

For our final measurement we therefore conservatively choose the redshift range $0.7 < z < 1.1$ for the RSD part of the likelihood, and keep $0.6 < z < 1.1$ for the BAO part, as decided in \Sec{results_bao_data}.
Following \Eq{effective_redshift}, the effective redshift for the redshift range $0.6 < z < 1.1$ is $0.845$ while it is $0.857$ for $0.7 < z < 1.1$. We choose the effective redshift as $0.85$ for the combined RSD~+~BAO measurement, and check that the expected variations of $\fsig$, $\DHu/\rdrag$ and $\DM/\rdrag$ over the redshift range $0.845 - 0.857$ within our fiducial cosmology are small ($0.3\%$, $0.7\%$ and $1.1\%$, respectively) compared to the statistical uncertainty on data ($24\%$, $9\%$ and $4\%$, respectively).

\begin{figure*}
\centering
\includegraphics[width=\columnwidth]{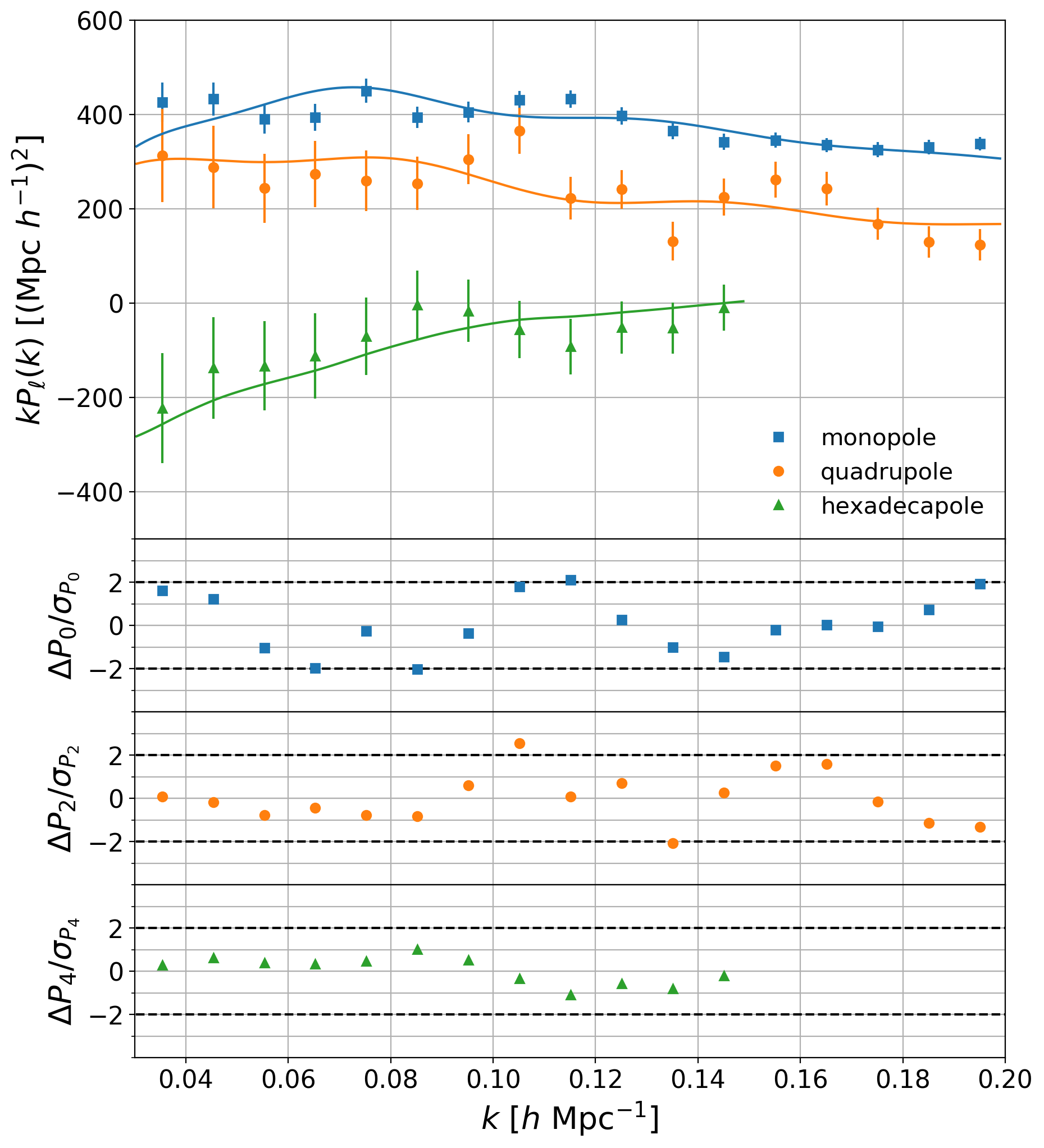}
\includegraphics[width=\columnwidth]{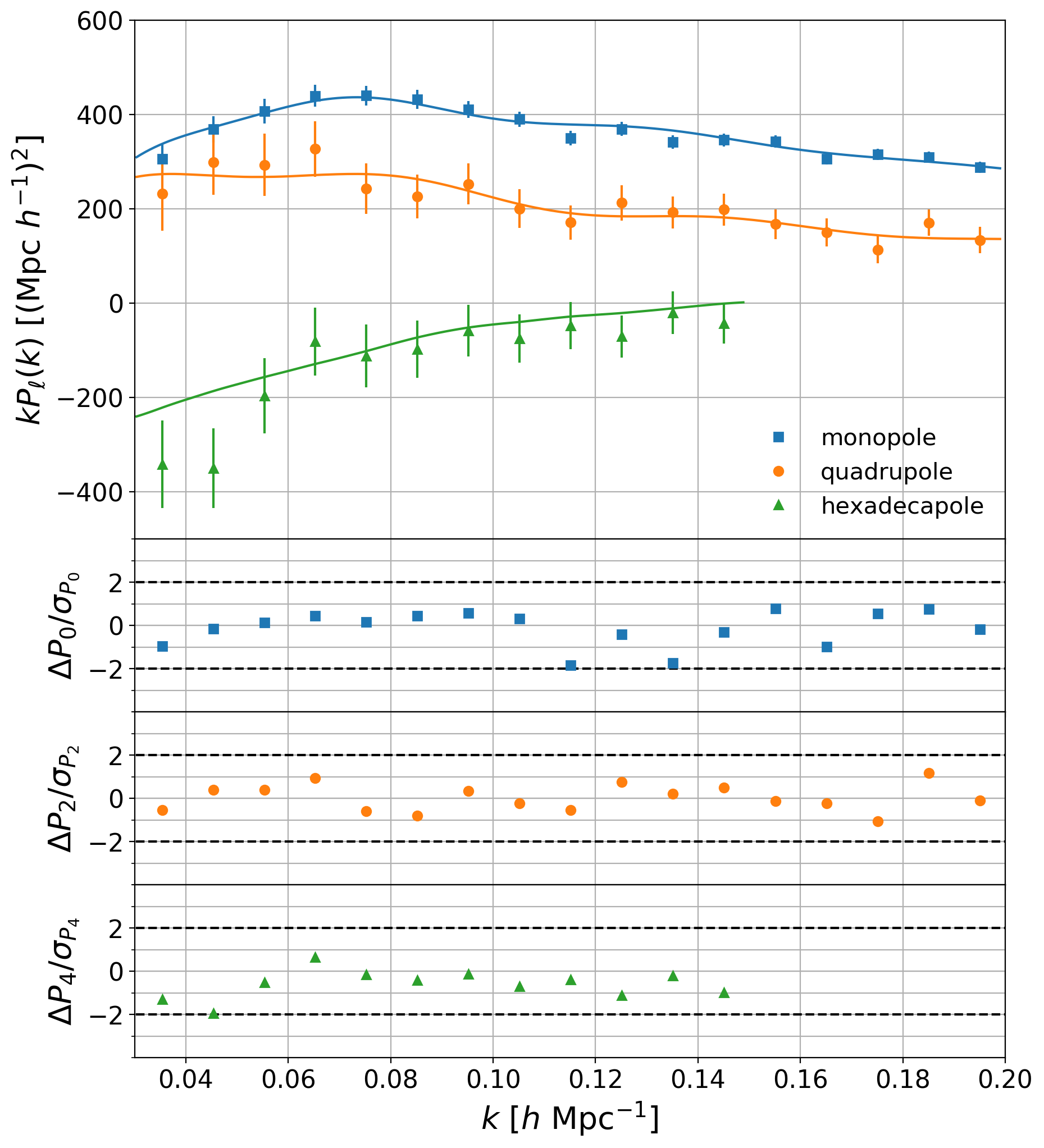}
\includegraphics[width=\columnwidth]{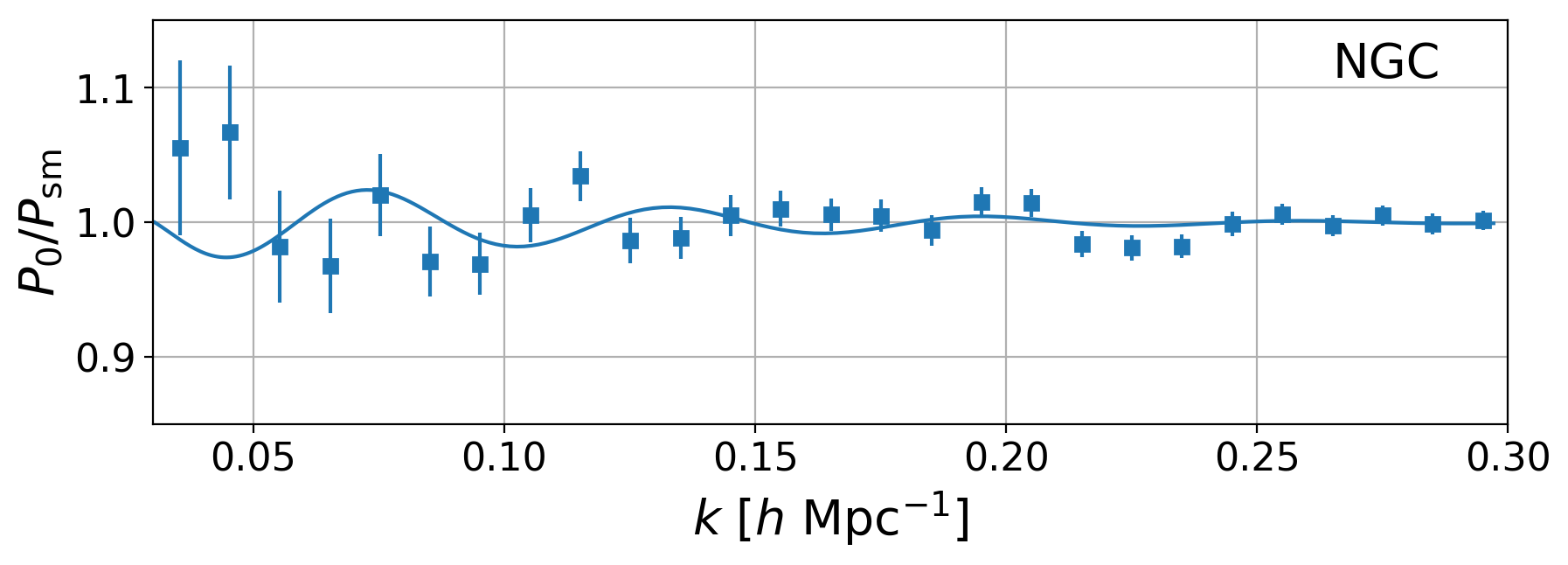}
\includegraphics[width=\columnwidth]{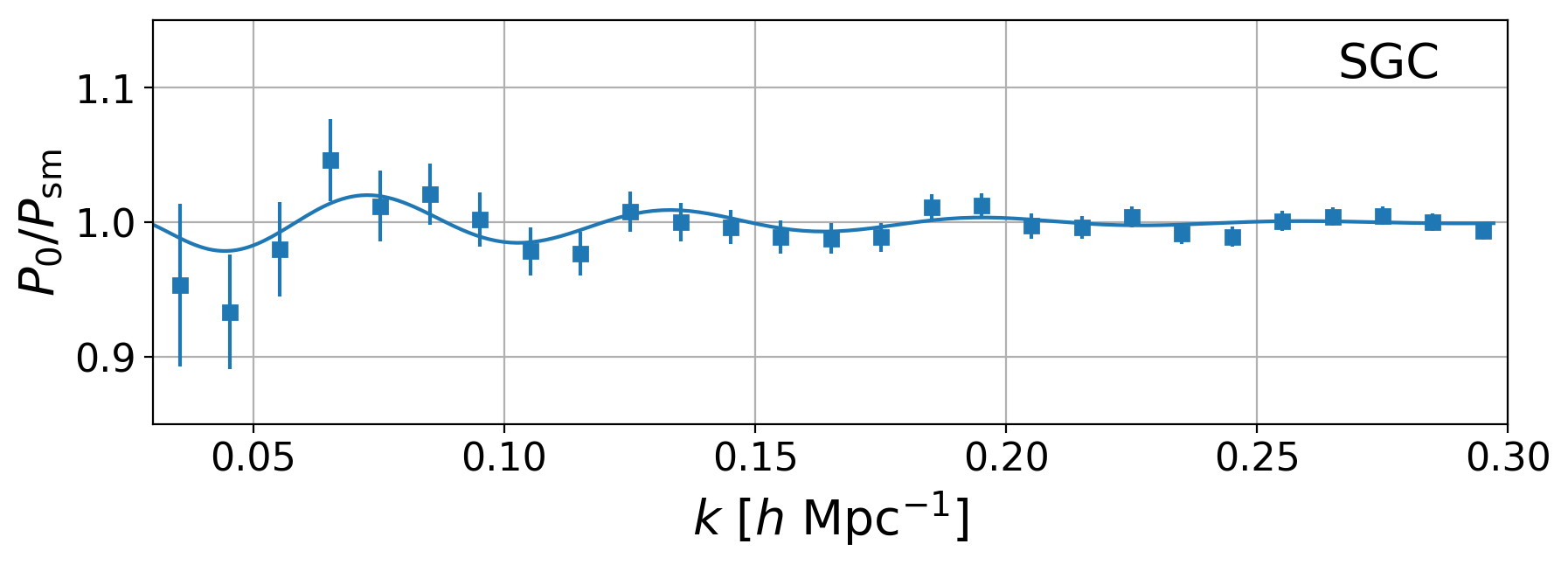}
\caption{Combined RSD~+~BAO fit (left: NGC, right: SGC): data points with error bars from the EZ mocks and best fit model as continuous line for the power spectrum multipoles (top) and the BAO oscillation pattern (bottom), normalised residuals for every power spectrum multipole (middle).} %
\label{fig:fit_rsdbao}
\end{figure*}

\begin{figure}
\centering
\includegraphics[width=\columnwidth]{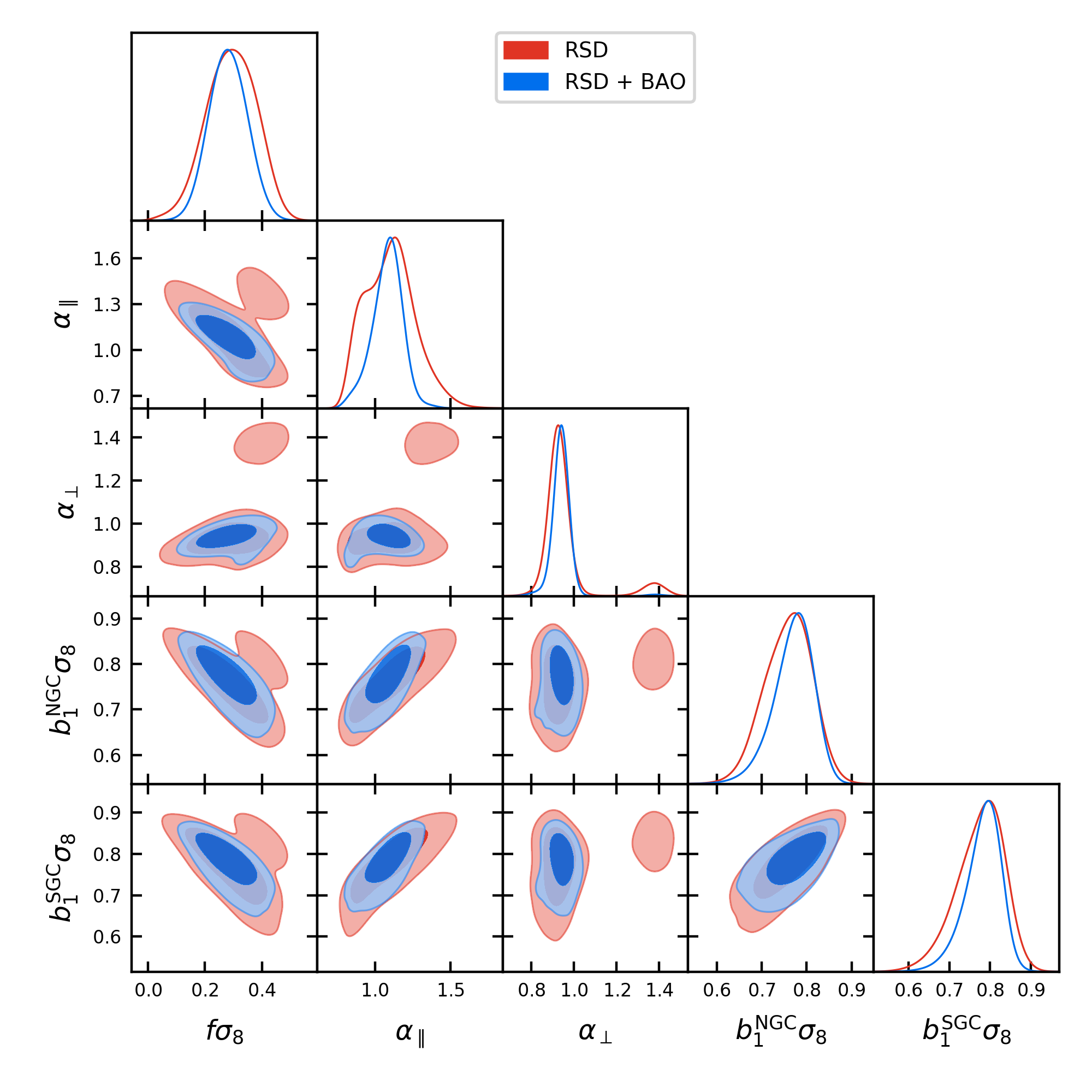}
\caption{Posteriors of the RSD and RSD~+~BAO measurements. Systematic uncertainties of \Tab{error_budget_rsdbao} are not included.}
\label{fig:corner_rsdbao}
\end{figure}

Best fit models of ELG power spectra in the NGC and SGC are compared with data in \Fig{fit_rsdbao}, while \Fig{corner_rsdbao} shows the posteriors of the RSD and RSD~+~BAO measurements. One would note that the RSD~+~BAO combination helps reducing the posterior tails while not changing the central values. In particular, the combination of RSD with BAO removes secondary local minima in the contours. An illustration of this ill-shaped RSD only posteriors is the large change for the RSD only best fit to the NGC~+~SGC data in the redshift range $0.7 < z < 1.1$ with respect to $0.6 < z < 1.1$, as reported in \Tab{results_rsdbao_data}. In particular, the bump seen on the left side of the peak in the $\apar$ marginal RSD only posterior in \Fig{corner_rsdbao} corresponds to the position of the best fit value $\apar = {0.871}_{-0.061}^{+0.109}$. We checked that the overall shift in the RSD only posterior contours when changing the redshift range is much smaller than that of the RSD only best fits and consistent with that of the RSD~+~BAO case. Note that in the following the systematic budget will be derived for the RSD~+~BAO combination, for which the shape of the posterior is more Gaussian.

As in \Sec{results_bao_data}, we test the robustness of our result with respect to the choice of the $\Sigma_{\mathrm{nl}}$ value: we find a $1.9\%$ shift on $\fsig$, $0.1\%$ on $\apar$ and $0.8\%$ on $\aper$ between $\Sigma_{\mathrm{nl}} = 4 \Mpch$ and $\Sigma_{\mathrm{nl}} = 6 \Mpch$, which we add to our uncertainty budget. Larger shifts are seen in $104/1000$ (both tails: $142/1000$), $402/1000$ ($861/1000$) and $211/1000$ ($345/1000$) EZ mocks for $\fsig$, $\apar$ and $\aper$.

The uncertainty in the modelling of the selection function variations with photometry is estimated similarly to \Sec{results_bao_data}: we compare the baseline correction, using the \emph{pixelated} scheme, to the best fit measurement (\emph{no \chunkz{}}) obtained without any angular mitigation scheme nor modelling of the variations of the radial selection function with \chunkz{}. The covariance matrix is also built from EZ mocks without angular nor radial photometric systematics. We measure shifts of $9.4\%$ on $\fsig$, $2.2\%$ on $\apar$ and $1.4\%$ on $\aper$, which we take as additional systematic uncertainty. %

Similarly, the impact of redshift failures is estimated by comparing the baseline measurement with the best fit obtained on the data without including the $\wnoz$ correction weight (\emph{no $\wnoz$}). We measure shifts of $6.1\%$ on $\fsig$, $0.3\%$ on $\apar$ and $0.3\%$ on $\aper$, which we take as systematic uncertainty. Larger variations happen for $176/1000$ (considering both tails: $368/1000$), $479/1000$ ($856/1000$) and $409/1000$ ($882/1000$) EZ mocks, respectively. If we rather consider EZ mocks where redshift failures are stochastic, based on the model fitted to the observed data, we find larger shifts for $146/1000$ (considering both tails: $318/1000$), $499/1000$ ($880/1000$) and $403/1000$ ($862/1000$) EZ mocks for $\fsig$, $\apar$ and $\aper$, respectively.

Dividing the number of mocks used to build the covariance matrix by $2$, we find a shift of $4.3\%$ on $\fsig$, $0.1\%$ on $\apar$ and $0.0\%$ on $\aper$ (\emph{500 mocks in cov} versus \emph{baseline}). This would lead to a $1.7\%$, $0.0\%$ and $0.0\%$  increase of the error if added in quadrature, less than $5.8\%$, the typical increase of error bars required to account for the change in covariance matrix between these two configurations. We thus conclude that the RSD~+~BAO covariance matrix is stable with respect to the number of mocks.

Changing the covariance matrix built from contaminated EZ mocks for the one obtained from GLAM-QPM mocks, we find a $5.6\%$ shift on $\fsig$, $1.5\%$ on $\apar$ and $1.6\%$ on $\aper$. Larger variations happen for $171/1000$, $218/1000$, $107/1000$ EZ mocks, respectively (considering both tails: $363/1000$, $402/1000$ and $290/1000$). Even larger shifts are seen when using a covariance matrix based on EZ mocks without systematics (only the \emph{shuffled} scheme, \emph{no syst. cov}): $12.9\%$ shift on $\fsig$, $4.7\%$ on $\apar$ and $1.2\%$ on $\aper$. These changes happen for $12/1000$, $18/1000$, $252/1000$ EZ mocks, respectively (considering both tails: $27/1000$, $50/1000$ and $428/1000$).

One would also notice a large change ($26.7\%$ on $\fsig$, $5.4\%$ on $\apar$ and $4.7\%$ on $\aper$) for the SGC only measurements with the GLAM-QPM covariance matrix (\emph{GLAM-QPM cov}). A larger shift in $\fsig$ happens for $4/1000$ EZ mocks (considering both tails: $18/1000$). One may be concerned by some possible coupling between the data and the covariance matrix from contaminated EZ mocks, as the map of angular systematics injected in these mocks was inferred directly from the smoothed observed data density (see \Sec{ezmocks}). In \App{coupling_data_covariance_matrix} we show that we cannot find any evidence for this coupling based on EZ mocks.

The variations in the best fit parameters with the covariance matrix being very untypical, for conservativeness we include in our systematic budget the largest shifts seen in the NGC~+~SGC fit, %
namely those obtained with a covariance matrix based on EZ mocks without systematics (only the \emph{shuffled} scheme, \emph{no syst. cov}), i.e. $12.9\%$ on $\fsig$, $4.7\%$ on $\apar$ and $1.2\%$ on $\aper$.

Moving the centre of the $k$-bin by half a bin ($0.005 \hMpc$) leads to a shift of $0.6\%$ for $\fsig$ and $0.2\%$ for $\apar$ and $\aper$. A larger $\fsig$ shift is observed for $393/1000$ (considering both tails: $931/1000$) EZ mocks, $463/1000$ ($884/1000$) for $\apar$ and $539/1000$ ($922/1000$) for $\aper$. As in \Sec{results_bao_data}, since these shifts are compatible with mocks, %
we do not include them in the systematic budget.

Changing the fiducial cosmology of \Eq{fiducial_cosmology} for the OuterRim cosmology of \Eq{outerrim_cosmology}, we find moderate shifts of $6.0\%$ on $\fsig$, $1.2\%$ on $\apar$ and $0.3\%$ on $\aper$ (which are rescaled to the fiducial cosmology \Eq{fiducial_cosmology} for comparison purposes).
Larger shifts are observed for $263/1000$ (considering both tails: $349/1000$) EZ mocks for $\fsig$, $283/1000$ ($474/1000$) for $\apar$ and $312/1000$ ($831/1000$) for $\aper$ and hence are fully compatible with a statistical fluctuation. Since we accounted for the change of fiducial cosmology in the systematic modelling budget, we do not quote any other related systematic uncertainty.

\begin{table}
\caption{Error budget for RSD~+~BAO measurements on the eBOSS DR16 ELG sample. Percentages are provided with respect to the parameter value. The last three lines (statistics, systematics and total) recap the absolute statistical error bar, the systematic contribution (total minus statistics), and the total error bar, respectively.}
\label{tab:error_budget_rsdbao}
\centering
\begin{tabular}{lccc} 
\hline
source & $\fsig$ & $\apar$ & $\aper$ \\
\hline
\hline
linear & & &\\
\hline
survey geometry (from EZ mocks) & $0.1\%$ & $0.5\%$ & $0.1\%$ \\
RSD~+~BAO combination (from EZ mocks) & $1.0\%$ & $0.3\%$ & $0.5\%$ \\
model non-linearity (from EZ mocks) & $1.6\%$ & $1.0\%$ & $0.4\%$ \\
ARIC modelling (from EZ mocks) & $0.8\%$ & $0.6\%$ & $0.1\%$ \\
\hline
\hline
quadrature & & &\\
\hline
modelling systematics (from mock challenge) & $3.0\%$ & $0.9\%$ & $0.8\%$ \\
damping term $\Sigma_{\mathrm{nl}}$ & $1.9\%$ & $0.1\%$ & $0.8\%$ \\
photometric systematics & $9.4\%$ & $2.2\%$ & $1.4\%$ \\
fibre collisions (from EZ mocks) & $1.1\%$ & $0.6\%$ & $0.2\%$ \\
redshift failures & $6.1\%$ & $0.3\%$ & $0.3\%$ \\
covariance matrix & $12.9\%$ & $4.7\%$ & $1.2\%$ \\
\hline
\hline
statistics & ${}_{-0.075}^{+0.060}$ & ${}_{-0.090}^{+0.104}$ & ${}_{-0.037}^{+0.036}$ \\
systematics & ${}_{-0.021}^{+0.024}$ & ${}_{-0.031}^{+0.029}$ & ${}_{-0.011}^{+0.012}$ \\
total & ${}_{-0.096}^{+0.085}$ & ${}_{-0.12}^{+0.13}$ & ${}_{-0.049}^{+0.048}$ \\
\hline
\end{tabular}
\end{table}

The final error budget is reported in \Tab{error_budget_rsdbao}. The systematic bias related to the analysis methodology, namely the survey geometry, the RSD and BAO combination, the model non-linearity and the modelling of the angular and radial integral constraints (ARIC) are summed together in quadrature and added linearly to the statistical error bars. Other contributions are uncertainties due to our limited understanding of the ELG small-scale clustering (modelling systematics) or the survey selection function (photometric systematics, fibre collisions, redshift failures) or consist in analysis choices (damping term, covariance matrix, fiducial cosmology). These other terms are added in quadrature to the  statistical error bars.

Including both statistical and systematic uncertainties, our final combined RSD~+~BAO measurement is (in the \emph{Bayesian} case):
$\fsig = 0.289_{-0.096}^{+0.085}$,
$\apar = 1.08_{-0.12}^{+0.13}$ and
$\aper = 0.941 \pm 0.049$.

In terms of angular distance and Hubble parameter, we find:
\begin{align}
\fsig(\zeff=0.85) &= 0.289_{-0.096}^{+0.085} \nonumber \\
\DHu(\zeff=0.85)/\rdrag &= 20.0_{-2.2}^{+2.4} \\
\DM(\zeff=0.85)/\rdrag &= 19.17 \pm 0.99 \nonumber
\label{eq:rsdbao_cosmo_result}
\end{align}

As can be seen in \Fig{corner_rsdbao}, the linear bias combination $b_{1}\sig$ is correlated with $\fsig$.
Fixing $\sig$ to the fiducial cosmology in \Eq{fiducial_cosmology} we find $b_{1}^{\mathrm{NGC}} = 1.49 \pm 0.10$ and %
$b_{1}^{\mathrm{SGC}} = 1.52_{-0.11}^{+0.10}$, in agreement with previous studies \cite[e.g.][]{Comparat2013:1302.5655v2}. Best fit values and errors for all parameters are given in \App{full_best_fit}. No discrepancy can be seen between NGC and SGC nuisance parameters ($b_{1}\sig$, $b_{2}\sig$, $\sigma_{v}$ and $A_{g}$). $A_{g}^{\mathrm{NGC}}$ and $A_{g}^{\mathrm{SGC}}$ are compatible with zero, as expected with nearly Poisson shot noise.

To include the systematic error budget in the above RSD~+~BAO posteriors, we rescale the distance of each MCMC sample $(\fsig,\apar,\aper)$ to the median values by the ratio of total to statistical only errors. For example, the new $\fsig$ position is:
\begin{equation}
\frac{\sigma_{\fsig,\mathrm{tot}}}{\sigma_{\fsig ,\mathrm{stat}}}\left[\fsig - \median(\fsig)\right] + \median(\fsig).
\end{equation}
$\sigma_{\fsig,\mathrm{tot}}$ and $\sigma_{\fsig,\mathrm{stat}}$ are the upper (lower) total and statistical error bars with respect to the median when $\fsig > \median(\fsig)$ ($\fsig < \median(\fsig)$). We proceed similarly for $\apar$ and $\aper$.

\section{Consensus}
\label{sec:consensus}

BAO and RSD analyses of the eBOSS ELG sample are also performed in configuration space, as detailed in~\citet{Raichoor2020,Tamone2020}.

\begin{figure*}
\centering
\includegraphics[width=0.8\textwidth]{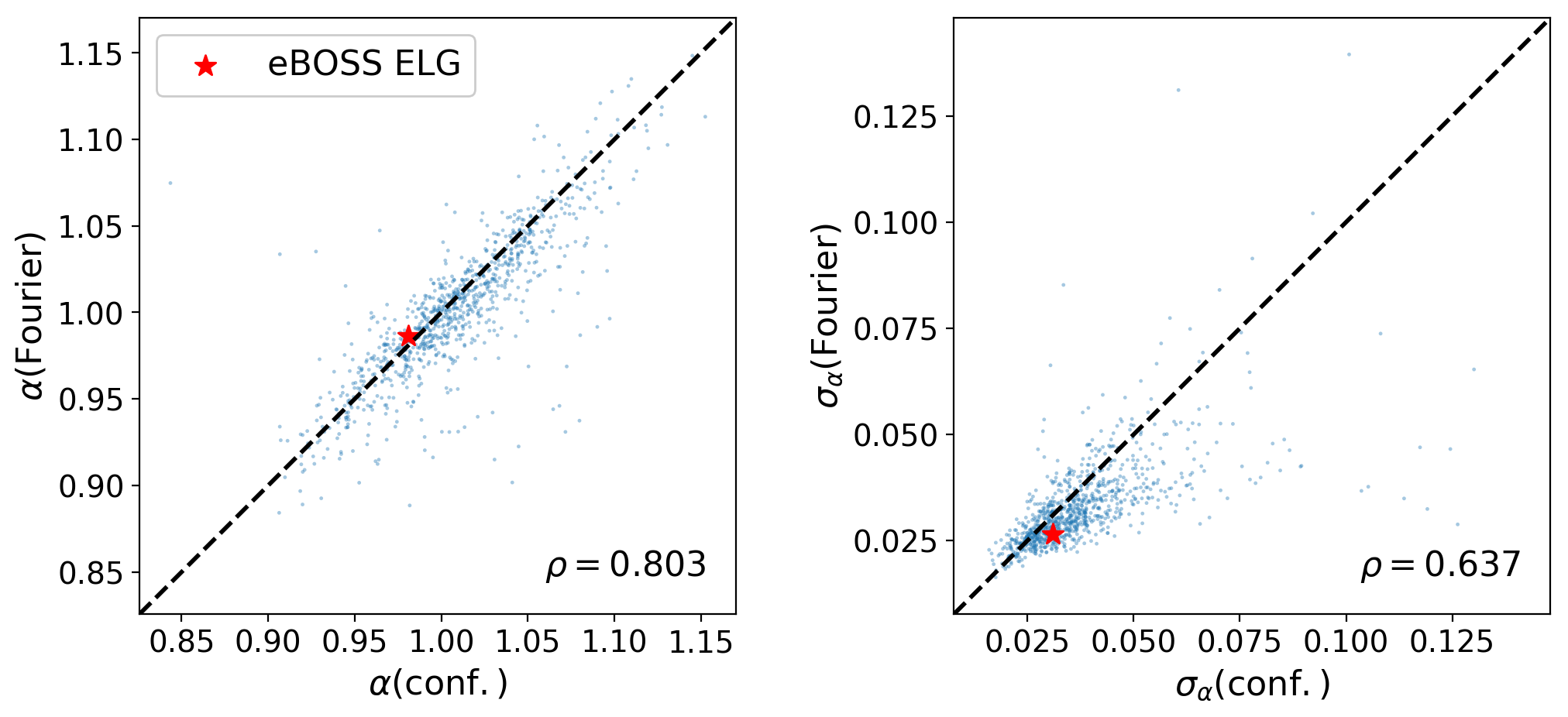}
\caption{Post-reconstruction isotropic BAO measurements in the EZ mocks including all systematics, in Fourier and configuration space (left: best fits, right: errors, taken to be the mean of the upper and lower error bars). The red cross shows the eBOSS ELG data.}
\label{fig:consensus_bao}
\end{figure*}

Post-reconstruction isotropic BAO measurements in Fourier and configuration space are compared for each mock in \Fig{consensus_bao}. Both measurements are well correlated ($\rho = 0.8$). \cite{Raichoor2020} measurement, %
$\DV(\zeff=0.845)/\rdrag = 18.23 \pm 0.58$ (statistical only)
is $<0.2\sigma$ away from our Fourier space measurement, $\DV(\zeff=0.845) = 18.33_{-0.52}^{+0.46}$ (statistical only). A larger difference in the best fits occurs for $223/956$ EZ mocks including all systematics (considering both tails: $684/1000$), while we find $387/956$ EZ mocks (considering both tails: $473/1000$) with a larger difference in the mean of the lower and upper error bars. The good agreement between configuration and Fourier space measurements is shown by the data cross in \Fig{consensus_bao}, lying close to the diagonal. We choose the Fourier space measurement as consensus as it has a lower statistical uncertainty.

This measurement, $\DV(\zeff=0.845)/\rdrag = 18.33_{-0.62}^{+0.57}$, is $0.6\sigma$ below the \citet{Planck2018:1807.06209v2} CMB-based (TT, TE, EE, lowE, lensing) prediction.

\begin{figure}
\centering
\includegraphics[width=0.9\columnwidth]{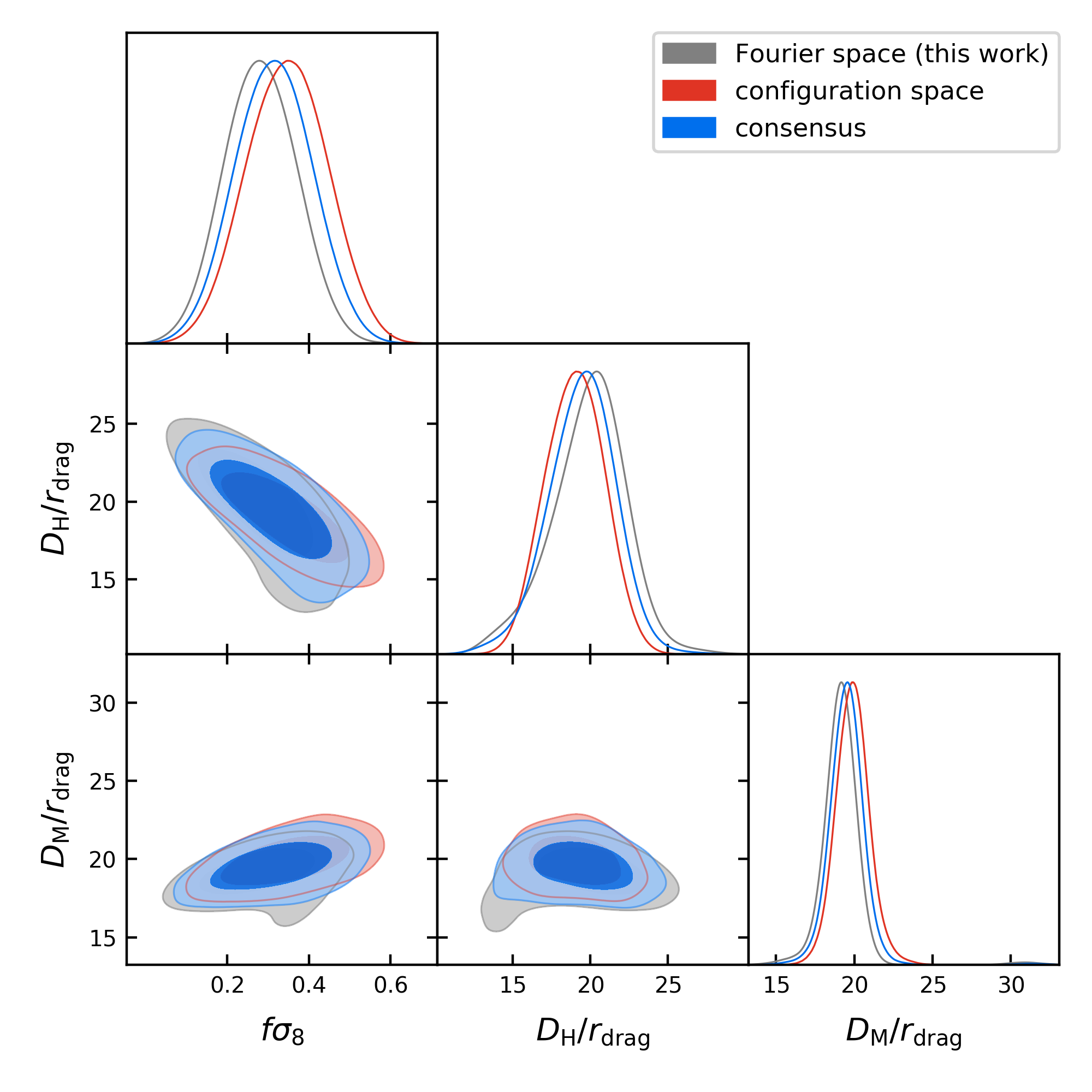}
\caption{Fourier and configuration space, and combined posteriors of the final eBOSS ELG RSD~+~BAO measurements (including systematic uncertainties).}
\label{fig:consensus_rsdbao}
\end{figure}

Combining RSD and BAO measurements in configuration space and including systematic uncertainties, \citet{Tamone2020} find:
\begin{align}
\fsig(\zeff=0.85) &= 0.35 \pm 0.10 \nonumber \\
\DHu(\zeff=0.85)/\rdrag &= 19.1_{-2.0}^{+1.9} \\
\DM(\zeff=0.85)/\rdrag &= 19.9 \pm 1.0. \nonumber
\end{align}
These values are $0.7\sigma$, $0.5\sigma$ and $0.7\sigma$ away from our Fourier space median $\fsig$, $\DHu/\rdrag$ and $\DM/\rdrag$ values, respectively. Comparing the best fits instead, differences are $0.4\sigma$, $0.4\sigma$ and $0.7\sigma$, respectively. Again comparing best fits, larger differences occur in $386/1000$ (considering both tails: $543/1000$), $193/1000$ ($358/1000$) and $246/1000$ ($311/1000$) EZ mocks including all systematics. Considering posterior medians instead, larger differences occur in $158/1000$ (considering both tails: $185/1000$), $119/1000$ ($234/1000$) and $256/1000$ ($324/1000$) best fits to EZ mocks including all systematics.
To combine these two measurements, since posteriors are not Gaussian but show comparable error bars, we translate them such that their two medians are located at the mean median and take the mean of the two posteriors as consensus. This method leads to an unbiased measurement if both measurements are unbiased, and is conservative about the final error bars. We show the two posteriors and their combination in \Fig{consensus_rsdbao}. The consensus RSD~+~BAO eBOSS ELG measurement is thus:
\begin{align}
\fsig(\zeff=0.85) &= 0.315 \pm 0.095 \nonumber \\
\DHu(\zeff=0.85)/\rdrag &= 19.6_{-2.1}^{+2.2} \\
\DM(\zeff=0.85)/\rdrag &= 19.5 \pm 1.0. \nonumber
\end{align}

These measurements are $1.4\sigma$, $0.5\sigma$ and $0.9\sigma$ from the \citet{Planck2018:1807.06209v2} CMB-based (TT, TE, EE, lowE, lensing) predictions for $\fsig$, $\DHu/\rdrag$ and $\DM/\rdrag$, respectively.

We interpolate the consensus MCMC posterior (including systematic error bars) on a grid of $\fsig$, $\DHu/\rdrag$ and $\DM/\rdrag$ which is used, together with the BAO likelihood profile (including systematic error bars), for the combined eBOSS cosmological constraints presented in~\cite{eBOSS2020}.

\section{Conclusions}
\label{sec:conclusions}

In the above, together with~\cite{Raichoor2020,Tamone2020}, we presented the first measurement of the pre-reconstruction RSD and post-reconstruction BAO signal in the eBOSS ELG sample. 

We started by testing our implementation of the RSD TNS~\citep{Taruya2010:1006.0699v1} model and isotropic BAO template using N-body based simulations. These models proved robust enough for our analysis. However, this analysis was complicated by various observational and analysis artefacts.

First, the fine-grained veto masks applied to the eBOSS ELG data led us to revise the way window functions are normalised in the model. We also noticed that measuring the expected redshift distribution from observed data itself (using the \emph{shuffled} scheme) led to a significant bias of cosmological parameters, which we corrected by modelling the induced radial integral constraint as in~\cite{deMattia2019:1904.08851v3}. Fibre collisions have been found to have a moderate impact, which we modelled following~\cite{Hahn2017:1609.01714v1}. Finally, residual angular systematics due to the inhomogeneous photometry, which could not be treated following the standard technique of template regression, were mitigated using the \emph{pixelated} scheme as discussed in~\cite{deMattia2019:1904.08851v3}.

For this study, to test our analysis pipeline, we implemented the survey geometry and realistic systematics into mock catalogues: depth-dependent radial selection function, angular photometric systematics, fibre collisions and redshift failures. We proved on mocks that our analysis is robust to these systematics when including all the corrections mentioned above. We derived residual systematic uncertainties from both our mock studies and robustness tests run on data.

As another extension over previous work, we combined RSD and post-reconstruction BAO measurements at the likelihood level to avoid assuming them Gaussian and strengthen our cosmological measurements.

Taking into account statistical and systematic uncertainties, the post-reconstruction isotropic BAO analysis in Fourier space provides a measurement of the ratio of the volume-averaged distance to the sound horizon at the drag epoch: 
\begin{equation*}
\DV(\zeff=0.845)/\rdrag = 18.33_{-0.62}^{+0.57},
\end{equation*}
which is the BAO consensus measurement of the eBOSS ELG sample.
The Fourier space RSD~+~BAO measurement, including statistical and systematic uncertainties, is:
\begin{align*}
\fsig(\zeff=0.85) &= 0.289_{-0.096}^{+0.085} \nonumber \\
\DHu(\zeff=0.85)/\rdrag &= 20.0_{-2.2}^{+2.4} \\
\DM(\zeff=0.85)/\rdrag &= 19.17 \pm 0.99. \nonumber
\end{align*}
Combined with configuration space results of~\citet{Tamone2020}, we find:
\begin{align*}
\fsig(\zeff=0.85) &= 0.315 \pm 0.095 \nonumber \\
\DHu(\zeff=0.85)/\rdrag &= 19.6_{-2.1}^{+2.2} \\
\DM(\zeff=0.85)/\rdrag &= 19.5 \pm 1.0. \nonumber
\end{align*}

Some observational systematics may remain, due to our incomplete understanding of the relation between the ELG target density and the imaging properties in the early DECaLS release used in the eBOSS ELG target selection. We note that DESI~\citep{DESI2016:1611.00036v2} will target ELG with the Legacy Imaging Surveys~\citep{Dey2019:1804.08657v2} that includes DECaLS. Though this latest photometric survey is deeper and more isotropic, fainter ELG targets are targeted, such that one may expect significant 3-dimensional fluctuations of the survey selection function due to photometric variations as seen in the eBOSS ELG sample. Understanding these fluctuations will be key to the clustering analysis of the DESI ELG sample.

In conclusion, this work was the opportunity to deal with analysis systematics (impact of fixing the template cosmology, integral constraints) and observational systematics (3-dimensional fluctuations of the survey selection function) which were not fully tackled in previous SDSS clustering analyses and are of importance for the next generation of spectroscopic surveys, including DESI~\citep{DESI2016:1611.00036v2} and Euclid~\citep{Euclid2011:1110.3193v1}.

\section*{acknowledgements}

AdM acknowledges support from the P2IO LabEx (ANR-10-LABX-0038) in the framework "Investissements d'Avenir" (ANR-11-IDEX-0003-01) managed by the Agence Nationale de la Recherche (ANR, France).

This work was supported by the ANR eBOSS project (ANR-16-CE31-0021) of the French National Research Agency.

AR acknowledges support from the ERC advanced grant LIDA.
AR, CZ, and AT acknowledge support from the SNF grant 200020\_175751.
AJR is grateful for support from the Ohio State University Center for Cosmology and Particle Physics.
S. Alam is supported by the European Research Council through the COSFORM Research Grant (\#670193).
S. Avila was supported by the MICUES project, funded by the European Union’s Horizon 2020 research programme under the Marie Sklodowska-Curie Grant Agreement No. 713366 (InterTalentum UAM).
VGP acknowledges support from the European Union's Horizon 2020 research and innovation programme (ERC grant \#769130).
YW and GBZ are supported by NSFC Grants 11925303, 11720101004, 11673025 and 11890691. GBZ is also supported by the National Key Basic Research and Development Program of China (No. 2018YFA0404503), and a grant of CAS Interdisciplinary Innovation Team. YW is also supported by the Nebula Talents Program of NAOC.
GR acknowledges support from the National Research Foundation of Korea (NRF) through Grants No. 2017R1E1A1A01077508 and No. 2020R1A2C1005655 funded by the Korean Ministry of Education, Science and Technology (MoEST), and from the faculty research fund of Sejong University.

Funding for the Sloan Digital Sky Survey IV has been provided by the Alfred P. Sloan Foundation, the U.S. Department of Energy Office of Science, and the Participating Institutions. SDSS acknowledges support and resources from the Center for High-Performance Computing at the University of Utah. The SDSS web site is www.sdss.org.

SDSS is managed by the Astrophysical Research Consortium for the Participating Institutions of the SDSS Collaboration including the Brazilian Participation Group, the Carnegie Institution for Science, Carnegie Mellon University, the Chilean Participation Group, the French Participation Group, Harvard-Smithsonian Center for Astrophysics, Instituto de Astrofísica de Canarias, The Johns Hopkins University, Kavli Institute for the Physics and Mathematics of the Universe (IPMU) / University of Tokyo, the Korean Participation Group, Lawrence Berkeley National Laboratory, Leibniz Institut für Astrophysik Potsdam (AIP), Max-Planck-Institut für Astronomie (MPIA Heidelberg), Max-Planck-Institut für Astrophysik (MPA Garching), Max-Planck-Institut für Extraterrestrische Physik (MPE), National Astronomical Observatories of China, New Mexico State University, New York University, University of Notre Dame, Observatório Nacional / MCTI, The Ohio State University, Pennsylvania State University, Shanghai Astronomical Observatory, United Kingdom Participation Group, Universidad Nacional Autónoma de México, University of Arizona, University of Colorado Boulder, University of Oxford, University of Portsmouth, University of Utah, University of Virginia, University of Washington, University of Wisconsin, Vanderbilt University, and Yale University.

This research used resources of the National Energy Research Scientific Computing Center, a DOE Office of Science User Facility supported by the Office of Science of the U.S. Department of Energy under Contract No. DE-AC02-05CH11231.

This work used resources from the Sciama High Performance Computing cluster, which is supported by the Institute of Cosmology and Gravitation and the University of Portsmouth.

This research made use of the MINUIT algorithm~\cite{Minuit1975} via the \cite{iminuit} Python interface.

The CosmoSim database used in this paper is a service by the Leibniz-Institute for Astrophysics Potsdam (AIP).
The MultiDark database was developed in cooperation with the Spanish MultiDark Consolider Project CSD2009-00064. The authors gratefully acknowledge the Gauss Centre for Supercomputing e.V. (www.gauss-centre.eu) and the Partnership for Advanced Supercomputing in Europe (PRACE, www.prace-ri.eu) for funding the MultiDark simulation project by providing computing time on the GCS Supercomputer SuperMUC at Leibniz Supercomputing Centre (LRZ, www.lrz.de).

We thank the anonymous referee for their report, which helped us improve the clarity of the paper.

\section*{Data availability}

The power spectrum, covariance matrices, and resulting likelihoods for cosmological parameters are available via the SDSS Science Archive Server (\url{https://sas.sdss.org/sas/dr16/eboss}).

\bibliographystyle{mnras}
\bibliography{references}

\appendix
\newcommand{\dxp}[2]{\frac{\partial x_{#1}^{t}}{\partial p_{#2}}}
\newcommand*{\dprime}{^{\prime\prime}\mkern-1.2mu}

\appendix

\section{Covariance corrections}
\label{app:covariance_corrections}

In this section we justify the corrections applied to the parameter covariance for our combined NGC and SGC cosmological fit. Formula \Eq{covariance_m_1_naive} is (formally) incorrect when combining two independent Gaussian likelihoods (e.g. NGC and SGC). Indeed, an intuitive use of \Eq{covariance_m_1_naive} would take $n_{b}$ as the number of bins in either NGC or SGC and $n_{p}$ the total number of parameters. Then, however, the constraint $n_{p} \leq n_{b}$ inherent to this formula appears artificial. Indeed, there is no issue with having $n_{p} > n_{b}$ as long as $n_{p}$ is less than the total number of bins. Another approach would be to take $n_{b}$ as the total number of bins in NGC and SGC; though this would be correct if we estimated the combined NGC and SGC covariance from mocks, this does not apply to our case where we impose the cross-covariance between NGC and SGC to be zero. Therefore, we have to revise \Eq{covariance_m_1_naive} in the context of a block-diagonal covariance matrix.

The Hartlap~\citep{Hartlap2007:astro-ph/0608064v2} correction is applied to the inverse NGC covariance matrix to obtain the precision matrix:
\begin{equation}
\vc{\Psi}^{\mathrm{NGC}} = \left(1-D^{\mathrm{NGC}}\right) \left(\vc{C}^{\mathrm{NGC}}\right)^{-1}, \quad D^{\mathrm{NGC}} = \frac{n_{b}^{\mathrm{NGC}}+1}{n_{m}^{\mathrm{NGC}}-1}
\label{eq:covariance_hartlap_ngc}
\end{equation}
with $n_{b}^{\mathrm{NGC}}$ the number of bins and $n_{m}^{\mathrm{NGC}}$ the number of mocks in NGC; similarly for SGC. The full precision matrix $\vc{\Psi}$ is a block-diagonal matrix built from $\vc{\Psi}^{\mathrm{NGC}}$ and $\vc{\Psi}^{\mathrm{SGC}}$.
To obtain the corrections to the parameter covariance the errors on each precision matrix must be propagated through the Fisher information.
The estimator for parameter $p_{\alpha}$, whose true value is assumed to be zero without loss of generality, is (Eq.~(24) in~\citealt{Dodelson2013:1304.2593v2}):
\begin{align}
\hat{p}_{\alpha} = \left[F + \Delta F\right]_{\alpha\alpha^{\prime}}^{-1} \dxp{i}{\alpha^{\prime}} \Psi_{ij} \left(x_{j}^{d}-x_{j}^{t}\right),
\end{align}
where $x_{i}^{d}$ and $x_{i}^{t} = \aver{x_{i}^{d}}$ are the data measurement and its true value. $F_{\alpha\beta}$ is the true Fisher matrix:
\begin{equation}
F_{\alpha\beta} = \dxp{i}{\alpha} \Psi_{ij}^{t} \dxp{j}{\beta},
\end{equation}
with $\vc{\Psi}^{t} = \left(\vc{C}^{t}\right)^{-1}$ the true inverse covariance matrix and similarly:
\begin{equation}
\Delta F_{\alpha\beta} = \dxp{i}{\alpha} \Delta \Psi_{ij}^{t} \dxp{j}{\beta}
\end{equation}
with $\Delta \Psi_{ij} = \Psi_{ij} - \Psi_{ij}^{t}$.
Then,~\cite{Dodelson2013:1304.2593v2} recall that the leading order parameter covariance is $\aver{\hat{p}_{\alpha} \hat{p}_{\beta}} \ni F_{\alpha\beta}^{-1}$ and derive the next-to-leading (second order) contribution:
\begin{align}
\aver{\hat{p}_{\alpha} \hat{p}_{\beta}} & \ni F_{\alpha\alpha^{\prime}}^{-1} \left[ \dxp{i}{\alpha^{\prime}} \dxp{i^{\prime}}{\beta^{\prime}} C_{j j^{\prime}}^{t} (\Delta \Psi)_{ij} (\Delta \Psi)_{i^{\prime}j^{\prime}} \right] F_{\beta\beta^{\prime}}^{-1} \label{eq:covariance_parameter_1}\\
& - \left[F^{-1} \Delta F F^{-1} \Delta F F^{-1}\right]_{\alpha\beta} \label{eq:covariance_parameter_2}
\end{align}
Similarly to~\cite{Taylor2013:1212.4359v1,Dodelson2013:1304.2593v2} we write the covariance of the precision matrix fluctuations as:
\begin{align}
\aver{\Delta \Psi_{ij} \Delta \Psi_{i^{\prime}j^{\prime}}} &= \left(A_{iji^{\prime}j^{\prime}}^{\mathrm{NGC}} + A_{iji^{\prime}j^{\prime}}^{\mathrm{SGC}}\right) \Psi_{ij}^{t} \Psi_{i^{\prime}j^{\prime}}^{t} \\
& + \left(B_{iji^{\prime}j^{\prime}}^{\mathrm{NGC}} + B_{iji^{\prime}j^{\prime}}^{\mathrm{SGC}}\right) \left( \Psi_{ii^{\prime}}^{t} \Psi_{jj^{\prime}}^{t} + \Psi_{ij^{\prime}}^{t} \Psi_{ji^{\prime}}^{t} \right)
\end{align}
where $A_{iji^{\prime}j^{\prime}}^{\mathrm{NGC}}$ is constant (equal to $A^{\mathrm{NGC}}$) if indices $iji^{\prime}j^{\prime}$ all lie in the same block NGC, zero elsewhere (i.e. cross-covariance terms between NGC and SGC are zero), and similarly for SGC and B terms. We recall that:
\begin{align}
A^{\mathrm{NGC}} &= \frac{2}{\left(n_{m}^{\mathrm{NGC}} - n_{b}^{\mathrm{NGC}} - 1\right)\left(n_{m}^{\mathrm{NGC}} - n_{b}^{\mathrm{NGC}} - 4\right)}, \\
B^{\mathrm{NGC}} &= \frac{n_{m}^{\mathrm{NGC}} - n_{b}^{\mathrm{NGC}} - 2}{\left(n_{m}^{\mathrm{NGC}} - n_{b}^{\mathrm{NGC}} - 1\right)\left(n_{m}^{\mathrm{NGC}} - n_{b}^{\mathrm{NGC}} - 4\right)},
\end{align}
and similarly for SGC.
Let us consider the contribution from the $A$ terms first. From \Eq{covariance_parameter_1} one gets the contribution:
\begin{equation}
\aver{\hat{p}_{\alpha} \hat{p}_{\beta}} \ni F_{\alpha\alpha^{\prime}}^{-1} \left[A^{\mathrm{NGC}} F_{\alpha^{\prime}\beta^{\prime}}^{\mathrm{NGC}} +  A^{\mathrm{SGC}} F_{\alpha^{\prime}\beta^{\prime}}^{\mathrm{SGC}} \right] F_{\beta\beta^{\prime}}^{-1}
\end{equation}
where we split the total Fisher information $F = F^{\mathrm{NGC}} + F^{\mathrm{SGC}}$ (since NGC and SGC are independent).
\Eq{covariance_parameter_2} gives:
\begin{multline}
\aver{\hat{p}_{\alpha} \hat{p}_{\beta}} \ni - F_{\alpha\alpha^{\prime}}^{-1} F_{\beta^{\prime}\alpha^{\dprime}}^{-1}F_{\beta^{\dprime}\beta}^{-1} \left( A^{\mathrm{NGC}} F_{\alpha^{\prime}\beta^{\prime}}^{\mathrm{NGC}} F_{\alpha^{\dprime}\beta^{\dprime}}^{\mathrm{NGC}} \right.\\
\left. + A^{\mathrm{SGC}} F_{\alpha^{\prime}\beta^{\prime}}^{\mathrm{SGC}} F_{\alpha^{\dprime}\beta^{\dprime}}^{\mathrm{SGC}} \right)
\end{multline}
Let us move to the $B$ terms. From \Eq{covariance_parameter_1} one gets the contribution:
\begin{multline}
\aver{\hat{p}_{\alpha} \hat{p}_{\beta}} \ni F_{\alpha\alpha^{\prime}}^{-1} \left[B^{\mathrm{NGC}} \left( n_{b}^{\mathrm{NGC}} + 1 \right) F_{\alpha^{\prime}\beta^{\prime}}^{\mathrm{NGC}} \right. \\
\left. + B^{\mathrm{SGC}} \left( n_{b}^{\mathrm{SGC}} + 1 \right) F_{\alpha^{\prime}\beta^{\prime}}^{\mathrm{SGC}} \right] F_{\beta\beta^{\prime}}^{-1},
\end{multline}
and from \Eq{covariance_parameter_2}:
\begin{multline}
\aver{\hat{p}_{\alpha} \hat{p}_{\beta}} \ni - F_{\alpha\alpha^{\prime}}^{-1} F_{\beta^{\prime}\alpha^{\dprime}}^{-1}F_{\beta^{\dprime}\beta}^{-1}
\left( B^{\mathrm{NGC}} F_{\alpha^{\prime}\alpha^{\dprime}}^{\mathrm{NGC}} F_{\beta^{\prime}\beta^{\dprime}}^{\mathrm{NGC}} \right. \\
\left. + B^{\mathrm{SGC}} F_{\alpha^{\prime}\alpha^{\dprime}}^{\mathrm{SGC}} F_{\beta^{\prime}\beta^{\dprime}}^{\mathrm{SGC}} + B^{\mathrm{NGC}} F_{\alpha^{\prime}\beta^{\dprime}}^{\mathrm{NGC}} F_{\beta^{\prime}\alpha^{\dprime}}^{\mathrm{NGC}} + B^{\mathrm{SGC}} F_{\alpha^{\prime}\beta^{\dprime}}^{\mathrm{SGC}} F_{\beta^{\prime}\alpha^{\dprime}}^{\mathrm{SGC}} \right).
\end{multline}
Formulae above could be evaluated numerically, but we will consider a simpler case in the following. We assume that $F_{\alpha\beta}^{\mathrm{NGC}}$ and $F_{\alpha\beta}^{\mathrm{SGC}}$ are block-diagonal, with specific parameters for NGC and SGC which are uncorrelated, and a set of common parameters $\alpha\beta$ for which the Fisher information content can be written $F_{\alpha\beta}^{\mathrm{NGC}} = f^{\mathrm{NGC}} F_{\alpha\beta}$ (respectively $F_{\alpha\beta}^{\mathrm{SGC}} = f^{\mathrm{SGC}} F_{\alpha\beta}$), with  $f^{\mathrm{NGC}} + f^{\mathrm{SGC}} = 1$. We also assume the common parameters to be uncorrelated to the NGC and SGC specific parameters\footnote{This is of course not the case in practice.}. 
Then, the contribution from the $A$ and $B$ terms to the covariance of the common parameters is simply the sum of terms $\mathcal{C}_{1}$ (from Eq.~\eqref{eq:covariance_parameter_1}) and $\mathcal{C}_{2}$ (from Eq.~\eqref{eq:covariance_parameter_2}):
\begin{equation}
\aver{\hat{p}_{\alpha} \hat{p}_{\beta}} \ni \left[\mathcal{C}_{1}^{\mathrm{NGC}} + \mathcal{C}_{1}^{\mathrm{SGC}} - \mathcal{C}_{2}^{\mathrm{NGC}} - \mathcal{C}_{2}^{\mathrm{SGC}}\right] F_{\alpha\beta}^{-1}
\end{equation}
where:
\begin{equation}
\mathcal{C}_{1}^{\mathrm{NGC}} = A^{\mathrm{NGC}} f^{\mathrm{NGC}} + B^{\mathrm{NGC}} f^{\mathrm{NGC}} \left(n_{b}^{\mathrm{NGC}} + 1\right)
\end{equation}
and
\begin{equation}
\mathcal{C}_{2}^{\mathrm{NGC}} = A^{\mathrm{NGC}} \left(f^{\mathrm{NGC}}\right)^{2} + B^{\mathrm{NGC}} f^{\mathrm{NGC}} \left(n_{\mathrm{eff}}^{\mathrm{NGC}} + f^{\mathrm{NGC}}\right)
\end{equation}
where we use the effective number of parameters $n_{\mathrm{eff}}^{\mathrm{NGC}} = n_{\mathrm{sp}}^{\mathrm{NGC}} + f^{\mathrm{NGC}} n_{\mathrm{co}}$, with $n_{\mathrm{sp}}^{\mathrm{NGC}}$ the number of parameters specific to NGC (similarly for SGC) and $n_{\mathrm{co}}$ the number of parameters in common. The contribution to the covariance of the specific parameters is obtained for e.g. NGC by forcing $f^{\mathrm{NGC}} = 1$ (then $f^{\mathrm{SGC}} = 0$) and keeping $n_{\mathrm{eff}}^{\mathrm{NGC}}$ fixed.
In this case, errors from the NGC precision matrix only contribute to the parameter covariance. In the simplified case where $n_{\mathrm{co}} = 0$, the known result of~\citet{Dodelson2013:1304.2593v2}:
\begin{equation}
\aver{\hat{p}_{\alpha} \hat{p}_{\beta}} \ni B^{\mathrm{NGC}} \left[ n_{b}^{\mathrm{NGC}} - n_{\mathrm{sp}}^{\mathrm{NGC}} \right] F_{\alpha\beta}^{-1}
\end{equation}
is recovered.
Adding up $A$ and $B$ contributions, the parameter covariance is:
\begin{equation}
V_{\alpha\beta} = \left[1 + \mathcal{C}_{1}^{\mathrm{NGC}} + \mathcal{C}_{1}^{\mathrm{SGC}} - \mathcal{C}_{2}^{\mathrm{NGC}} - \mathcal{C}_{2}^{\mathrm{SGC}}\right] F_{\alpha\beta}^{-1}.
\end{equation}
The parameter variance estimated from the likelihood is (Eq.~(16) in~\citealt{Percival2014:1312.4841v1}):
\begin{equation}
\sigma_{\alpha\beta}^{2} = \left[F +\Delta F\right]_{\alpha\beta}^{-1},
\end{equation}
whose second order term is just \Eq{covariance_parameter_2} (Eq.~(15) in~\citealt{Percival2014:1312.4841v1}). Then:
\begin{equation}
\sigma_{\alpha\beta}^{2} = \left[1 + \mathcal{C}_{2}^{\mathrm{NGC}} + \mathcal{C}_{2}^{\mathrm{SGC}} \right] F_{\alpha\beta}^{-1}.
\label{eq:covariance_likelihood}
\end{equation}
Therefore, the full correction to apply to the parameter covariance estimated from the likelihood is:
\begin{equation}
m_{1} = \frac{V_{\alpha\beta}}{\sigma_{\alpha\beta}^{2}} = \frac{1 + \mathcal{C}_{1}^{\mathrm{NGC}} + \mathcal{C}_{1}^{\mathrm{SGC}} - \mathcal{C}_{2}^{\mathrm{NGC}} - \mathcal{C}_{2}^{\mathrm{SGC}}}{1 + \mathcal{C}_{2}^{\mathrm{NGC}} + \mathcal{C}_{2}^{\mathrm{SGC}}}.
\label{eq:covariance_m_1}
\end{equation}
As noted by~\citet{Percival2014:1312.4841v1}, in case the mocks used to produce the covariance matrix are fitted, the covariance of the measurements $x_{i}$ is just $C_{ij}$:
\begin{align}
\aver{\left(x_{i}^{d}-x_{i}^{t}\right)\left(x_{j}^{d}-x_{j}^{t}\right)} &= C_{ij}\\
 &= \left(1 - D_{ij}^{\mathrm{NGC}}\right) \Psi_{ij}^{-1} + \left(1 - D_{ij}^{\mathrm{SGC}}\right) \Psi_{ij}^{-1}
\end{align}
where $D_{ij}^{\mathrm{NGC}} = D^{\mathrm{NGC}}$ if $ij$ lie in the NGC block, zero otherwise, and similarly for SGC. Then:
\begin{align}
\aver{\hat{p}_{\alpha} \hat{p}_{\beta}} &= \left[F +\Delta F\right]_{\alpha\alpha^{\prime}}^{-1}
\left[F +\Delta F\right]_{\beta\beta^{\prime}}^{-1} \dxp{i}{\alpha^{\prime}} \dxp{i^{\prime}}{\beta^{\prime}} \\
& \Psi_{ii^{\prime}} \Psi_{jj^{\prime}} \aver{\left(x_{i}^{d}-x_{i}^{t}\right)\left(x_{j}^{d}-x_{j}^{t}\right)} \\
&= \left[F +\Delta F\right]_{\alpha\alpha^{\prime}}^{-1}
\left[F +\Delta F\right]_{\beta\beta^{\prime}}^{-1} \\
& \left\lbrace \left(1 - D^{\mathrm{NGC}}\right) \left[F + \Delta F\right]_{\alpha^{\prime}\beta^{\prime}}^{\mathrm{NGC}} \right.\\
& \left. + \left(1 - D^{\mathrm{SGC}}\right) \left[F + \Delta F\right]_{\alpha^{\prime}\beta^{\prime}}^{\mathrm{SGC}} \right\rbrace \\
&= \left[F + \Delta F\right]_{\alpha\beta}^{-1} \left[\left(1 - D^{\mathrm{NGC}}\right) f^{\mathrm{NGC}} + \left(1 - D^{\mathrm{SGC}}\right) f^{\mathrm{SGC}} \right].
\end{align}
Hence, the covariance of best fit parameter values obtained from the mocks should be rescaled by:
\begin{align}
m_{2} &= \frac{V_{\alpha\beta}}{\aver{\hat{p}_{\alpha} \hat{p}_{\beta}}} = \frac{V_{\alpha\beta}}{\sigma_{\alpha\beta}^{2}} \frac{\sigma_{\alpha\beta}^{2}}{\aver{\hat{p}_{\alpha} \hat{p}_{\beta}}} \\
&= m_{1} \left[\left(1 - D^{\mathrm{NGC}}\right) f^{\mathrm{NGC}} + \left(1 - D^{\mathrm{SGC}}\right) f^{\mathrm{SGC}} \right]^{-1}.
\label{eq:covariance_m_2}
\end{align}
to be compared with the parameter covariance derived from the likelihood (applying the Harlap factor of \Eq{covariance_hartlap_ngc} and $m_{1}$ of \Eq{covariance_m_1}).

\section{Effective redshift}
\label{app:effective_redshift}

In this section we derive another definition of the effective redshift, more specific to the Fourier space analysis.

With the Yamamoto estimator, in the local plane parallel approximation, assuming the (perfectly known) survey selection function varies slowly compared to the correlation function, and with infinite $d\Omega_{k}$ sampling we measure (in average):
\begin{equation}
\aver{\hat{P}_{\ell}(k)} = \frac{\int d^{3} r \bar{n}^{2}(r) P_{\ell}(k,z(r))}{\int d^{3} r \bar{n}^{2}(r)}
\label{eq:expected_yamamoto_estimator}
\end{equation}
where $\bar{n}(r)$ is the expected mean density of weighted galaxies in the absence of clustering and $P_{\ell}(k,z)$ is the true galaxy power spectrum at redshift $z$. Taylor expanding the power spectrum about the effective redshift $\zeff$:
\begin{equation}
P_{\ell}(k,z) = P_{\ell}(k,\zeff) + P_{\ell}^{\prime}(k,\zeff) (z - \zeff) + \cdots
\end{equation}
and substituting this expression into \Eq{expected_yamamoto_estimator}, we find that the expected value of the Yamamoto estimator can be approximated at first order by the quantity that we actually model:
\begin{equation}
\tilde{P}_{\ell}(k) = \frac{\int d^{3} r \bar{n}^{2}(r) P_{\ell}(k,\zeff)}{\int d^{3} r \bar{n}^{2}(r)}
\end{equation}
if we use the effective redshift:
\begin{equation}
z_{\eff} = \frac{\int d^{3} r \bar{n}^{2}(r) z(r)}{\int d^{3} r \bar{n}^{2}(r)} \simeq \frac{\sum_{i=1}^{N_{g}} \wtot n_{g,i} z_{g,i}}{\sum_{i=1}^{N_{g}} \wtot n_{g,i}}
\end{equation}
We checked that the value obtained with this definition of the effective redshift agrees with \Eq{effective_redshift} to the $0.5 \%$ level.

\section{Coupling between data and covariance matrix}
\label{app:coupling_data_covariance_matrix}

Some concern may be raised about a possible coupling between the data power spectrum measurements and the covariance matrix built from the $1000$ EZ mocks. Indeed, angular systematics were implemented in these mocks based on a map of the observed ELG density (smoothed by a Gaussian beam of radius $1\deg$). By construction, this systematic map includes data clustering angular modes which diffuse in the clustering of the contaminated EZ mocks. To check the importance of this effect, we generate a new systematic map based on the angular target density measured on one contaminated EZ \emph{reference mock}, and contaminate the other $999$ EZ mocks with this new map. Best fits to the \emph{reference mock} using the original covariance matrix and using the new one are compared. 
Performing RSD fits to the SGC, we find small shifts of $3.5\%$ for $\fsig$, $0.2\%$ for $\apar$ and $0.3\%$ for $\aper$, which are not significant since larger shifts are seen in $116/999$, $471/999$ and $440/999$ EZ mocks (considering both tails: $219/999$, $901/999$ and $783/999$).

Redshift failures are also implemented in the mocks based on the observed data. We check the potential bias due to this tuning following the same procedure as for photometric systematics. %
With both photometric systematics and redshift failures coming from one mock, we find small shifts in SGC RSD fits of $1.9\%$ for $\fsig$, $1.2\%$ for $\apar$ and $0.3\%$ for $\aper$, which are again not significant since larger shifts are seen in $279/999$, $257/999$ and $418/999$ EZ mocks (considering both tails: $498/999$, $463/999$ and $783/999$).
We therefore see no evidence for a systematic bias due to coupling between data measurements and the covariance matrix built from contaminated EZ mocks.

\section{RSD~+~BAO full best fit}
\label{app:full_best_fit}

We report all RSD~+~BAO best fit parameters to eBOSS ELG NGC~+~SGC in \Tab{results_full_rsdbao_data}.

\begin{table*}
\caption{Parameters for the eBOSS ELG combined RSD~+~BAO likelihood. We provide the best fit, as well as the mean and the median of the MCMC samples. Error bars are given around the best fit. The $\Delta \chi^{2} = 1$ errors in parameter $x$ are given by the $x$ values for which $\chi^{2}$ is increased by $1$ compared to the best fit, when minimising with respect to all other parameters. The $68\%$ interval for parameter $x$ corresponds to the smallest interval of $x$ which contains $68\%$ of the MCMC samples. We also quote the standard deviation of the MCMC samples. The $[16\%,84\%]$ interval is given by the $16\%$ and $84\%$ percentiles of the MCMC samples. Systematic uncertainties of \Tab{error_budget_rsdbao} are not included.}
\label{tab:results_full_rsdbao_data}
\centering
\begin{tabular}{lccccccc} 
\hline
parameter & best fit & mean & median & $\Delta \chi^{2} = 1$ & $68\%$ interval & standard deviation & $[16\%,84\%]$ interval \\
\hline
\hline
$\fsig$ & $0.289$ & $0.282$ & $0.281$ & ${}_{-0.066}^{+0.068}$ & $0.289_{-0.075}^{+0.060}$ & $0.067$ & $0.289_{-0.074}^{+0.061}$\\
$\apar$ & $1.085$ & $1.082$ & $1.089$ & ${}_{-0.107}^{+0.087}$ & $1.085_{-0.090}^{+0.104}$ & $0.103$ & $1.085_{-0.102}^{+0.094}$\\
$\aper$ & $0.941$ & $0.942$ & $0.940$ & ${}_{-0.034}^{+0.035}$ & $0.941_{-0.037}^{+0.036}$ & $0.061$ & $0.941_{-0.039}^{+0.035}$\\
$b_{1}^{\mathrm{NGC}}\sig$ & $0.779$ & $0.770$ & $0.774$ & ${}_{-0.066}^{+0.048}$ & $0.779_{-0.044}^{+0.042}$ & $0.046$ & $0.779_{-0.053}^{+0.035}$\\
$b_{1}^{\mathrm{SGC}}\sig$ & $0.795$ & $0.782$ & $0.787$ & ${}_{-0.079}^{+0.046}$ & $0.795_{-0.046}^{+0.038}$ & $0.045$ & $0.795_{-0.057}^{+0.029}$\\
$b_{2}^{\mathrm{NGC}}\sig$ & $-0.23$ & $-0.14$ & $-0.18$ & ${}_{-0.75}^{+0.97}$ & $-0.23_{-0.76}^{+0.77}$ & $0.72$ & $-0.23_{-0.66}^{+0.87}$\\
$b_{2}^{\mathrm{SGC}}\sig$ & $-0.1$ & $-0.01$ & $-0.02$ & ${}_{-1.0}^{+1.2}$ & $-0.06_{-0.85}^{+0.89}$ & $0.79$ & $-0.06_{-0.81}^{+0.93}$\\
$A_{g}^{\mathrm{NGC}}$ & $0.02$ & $0.04$ & $0.01$ & ${}_{-0.15}^{+0.24}$ & $0.02_{-0.17}^{+0.12}$ & $0.17$ & $0.02_{-0.13}^{+0.20}$\\
$A_{g}^{\mathrm{SGC}}$ & $-0.04$ & $0.02$ & $-0.04$ & ${}_{-0.15}^{+0.35}$ & $-0.04_{-0.16}^{+0.14}$ & $0.20$ & $-0.04_{-0.11}^{+0.26}$\\
$\sigma_{v}^{\mathrm{NGC}}$ & $2.53$ & $2.14$ & $2.23$ & ${}_{-0.93}^{+0.73}$ & $2.53_{-1.18}^{+0.72}$ & $0.94$ & $2.53_{-1.42}^{+0.54}$\\
$\sigma_{v}^{\mathrm{SGC}}$ & $3.05$ & $2.73$ & $2.81$ & ${}_{-0.82}^{+0.73}$ & $3.05_{-0.98}^{+0.66}$ & $0.88$ & $3.05_{-1.16}^{+0.51}$ \\
\hline
\end{tabular}
\end{table*}

\end{document}